\documentclass[11pt,3p,review,authoryear]{elsarticle}

\usepackage{amssymb, amsmath, amsfonts}
\usepackage{adjustbox} 
\usepackage{subcaption}
\usepackage{xurl}
\usepackage{graphicx}
\usepackage{textcomp}
\usepackage{eurosym}
\usepackage{booktabs}
\usepackage{multirow}
\usepackage{rotating}
\usepackage{comment}
\usepackage{pifont}
\usepackage{makecell}
\usepackage{dsfont}
\usepackage[table]{xcolor}

\usepackage{bm}

\newcommand{\rev}[1]{\textcolor{black}{#1}}

\journal{International Journal of Forecasting}

\begin{document}

\begin{frontmatter}






\title{\rev{Day-Ahead Electricity Price Forecasting Using Merit-Order Curves Time Series}}

\author[mox]{Guillaume Koechlin\corref{cor1}}
\cortext[cor1]{Corresponding author}
\ead{guillaume.koechlin@polimi.it}

\author[deng]{Filippo Bovera}

\author[mox]{Piercesare Secchi}

\affiliation[mox]{organization={MOX, Department of Mathematics, Politecnico di Milano},
            addressline={Piazza Leonardo da Vinci 32}, 
            city={Milano},
            postcode={20133}, 
            state={MI},
            country={Italy}}

\affiliation[deng]{organization={Department of Energy, Politecnico di Milano},
            addressline={Via Lambruschini 4a}, 
            city={Milano},
            postcode={20156}, 
            state={MI},
            country={Italy}}

\begin{abstract}
\rev{We introduce a general, simple, and computationally efficient functional data analysis framework for forecasting day-ahead supply and demand merit-order curves, and the resulting electricity market price. 
We conduct a rigorous empirical comparison on data from the Italian (GME), German (EPEX-DE-LU), and French (EPEX-FR) day-ahead markets over the 2023--2024 period, analyzing curve forecasting performance, price forecasting performance, and the relationship between the two. We find that strong curve forecasting performance does not necessarily translate into strong price forecasting performance, with important implications for curve model evaluation and selection when price forecasting is among the objectives.
We also show that this functional data representation approach consistently outperforms the original discretization-based approach of \citet{ziel_electricity_2016} on price forecasting across all three markets.
Finally, the proposed curve-based approach is competitive with state-of-the-art price-based models for two out of three markets (GME and EPEX-FR), and substantially improve accuracy during midday hours (when prices frequently drop due to the combined effect of high renewable generation and low demand) with MAE reductions of up to 27\% in those windows. For EPEX-DE-LU, however, price-based models retain a clear and significant advantage.}

\end{abstract}




\end{frontmatter}



\section{Introduction}
\label{intro}

\subsection{Context}

The rising integration of non-programmable renewable energy sources and the growing uncertainty surrounding fossil fuel supply have made electricity markets more volatile and difficult to forecast. As a result, the development of sophisticated and reliable market forecasting tools has become even more essential for all actors of the electric power industry.

\begin{figure}[h]
    \centering
    \includegraphics[width=0.9\linewidth]{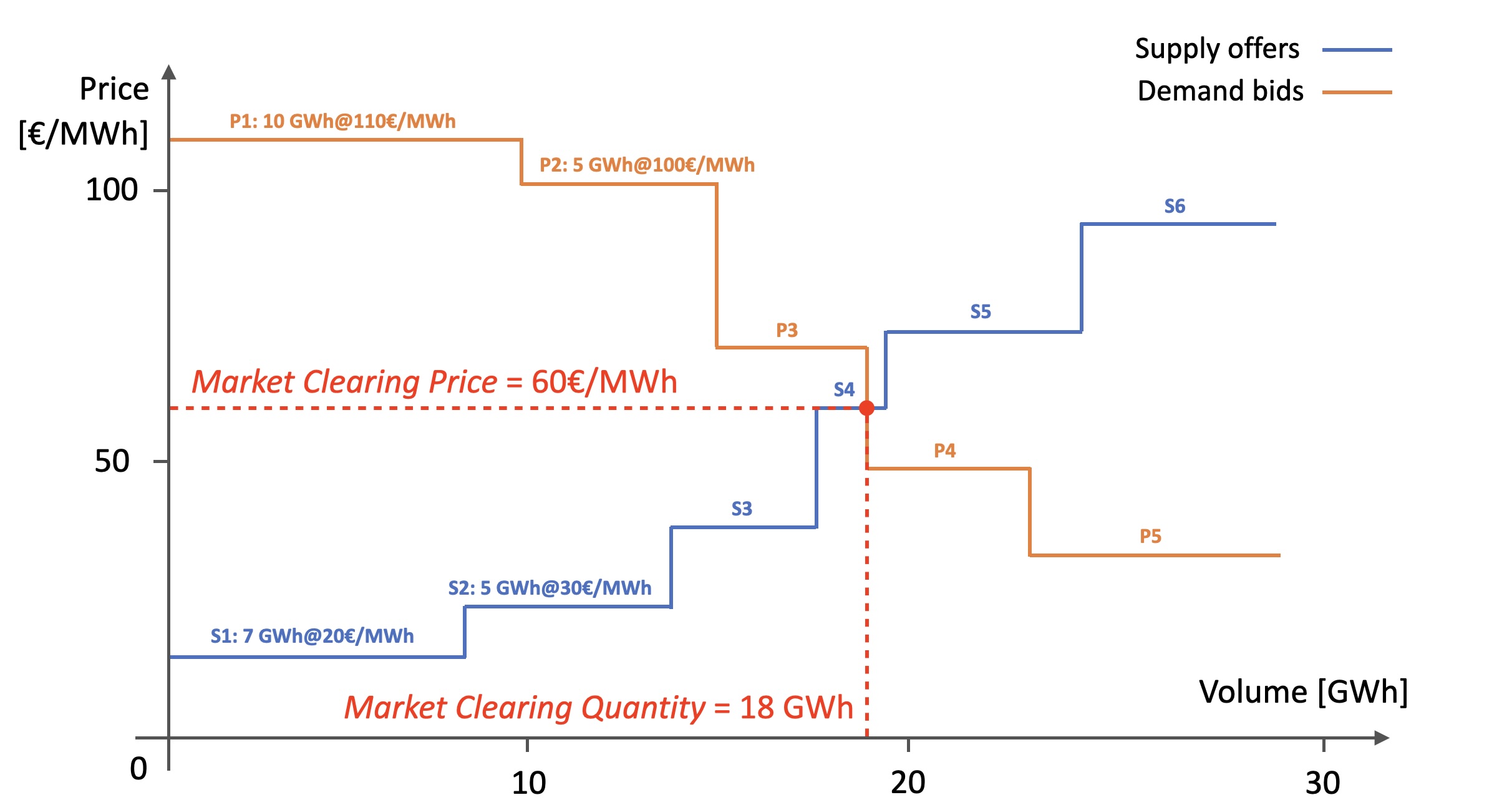}
    \caption{(\textit{Color optional}) Example of a double uniform price electricity auction with supply and demand merit-order curves.}
    \label{fig:moc}
\end{figure}

Most wholesale electricity trading in Europe takes place on day-ahead markets, where producers and retailers exchange power one day before delivery. These markets generally operate under a double uniform-price (pay-as-clear) auction mechanism, where the intersection of the supply and demand \textit{merit-order curves} sets the market clearing price for all accepted offers and bids. These curves are formed from the aggregation of individual supply offers and demand bids following the merit order, i.e. increasing price for supply and decreasing price for demand (Fig. \ref{fig:moc}). Offers (or bids) priced below (resp. above) the clearing price are accepted\footnote{The intersection point may split a supply offer or demand bid which in that case would be partially accepted} while the remainder are rejected. For market participants, accurately estimating the clearing price is therefore sufficient to anticipate whether an offer/bid will be accepted and at which price.

Because of this single-price property, a large body of research has focused on forecasting the day-ahead clearing price itself, a discipline known as \textit{electricity price forecasting} (EPF). The consensus is that data-driven approaches based on historical prices tend to outperform \textit{fundamental} (or structural) ones which try to model exact market mechanisms, especially for short-term hourly forecasts \citep{weron_electricity_2014}.
In particular, a highly competitive day-ahead EPF benchmark based on regularized autoregressive models and feedforward neural networks was consolidated by \citet{lago_forecasting_2021}. This framework is still state-of-the-art and has been widely employed in recent studies \citep{olivares_neural_2023, lipiecki_postprocessing_2024, oconnor_review_2025, cerasa_enhancing_2025}.

Nevertheless, the problem of forecasting the entire merit-order curves underlying price formation is gaining momentum in the academic community \citep{petropoulos_forecasting_2022}. The seminal work of \citet{ziel_electricity_2016} demonstrated the capacity of "data-driven fundamental" approaches based on historical auction data and their potential for generating competitive clearing price forecasts. Forecasting entire curves offers several advantages. For participants with market power, it enables precise quantification of how marginal changes in offers may influence the clearing price and allows realistic market simulations from which optimal bidding strategies can be derived \citep{pelagatti_supply_2013}. For price-takers and other actors exposed to the day-ahead price, curve-based models offer a more comprehensive view of market results, enhances the interpretability of price predictions, and may improve the quantification of price predictions' uncertainty. Additionally, complex nonlinear dynamics observed at the price level may become more tractable when modeled at the curve level.

\begin{table}[tb]
\centering
\begin{adjustbox}{max width=\textwidth, center}
\begin{tabular}{l l l l c}

\toprule
\textbf{Work} & \textbf{Market} & \textbf{Model typology} & \textbf{Dim. reduction} &
\textbf{EPF} \\

\midrule
\citet{pelagatti_supply_2013} & GME & VAR & FPCA &  \\
\citet{ziel_electricity_2016} & EPEX-DE-LU & VAR & Discretization & \ding{51} \\
\citet{canale_constrained_2016} & GME (Gas market) & Functional AR & \ding{55} &  \\
\citet{mestre_forecasting_2020} & GME & Functional AR & \ding{55} &  \\
\citet{kulakov_xmodel_2020} & EPEX-DE-LU & VAR & Discretization & \ding{51} \\
\citet{shah_forecasting_2020} & GME & Functional AR & \ding{55} & \ding{51} \\
\citet{guo_forecast_2021} & MISO & Neural Network & PCA &  \\
\citet{soloviova_efficient_2021} & GME & Functional AR & \ding{55} & \ding{51} \\
\citet{yildirim_supply_2023} & EPIAS & State-space & Parametrization &  \\
\citet{ciarreta_forecasting_2023} & OMIE & VAR & Parametrization & \ding{51} \\
\citet{tang_forecasting_2024} & ANEM & Neural Network & PCA, NMF, SDL &  \\
\citet{vivo_prediction_2024} & OMIE (Intraday market) & Neural Network & \ding{55} &  \\
\citet{ghelasi_hierarchical_2024} & EPEX-DE-LU & VAR & Discretization &  \\
\citet{li_clustering_2024} & ANEM, OMIE, PJM & k-NN & \ding{55} &  \\
\citet{ghelasi_datadriven_2025} & EPEX-DE-LU & Fundamental & \ding{55} & \ding{51} \\
\citet{sinha_demand_2025} & GME & Neural Network & Autoencoder &  \\
\citet{li_predicting_2025} & OMIE & ARIMA, Functional AR & FPCA & \ding{51} \\

\midrule
\rev{\textit{This work}} & \rev{GME, EPEX-DE-LU, EPEX-FR} & \rev{VAR} & \rev{FPCA} & \rev{\ding{51}} \\

\bottomrule
\end{tabular}
\end{adjustbox}
\caption{Tabular view of (main) contributions addressing merit-order curves forecasting, sorted by publication year.}
\label{tab:works}
\end{table}

\subsection{\rev{Prior work}}

\noindent \rev{Table \ref{tab:works} provides an overview of main contributions on the topic. A central aspect of the problem regards the fact that merit-order curves are essentially infinite-dimensional functional objects, in the sense that they represent quantities that vary over a continuum. It is therefore necessary to consider either an appropriate statistical framework, e.g. functional data analysis (FDA) \citep{ramsay_functional_2005, ferraty_nonparametric_2006} or a finite-dimension representation of the curves to which classic multivariate time series methods can be applied.}

\rev{Within the FDA approach, several generalizations of the autoregressive (AR) models \citep{box_time_1994} to the functional data setting have been employed: Functional AR (FAR) models \citep{canale_constrained_2016, shah_forecasting_2020, soloviova_efficient_2021, li_predicting_2025}, functional SARMA \citep{mestre_forecasting_2020} and nonparametric FAR models \citep{shah_forecasting_2020}. Methodological reference on functional time series include \citet{bosq_linear_2000} and \citet{hormann_7_2012}. The advantage of the FDA approaches, beyond elegantly tackling the problem from a mathematical standpoint, is that they avoid the information loss inherent in finite dimension representations. Their main drawback, however, lies in the limited theoretical development and the lack of efficient, easy-to-use, open-access implementations of the models proposed in the literature \citep{aue_prediction_2015}.}

\rev{Aside of the FDA framework, we distinguish essentially three strategies to address the dimensionality problem: discretization, parametrization and dimensionality reduction. For all three, the overall workflow is similar: (i) transform the curves into a finite-dimension vector form, (ii) forecast this vector using multivariate methods --- e.g., a vector autoregressive (VAR) model --- and (iii) invert the transformation to recover the predicted curves.}

\rev{Although any software ultimately relies on discrete representations, we refer to the \emph{discretization} approach when the vector representation consists of curve evaluations over a grid of reasonable size (no more than a few dozen points), enabling multivariate modeling. The functional form is then recovered using a chosen interpolation method. This strategy was adopted in the seminal work of \citet{ziel_electricity_2016} and its extensions in \citet{kulakov_xmodel_2020} and \citet{ghelasi_hierarchical_2024}. \textit{Parametrization} refers to methods that assume a parametric form (e.g. linear, logarithmic or polynomial) for the curves, with the finite-dimensional representation given by the fitted parameters; see for instance \citet{yildirim_supply_2023} and \citet{ciarreta_forecasting_2023}. Finally, \textit{dimensionality reduction} refers to approaches where the finite-dimension representation is obtained by means of a low-rank approximation of the data through standard matrix factorization techniques \citep{guo_forecast_2021, tang_forecasting_2024}, neural networks \citep{sinha_demand_2025} or functional PCA \citep{pelagatti_supply_2013, li_predicting_2025}.}

\medskip

\rev{Functional PCA (FPCA) is a well-known, simple, and effective method widely used when forecasting functional time series \citep{hyndman_robust_2007, hyndman_forecasting_2009, shang_functional_2013, aue_prediction_2015}. Although FPCA can be viewed as a standard dimensionality reduction technique, it is more appropriately regarded as a fundamental method within functional data analysis (FDA). Indeed, \citet{aue_prediction_2015} show that a vector autoregressive model of order $p$ -- VAR($p$) -- applied to the $K$-dimensional vector representation given by the scores on the $K$ first functional principal components is asymptotically equivalent to a functional autoregressive model of the same order --- FAR($p$), as defined in \citet{bosq_linear_2000}, when $K \rightarrow\infty$. FAR models may even be estimated via this method, for instance in the \texttt{far} R package \citep{shah_forecasting_2020}. Therefore, \citet{aue_prediction_2015} argue that forecasting the functional time series by forecasting the vector time series of the FPC scores (given that it provides a satisfying dimensionality reduction) with any multivariate forecasting method --- being a VAR model or not --- is theoretically justified and is easier than devising complex FDA models. Following Aue et al.'s guidance, we adopt this approach in this work.}

\subsection{Knowledge gap}

Given the absence of a unified evaluation framework (\rev{like that in EPF}) and the scarcity of direct and fair comparisons between proposed approaches, identifying top-performing merit-order curves forecasting techniques \rev{on both curve and clearing price prediction} remains challenging.
In particular, aside from a few direct extensions \citep{kulakov_xmodel_2020, ghelasi_hierarchical_2024} and despite a large number of subsequent contributions, no alternative method has been conclusively shown to outperform the original approach of \citet{ziel_electricity_2016}. We identified only one study, \citet{yildirim_supply_2023}, that provides a direct comparison, though limited to supply curve forecast accuracy. Their results suggest that their method and that of Ziel and Steinert perform similarly. A comparison of curve forecast performance is also reported in \citet{sinha_demand_2025}, although the authors \rev{(i)} state that they employ a modified version of Ziel and Steinert’s model without providing further details \rev{(ii) use a single non-standard error metric of their own design}. \rev{None of these two studies, however, test for clearing price forecasting performance.}
As a result, Ziel and Steinert’s framework remains the de facto state-of-the-art, and whether it can be consistently improved remains an open question.

\medskip

\rev{An additional issue, related to the first, concerns the generalization of proposed methods across different markets. As shown in Table \ref{tab:works}, all works except one test their models on a single market, which provides no guarantee of performance on arbitrary markets. The only multi-market analysis is provided by \citet{li_clustering_2024}, although it is limited to supply curves.}

\medskip

\rev{Finally,} beyond the problem of curve forecasting itself, an essential question remains unanswered: are merit-order curve forecasting models competitive with state-of-the-art price-centric EPF models in terms of clearing price prediction accuracy?

While Ziel and Steinert demonstrated ten years ago that their approach outperformed existing price-based models for point forecasting, this result may no longer hold in light of recent advances in EPF.
Since then, several studies have assessed the effectiveness of curve-based approaches for clearing price prediction by comparing them with price-based methods; however, the latter are often overly simplistic and not representative of top-performing EPF models. In addition, the rigorous forecast evaluation practices recommended by \citet{lago_forecasting_2021}, such as daily recalibration, are seldom followed.
Furthermore, none of the post-2021 contributions compare curve-based price forecasts against the benchmark models of Lago et al., despite their availability through an open-access software implementation (the \texttt{epftoolbox} Python package).
For instance, \citet{shah_forecasting_2020} compare their curve-based forecasts only against basic autoregressive models without exogenous covariates, even though variables such as load and renewable generation forecasts are known to be essential for achieving top performance \citep{weron_electricity_2014}. Similar limitations apply to \citet{soloviova_efficient_2021}, \citet{ciarreta_forecasting_2023}, and \citet{li_predicting_2025}. \citet{ghelasi_datadriven_2025} consider an "expert" autoregressive model with exogenous covariates, but do not include regularized, parameter-rich autoregressive models with cross-hour dependence, despite evidence that such models consistently outperform expert ARX specifications \citep{ziel_dayahead_2018}.


\subsection{Contributions}



\medskip

\noindent In this context, this work has three main objectives:

\begin{enumerate}
    \item Introduce a general, simple, and computationally efficient \rev{FDA framework combining FPCA with regularized high-dimensional VARX models, for predicting supply and demand merit-order curves, from which accurate electricity price forecasts can be derived, on any day-ahead market.}
    \item \rev{Rigorously compare this FDA approach with the original state-of-the-art discretization-based approach of \citet{ziel_electricity_2016}.}
    \item Rigorously compare the price forecasting performance of the proposed \textit{curve-based} models, i.e. derived from merit-order curves predictions, with that of state-of-the-art \textit{price-based} models, which directly model the (clearing) price.
\end{enumerate}

The empirical comparison is run on the \rev{Italian (GME), German (EPEX-DE-LU) and French (EPEX-FR)} day-ahead markets, during the 2023-2024 period.

\bigskip


The remainder of this paper is structured as follows: section \ref{sec:methods} details the forecasting and testing methodology, section \ref{sec:results} presents the empirical results, and section \ref{sec:conclusion} provides discussion and concluding remarks.

\section{Methodology}
\label{sec:methods}

\subsection{Curves definition}
\label{sec:curves}

Formally, a merit-order curve represents price as a function of quantity $q \rightarrow P(q)$ and is defined from a set of \rev{\textit{bids}} characterized by a price and a quantity $(p_1, q_1), (p_2, q_2),\dots, (p_n, q_n)$ sorted following the merit-order, i.e., increasing order of price  $p_1 \leq p_2 \leq \dots \leq p_n$ for the supply curve, decreasing order of price $p_1 \geq p_2 \geq \dots \geq p_n$ for the demand curve. Mathematically, a merit-order curve can be defined as follows:
\begin{equation*}
P(q) =
p_k, \quad \text{if}\; \sum_{i=0}^{k-1}q_i \leq q < \sum_{i=0}^{k}q_i, \quad k = 1, \dots, n
\end{equation*}
where we set $q_0=0$, for notational convenience. For supply (or demand), $P(q)$ is said to be the marginal price associated to a total offered (resp. demanded) quantity $q$.
    




From the definition we can notice that merit-order curves are \rev{right}-continuous monotone step functions, \rev{non-decreasing for supply and non-increasing for demand}.
A direct consequence is that $q \rightarrow P(q)$ admits a \rev{\textit{right-continuous generalized inverse}} $p \rightarrow Q(p)$, which is itself a \rev{right-continuous} monotone step function. It is interesting to note that this \rev{"pseudo-inverse"} has an easier mathematical notation and interpretation. Indeed, its expression matches (for supply curves) the unnormalized\footnote{Given that the $q_i$'s don't sum to 1.} cumulative distribution function (CDF) of a discrete distribution over the support $\{p_1, \dots, p_n\}$. For demand, the only difference is that it is the unnormalized reverse CDF (or survival function). Using the superscript $(s)$ for supply and $(d)$ for demand, we have:
$$Q^{(s)}(p) = \sum_{i=1}^nq_i \mathds{1}_{\{p_i \leq p\}}\, , \quad
  Q^{(d)}(p) = \sum_{i=1}^nq_i \mathds{1}_{\{p_i \geq p\}} $$

\noindent That is, $Q(p)$ is the quantity that is offered (or demanded) at a price below (resp. above) $p$. Note that the definition above does not require the sequence of offers (or bids) to be price-ordered.
It is more intuitive to work with the quantity function $Q$ as it represents a more familiar mathematical object than the price function $P$. The quantity function simply tells us about the total offered quantity and how this offered quantity is distributed on the price domain. It also conveniently changes the perspective on the curves variability which is mainly attributed to their inelastic component: seasonal power demand variations for demand curves and variable renewable energy generation for supply curves.

From now on, unless otherwise specified, references to supply or demand curves will refer to the quantity function.


\subsection{Curves representation} \label{subsec:repr}

The goal is to model and forecast the time series of curves $Q^{(s)}_1, Q^{(s)}_2, \dots, Q^{(s)}_T$ and $Q^{(d)}_1, Q^{(d)}_2, \dots, \\ Q^{(d)}_T$ over a time period composed of $T$ hourly intervals.
Functional data analysis \citep{ramsay_functional_2005, ferraty_nonparametric_2006} is a framework that extends traditional statistical methods to handle data that vary over a continuum, such as curves, surfaces, or other functional forms.
In order to forecast these functional time series, we follow the guidance of \citet{aue_prediction_2015} that recommend the use of \textit{functional principal component analysis} (FPCA) to derive an uncorrelated vector representation of the curves and apply standard multivariate time series forecasting methods to this representation, finally obtaining a curves forecast using the Karhunen-Loève (KL) expansion \citep[pp.~37--43]{horvath_inference_2012}. FPCA is a dimensionality reduction technique for functional data, extending classical PCA to infinite-dimensional spaces. It decomposes a set of observed functions into an orthonormal basis that captures the dominant modes of variations in the data. This decomposition yields the truncated KL expansion:
$$Q_t(p) \approx \overline{Q}(p) + \sum_{k=1}^K \beta_{tk} \xi_k(p)$$
where  $\overline{Q}(p)$ is the mean function,  $\xi_1(p), \xi_2(p), \dots, \xi_K(p)$ are the first $K$ \textit{functional principal components} (FPCs) and $\beta_{t1}, \beta_{t2}, \dots, \beta_{tK}$ are the \textit{scores} of observation $t$ on the first $K$ FPCs.
Intuitively, the FPCs can be seen as specific features common to all curves while the scores are measures of how pronounced are these features in a specific curve. The perfect equality holds when $K=\infty$ but in practice a satisfying approximation can be found with a finite $K$, typically chosen -- as in traditional PCA -- as the elbow point of the scree plot or such that the ratio of explained variance surpasses a certain threshold \citep[pp.~444--447]{johnson_applied_2007}.

Since FDA assumes smooth functions, the estimated FPCs should also be smooth functions. This smoothness can be achieved either by incorporating a roughness penalty directly into the FPCA estimation procedure or by pre-smoothing the functional data curves prior to applying FPCA. We adopt the latter approach for computational simplicity, utilizing kernel smoothing with the \textit{Nadaraya-Watson} kernel estimator \citep[p.~71]{wasserman_all_2006}. The bandwidth parameter of the kernel function, which controls the level of smoothness, is chosen using the generalized cross-validation criterion \citep[p.~97]{ramsay_functional_2005}.


\medskip
In our case, we would like to account for possible dependence between the supply and demand curves time series. To do so, we perform FPCA separately for the supply and demand curves and by concatenating the $K_s$ scores for the supply curves with the $K_d$ scores of the demand curve, we get a $K=K_s + K_d$ -dimensional vector representation $\mathbf{y}_1, \dots, \mathbf{y}_T$ of the paired functional series $(Q^{(s)}_t, Q^{(d)}_t)_{t \in \{1, \dots, T\}}$, with 
\begin{equation*}
    \mathbf{y}_t =
\begin{bmatrix}
\beta^{(s)}_{t1}, \beta^{(s)}_{t2}, \dots, \beta^{(s)}_{tK_s}, \, \beta^{(d)}_{t1}, \beta^{(d)}_{t2}, \dots, \beta^{(d)}_{tK_d}
\end{bmatrix}^\top
\end{equation*}



We forecast $\mathbf{y}_t$ with some multivariate forecasting model (detailed in the next section), with possibly a set of exogenous covariates $\mathbf{x}_t$. We then inverse-transform any
$h$-step-ahead forecast $\hat{\mathbf{y}}_{t+h}$ using the truncated KL expansion to get a forecast for the curves pair $(\hat{Q}^{(s)}_{t+h}, \hat{Q}^{(d)}_{t+h})$.

\medskip

It can happen that curves predictions are not perfectly monotonic. Though preliminary experiments suggested the occurence of little deviations only and low impact on clearing price predictions, it is still desirable to respect this structural constraint. To do so, we post-process the curves predictions using isotonic regression with the pool-adjacent violators algorithm \citep{leeuw_isotone_2009}. Being a highly efficient $O(n)$ algorithm, it has a negligible impact on computation times.

\subsection{Forecasting the curves' vector representation}

We consider here the day-ahead forecasting problem, regarding the vector hourly time series as 24 separate daily time series, one for each hour of the day. This is common practice in EPF as the 24 hours of the next day are simultaneously settled the day before. In addition, market dynamics vary a lot depending on the hour of the day, justifying to treat them separately. For a detailed motivation of this choice, the reader can refer to \citet{ziel_dayahead_2018}. We hence change the time indexing of $\mathbf{y}_t$ to consider the value at day $d$ and hour $h$, $\mathbf{y}_{d,h}$ and we solve 24 one-step-ahead forecasting problems. 

\medskip

We consider four variants of the popular parameter-rich ARX model estimated with LASSO widely used in the context of electricity price forecasting \citep{ziel_forecasting_2016, uniejewski_automated_2016, ziel_dayahead_2018, lago_forecasting_2021, uniejewski_regularization_2024}. First this framework was shown to be highly performing for price forecasting and second, it was used in \citet{ziel_electricity_2016}.
Specifically, we consider two \textit{univariate} models treating each component of $\mathbf{y}_{d,h}$ separately as suggested by \cite{hyndman_forecasting_2009} -- the concurrent ARX and the full ARX (equations \ref{eq:arx} and \ref{eq:farx}) -- and two \textit{multivariate} models jointly modeling the vector $\mathbf{y}_{d,h}$, the concurrent VARX and the full VARX (equations \ref{eq:varx} and \ref{eq:fvarx}). The term \textit{concurrent} means that the target variable at hour $h$ is influenced \rev{only by past values observed at the same hour $h$ on previous days. Conversely, \textit{full} means that the target variable is also influenced by past values observed at different hours $j \neq h$ on previous days.} The comparison between the univariate and the multivariate approaches allow to test the added value of modeling the cross-dependence between scores while the comparison between the concurrent and full approaches allow to test the added value of modeling the cross-dependence between hours.

The *AR* part refers to the fact that lagged values are used to predict future values. In our case, similarly to state-of-the-art EPF 
models \citep{lago_forecasting_2021} we include lags 1, 2, 3 and 7, considering therefore information up to 7 days before. Additionally, each model has the suffix *X which means that it makes use of exogenous variables which refer to day $d$ and available on $d-1$ (i.e., they are known in anticipation or they are themselves one-day-ahead forecasts). This set of $r$ exogenous variables is represented by the $r$-dimensional vector $\mathbf{x}_{d,h}$. Again, like \citet{lago_forecasting_2021}, we also include lags 1 and 7 of these exogenous variables, and a three-dimensional vector $\mathbf{z}_{d}$ of dummy variables flagging the day type: Mondays, Working days (from Tuesday to Friday), Saturdays and Holidays (Sundays and bank holidays)\footnote{Note that this is slightly different from \citet{lago_forecasting_2021} who consider one dummy for each day of the week but similar to \citet{mestre_forecasting_2020} and addresses the calendar effects issues raised by \citet{ziel_electricity_2016}.}.

Finally, still following Lago et al.'s guidance, the $L^1$-regularization parameter $\lambda$ is selected as that minimizing the Akaike information criterion (AIC), exploiting the least angle regression (LARS) algorithm \citep{efron_least_2004} for the parameter search. Once $\lambda$ is selected, the model is estimated using the traditional coordinate descent algorithm \citep{tibshirani_regression_1996}.


\subsubsection*{Concurrent ARX \textup{(\textbf{ARX})}}
\noindent The concurrent ARX
models each component $y_{d,h}$ of $\mathbf{y}_{d,h}$ as:
\begin{equation} \label{eq:arx}
\begin{split}
y_{d,h} &= \phi_{1,h}y_{d-1,h} + \phi_{2,h}y_{d-2,h} + \phi_{3,h}y_{d-3,h} + \phi_{7,h}y_{d-7,h} \\
&\quad + \boldsymbol{\beta}^{\top}_{0,h} \mathbf{x}_{d,h} + \boldsymbol{\beta}_{1,h}^{\top} \mathbf{x}_{d-1,h} +  \boldsymbol{\beta}_{7,h}^{\top} \mathbf{x}_{d-7,h} + \boldsymbol{\theta}^{\top}_{h} \mathbf{z}_{d} + \epsilon_{d, h}
\end{split}
\end{equation}
where $\phi_{\cdot,h}$ are the autoregressive coefficients,  \rev{$\boldsymbol{\beta}_{\cdot, h}$ the $r$-dimensional vectors of lagged exogenous variables coefficients and $\boldsymbol{\theta}_{h}$ the 3-dimensional vector of dummy variables coefficients.}


\subsubsection*{Full ARX \textup{(\textbf{fARX})}}
\noindent The full ARX
, very similar to its homonym in \citet{uniejewski_automated_2016}, models each component $y_{d,h}$ of $\mathbf{y}_{d,h}$ as:
\begin{equation} \label{eq:farx}
\begin{split}
y_{d,h} &= \sum_{j=1}^{24}\phi^{(j)}_{1,h}y_{d-1,j} + \sum_{j=1}^{24}\phi^{(j)}_{2,h}y_{d-2,j} + \sum_{j=1}^{24}\phi^{(j)}_{3,h}y_{d-3,j} + \sum_{j=1}^{24}\phi^{(j)}_{7,h}y_{d-7,j} \\
&\quad + \boldsymbol{\beta}^{\top}_{0,h} \mathbf{x}_{d,h} + \boldsymbol{\beta}_{1,h}^{\top} \mathbf{x}_{d-1,h} +  \boldsymbol{\beta}_{7,h}^{\top} \mathbf{x}_{d-7,h} + \boldsymbol{\theta}^{\top}_{h} \mathbf{z}_{d} + \epsilon_{d, h}
\end{split}
\end{equation}
where $\phi_{\cdot, h}^{(j)}$ is the linear effect of the lagged hour $j$ on the hour $h$.


\subsubsection*{Concurrent VARX \textup{(\textbf{VARX})}}
\noindent The concurrent VARX models the full vector $\mathbf{y}_{d,h}$ as:
\begin{equation} \label{eq:varx}
    \begin{split}
    \mathbf{y}_{d,h} &= \boldsymbol{\Phi}_{1,h}\mathbf{y}_{d-1,h} + \boldsymbol{\Phi}_{2,h}\mathbf{y}_{d-2,h} +
    \boldsymbol{\Phi}_{3,h}\mathbf{y}_{d-3,h} + \boldsymbol{\Phi}_{7,h}\mathbf{y}_{d-7,h} \\
    &\quad + \mathbf{B}_{0, h}\mathbf{x}_{d,h} + \mathbf{B}_{1, h}\mathbf{x}_{d-1,h} +
    \mathbf{B}_{7, h}\mathbf{x}_{d-7,h} + \boldsymbol{\Theta}_{h} \mathbf{z}_{d} + \boldsymbol{\epsilon}_{d, h}
    \end{split}
\end{equation}
Where $\boldsymbol{\Phi}_{\cdot,h}$ are $K\times K$ matrices of autoregressive coefficients and $\mathbf{B}_{\cdot, h}$ and $\boldsymbol{\Theta}_{\cdot, h}$ are the $K \times r$ and $K \times 3$ matrices of (lagged) exogenous and dummy variables coefficients, respectively.
Contrarily to the two previous models, we allow for cross-dependence -- described by the off-diagonal coefficients of $\boldsymbol{\Phi}_{\cdot,h}$ -- between the components of $\mathbf{y}_{d,h}$.


\subsubsection*{Full VARX \textup{(\textbf{fVARX})}}

\noindent The full VARX models the full vector $\mathbf{y}_{d,h}$ as:
\begin{equation} \label{eq:fvarx}
    \begin{split} 
    \mathbf{y}_{d,h} &= \sum_{j=1}^{24}\boldsymbol{\Phi}^{(j)}_{1,h}\mathbf{y}_{d-1,j} + \sum_{j=1}^{24}\boldsymbol{\Phi}^{(j)}_{2,h}\mathbf{y}_{d-2,j} +
    \sum_{j=1}^{24}\boldsymbol{\Phi}^{(j)}_{3,h}\mathbf{y}_{d-3,j} +
    \sum_{j=1}^{24}\boldsymbol{\Phi}^{(j)}_{7,h}\mathbf{y}_{d-7,j} \\
    &\quad + \mathbf{B}_{0, h}\mathbf{x}_{d,h} + \mathbf{B}_{1, h}\mathbf{x}_{d-1,h} +
    \mathbf{B}_{7, h}\mathbf{x}_{d-7,h} + \boldsymbol{\Theta}_{h} \mathbf{z}_{d} + \boldsymbol{\epsilon}_{d, h}
    \end{split}
\end{equation}
Where we impose the restriction that the off-diagonal terms of $\boldsymbol{\Phi}^{(j)}_{\cdot,h}$ must be zero whenever $j \neq h$, that is, cross-dependence is allowed between hours, but only within the same component\footnote{In that sense, "\textit{semi}-full VARX" could be a more precise denomination but we keep "full VARX" for simplicity.}. This restriction is imposed to have a manageable number of parameters. Indeed, without this condition we would have $4 \times [24\times K] + 3r + 3 = 96K + 3r + 3$ parameters for each component of $\mathbf{y}_{d,h},$ while with this condition the number of parameters is only $4 \times [23 + K] + 3r + 3 = 4K + 3r + 95$. For instance, for $K=10$ and $r=4$, the condition implies 147 parameters instead of  975.

\bigskip

\rev{To make the equivalence proven by \citet{aue_prediction_2015} explicit for our case, we show in \ref{sec:far} that these models can be reformulated as \textit{univariate} --- for \textbf{ARX} and \textbf{fARX} --- and \textit{bivariate} --- for \textbf{VARX} and \textbf{fVARX} --- functional autoregressive (FAR) models.}


\medskip

\label{subsec:bench}
\subsection{Benchmark curves and price forecasting models}

\subsubsection*{Naive model}
\noindent The naive model \rev{(\textbf{Naive}) \citep{weron_electricity_2014}} works as follows:
\begin{equation}
    \hat{Q}^{\text{naive}}_{d,h} =
\begin{cases}
    Q_{d-7,h} & \text{if } d \text{ corresponds to a Monday, Saturday or Sunday}  \\
    Q_{d-1,h} & \text{otherwise}
\end{cases}
\end{equation}

\subsubsection*{\citet{ziel_electricity_2016}}

\noindent While our approach differs from that of \citet{ziel_electricity_2016} regarding the curves vector representation, the multivariate forecasting techniques applied to that representation are essentially the same. For that reason, to guarantee a fair comparison between the two approaches, we apply the models detailed in the previous section to (i) our vector of supply and demand FPCA scores (ii) the vector of supply and demand price class volumes obtained with the method described in \citet{ziel_electricity_2016}. We invite the reader to refer to the authors' paper for a rigorous and detailed description of this procedure (as well as for curves reconstruction), which we refer to as \textit{Ziel-Steinert Transformation} (\textbf{ZST}), as opposed to our \textit{functional principal component analysis} (\textbf{FPCA}).

\medskip

\begin{figure}[h]
    \centering
    \includegraphics[width=0.6\linewidth]{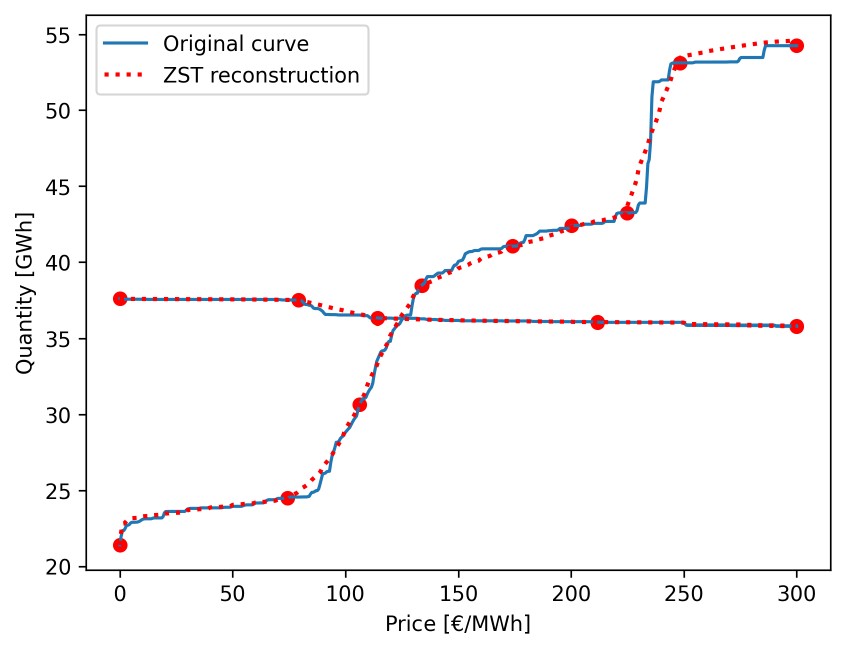}
    \caption{(\textit{Color optional}) Example of \textbf{ZST} reconstruction for a supply-demand curves pair with $K_s=9$ dimensions for supply and $K_d=5$ for demand. The vector representation is given by the first-order differences of the $y$-coordinates of the red dots. The fixed price grid corresponds to their $x$-coordinates.}
    \label{fig:zst}
\end{figure}

Briefly, \textbf{ZST} consists of taking the first-order differences of discrete evaluations of the quantity curve on a specific price grid.
This price grid is constructed by defining an equispaced \textit{quantity} (or volume) grid and transforming it into a non-uniform \textit{price} grid via the mean price curve $\bar{P}(q)$. For inverse transformation, the differences are cumulatively summed to recover the discrete values and the continuous curves are reconstructed using interpolation based on the mean quantity curve\footnote{\rev{Note that we set the $R_S(P)$ and $R_S(P)$ Bernoulli variables, appearing in eq. 10 and 11 of the original paper, to 1 with a probability threshold of 0 instead of 1/12. This approach looks more natural to us as the authors do not explain the reasons behind this choice (which may suggest it is tuned on their data) and simply setting $R_S(P)$ and $R_D(P)$ to 1 makes the reconstruction more interpretable: the prices are simply reconstructed by rescaling the mean curve with the total class quantity. In any case, the effect is most likely negligible.}}. Fig. \ref{fig:zst} shows an example of a transformed and reconstructed supply-demand curves pair.

\subsubsection*{Price-based point forecasting models}
\noindent We consider three benchmark \textit{price-based} clearing price forecasting models: \textbf{ARX}, \textbf{fARX} and the Lasso-Estimated AutoRegressive model (\textbf{LEAR}) of \citet{lago_forecasting_2021}. The first two are already described in the previous section (equations \ref{eq:arx} and \ref{eq:farx}). When both curve-based and price-based models are involved, to distinguish the clearing price forecasting model deriving from the curves model (\textit{curve-based}) from that directly forecasting the price (\textit{price-based}), we prefix the curve-based models with the curves representation used, e.g, \textbf{FPCA-ARX} or \textbf{ZST-ARX}.

Finally the \textbf{LEAR} model (equation \ref{eq:lear}) is very similar to \textbf{fARX} with the difference that it adds cross-hour dependence in the exogenous variables effects and accepts only two of them $\mathbf{x}_{d, h}=[x^1_{d,h}, x^2_{d,h}]^\top$ (usually \rev{and for this paper}, the load and combined wind-solar generation day-ahead forecasts):
\begin{equation} \label{eq:lear}
\begin{split}
y_{d,h} &= \sum_{j=1}^{24}\phi^{(j)}_{1,h}y_{d-1,j} + \sum_{j=1}^{24}\phi^{(j)}_{2,h}y_{d-2,j} + \sum_{j=1}^{24}\phi^{(j)}_{3,h}y_{d-3,j} + \sum_{j=1}^{24}\phi^{(j)}_{7,h}y_{d-7,j} \\
&\quad + \sum_{j=1}^{24}\boldsymbol{\beta}^{(j)\top}_{0,h} \mathbf{x}_{d,j} + \sum_{j=1}^{24}\boldsymbol{\beta}^{(j)\top}_{1,h} \mathbf{x}_{d-1,j} +
\sum_{j=1}^{24}\boldsymbol{\beta}^{(j)\top}_{7,h} \mathbf{x}_{d-7,j} +
 \boldsymbol{\theta}_{h}^\top \mathbf{z}_{d} + \epsilon_{d, h}
\end{split}
\end{equation}

As in \citet{lago_forecasting_2021}, for all three models, prices are transformed with median-based scaling and the \textit{area hyperbolic sine} variance stabilizing transformation.

\subsection{Forecasts evaluation}

\medskip
\subsubsection*{Curve forecasts}
\noindent We measure curves prediction accuracy with \rev{standard functional regression error metrics: \textit{pointwise} mean absolute error (MAE), root mean squared error (RMSE) and mean absolute percentage error (MAPE)}:
\begin{equation*}
\rev{
\mathrm{MAE}(p)
=
\frac{1}{T}
\sum_{t=1}^{T}
\left|
\hat{Q}_{t}(p)-Q_{t}(p)
\right|
}
\end{equation*}

\begin{equation*}
\rev{
\mathrm{RMSE}(p)
=
\left[
\frac{1}{T}
\sum_{t=1}^{T}
\bigl(
\hat{Q}_{t}(p)-Q_{t}(p)
\bigr)^2
\right]^{1/2}
}
\end{equation*}

\begin{equation*}
\rev{
\mathrm{MAPE}(p)
=
\frac{100}{T}
\sum_{t=1}^{T}
\left|
\frac{\hat{Q}_{t}(p)-Q_{t}(p)}
{Q_{t}(p)}
\right|
}
\end{equation*}
\rev{and their integration, called \textit{functional} (e.g., Functional MAE or FMAE), over the price domain $\mathcal{D}=\left[p_{ \text{min}}, p_{ \text{max}}\right]$:}
\begin{equation*}
\rev{
    \mathrm{FMAE}=\frac{1}{\left| \mathcal{D}\right|}\int_{\mathcal{D}}\mathrm{MAE}(p)dp\ ,
    \quad
    \mathrm{FRMSE}=\frac{1}{\left| \mathcal{D}\right|}\int_{\mathcal{D}}\mathrm{RMSE}(p)dp\ ,
    \quad
    \mathrm{FMAPE}=\frac{1}{\left| \mathcal{D}\right|}\int_{\mathcal{D}}\mathrm{MAPE}(p)dp
}
\end{equation*}

\rev{Similarly to EPF literature \citep{lago_forecasting_2021}, we also consider the \textit{relative} FMAE (rFMAE) in order to have a relative performance measure --- beyond FMAPE --- which can be compared across markets.}
\begin{equation*}
\rev{
    \mathrm{rFMAE}=\frac{\mathrm{FMAE}}{\mathrm{FMAE}_{\text{naive}}}
}
\end{equation*}
\rev{where $\mathrm{FMAE}_{\text{naive}}$ is the FMAE of a \textit{naive} model (in our case the \textbf{Naive} model defined in section \ref{subsec:bench}).}

Additionally, to test whether a model significantly outperforms another, we run the \textit{Diebold-Mariano} (DM) test \citep{diebold_comparing_1995} \rev{at the \textit{day} level on the daily average \rev{$L_1$}-norm of functional errors $\hat{Q}_t-Q_t$, i.e., considering the FMAE across the 24 hours of the day}\footnote{Considering daily forecast errors \rev{obviously results in a loss of power} but avoids the problem of the strong intraday error autocorrelation, typical in day-ahead price forecasting, which would violate DM tests assumptions \citep{weron_electricity_2014}.}:
\begin{equation*}
\rev{
    L_{d} = \frac{1}{24}\sum_{h=1}^{24}\frac{1}{\left| \mathcal{D}\right|}\int_{\mathcal{D}}\left| \hat{Q}_{d,h}(p)-Q_{d,h}(p) \right|dp
}
\end{equation*}
Two one-sided tests are run for each pair of models.

\bigskip
\subsubsection*{\rev{Clearing price forecasts}}
\noindent \rev{We evaluate clearing price predictions with the MAE, RMSE and rMAE (as defined in \citet{lago_forecasting_2021} and above in its functional version).}

\rev{
Similarly to the curves forecasts, we perform DM tests at the day level for clearing price forecasts considering absolute ($\ell_1$) errors. We run the test on the daily average errors (as for curve forecasts) but we also run, for each pair of models, 24 separate tests for each hour of the day. This approach is similar to that of \citet{lago_forecasting_2021} and enables inference at the hour level.}

\subsection{Daily recalibration}
\label{subsec:recal}

As advised by \citet{lago_forecasting_2021} and in order to simulate a live forecasting setting, the entire modeling pipeline is recalibrated on a daily basis using a rolling window of 364 days (52 weeks), since $\approx1$ year appears to the best single-window choice for EPF according to \citet{marcjasz_selection_2018} and \citet{hubicka_note_2019}. 
This means that for every day $d$ of the testing period:

\begin{enumerate}[(1)]
    \item We extract the vector representation of the $364 \times 24=8744$ curves pairs observed on $d-1, \ d-2,  \dots, \ d-364$  
    \item We fit the multivariate forecasting model on these vector observations
    \item We predict the 24 curves vector representations for day $d$
    \item We inverse transform the predicted vectors to obtain the 24 curves pairs forecasts
\end{enumerate}

\medskip
\noindent Note that this same daily recalibration procedure applies to the price-based models.

\subsection{Choosing the dimension of the curves' vector representation} \label{subsec:K}

Regarding step (1) in section \ref{subsec:recal}, this implies that the \textbf{FPCA} or \textbf{ZST} transformation is re-run from scratch every day. This dynamic "representation learning" procedure is clearly preferable to a time-invariant representation learned on a fixed initial time window but adds a layer of complexity: \rev{We must have a procedure to choose the supply and demand dimensions $(K_s, K_d)$ on a daily basis}.

\rev{To simplify the analysis, obtain more stable results and enable sensitivity analysis on $K_s$ and $K_d$, we choose to keep them fixed over the test period:}
\rev{$K_s$ and $K_d$ are chosen based both on curve and clearing price approximation quality on the initial time window, using the elbow method. Curve approximation quality is assessed with the cumulative explained variance ratio}\footnote{\rev{For FPCA, this is efficiently computed from the eigenvalues of the covariance operator while for ZST, we compute the \textit{functional} explained variance ratio for each $K_s$ and $K_d$ with the classical pointwise metric (see \url{https://fda.readthedocs.io/en/stable/modules/misc/autosummary/skfda.misc.scoring.explained_variance_score.html}), which equals the eigenvalue-based computation in the case of FPCA.}} \rev{while clearing price approximation quality is assessed with the MAE. Note that to provide a fair comparison between \textbf{FPCA} and \textbf{ZST}, we choose $K_s$ and $K_d$ such that both representations yield the same approximation quality. This implies that $(K_s, K_d)$ may not necessarily be the same for \textbf{FPCA} and \textbf{ZST}.}

\medskip

\rev{Hence, while the dimensions $K_s$ and $K_d$ are fixed, the $K_s$- and $K_d$-dimensional representations are recalibrated on a daily basis: the mean curve and (i) functional principal components for FPCA (ii) price classes for ZST, are re-estimated every day.}

\rev{We provide a comparison of this \textit{dynamic} approach with a \textit{static} approach where the representation is learnt once and for all on the initial time window, in \ref{sec:stat_vs_dyna}.}

\section{\rev{Results}}
\label{sec:results}

\subsection{Data}

\noindent We analyze the \rev{Italian (\textbf{GME}), German (\textbf{EPEX-DE-LU}) and French (\textbf{EPEX-FR})} day-ahead markets over the 2023-2024 period. The testing period covers the whole year 2024 (366 days). The quantity curves are built according to the procedure described in section \ref{sec:curves}, using:
\rev{
\begin{itemize}
    \item For \textbf{EPEX-DE-LU} and \textbf{EPEX-FR} (both managed by the EPEX market operator), we use the so-called \texttt{AggregatedCurves} dataset\footnote{See for instance \url{https://www.epexspot.com/en/market-results?market_area=DE-LU&auction=MRC&trading_date=2026-06-19&modality=Auction&sub_modality=DayAhead&technology=&data_mode=aggregated&period=&production_period=&product=15}} retrieved from a private data provider \citep{lsegdata&analytics_epexcurves_2025}. This dataset contains the aggregated supply and demand curves published by EPEX after each auction, which already incorporate, in addition to simple bids, block bids and import/export volumes with neighboring bidding zones participating in the single day-ahead coupling (SDAC). Specifically, buy block bids and imports are added to the supply curve at the minimum price (--500\euro/MWh), while sell block bids and exports are added to the demand curve at the maximum price (4000\euro/MWh). These curves are therefore "analysis-ready" in the sense that their intersection directly yields the official day-ahead clearing price.
    \item For \textbf{GME}, the \texttt{DemandSupply} dataset\footnote{\url{https://www.mercatoelettrico.org/en-us/Home/Results/Electricity/MGP/Results/DemandSupply}} retrieved from GME's FTP server, which contains all anonymized simple and block bids submitted on the Italian day-ahead market (MGP). Note that block bids were officially introduced in the Italian market on January 1st, 2025 so the dataset contains simple bids only during the period of study.
    However, unlike EPEX's aggregated curves, this dataset does not include SDAC import/export volumes. These volumes are incorporated using an auxilliary dataset, called \texttt{MarketCoupling}\footnote{\url{https://www.mercatoelettrico.org/en-us/Home/Results/Electricity/MGP/Results/MarketCoupling}}, also available on GME's FTP server. We however follow a slightly different logic than for EPEX's curves: instead of adding imports to supply curves and exports to demand curves, as done for EPEX, we add the net position (difference between imports and exports) to supply curves only. For either approach, the clearing price is found at the intersection of the two curves. Note that for GME, this clearing price does not exactly match the official day-ahead price paid or received by market participants, due to the zonal pricing mechanism. Additional details are provided in \ref{sec:copper}.
\end{itemize}
}

\rev{Finally, we note that both EPEX and GME\footnote{Note that, for GME, unlike the \texttt{PublicDomain} dataset which is published with a one-week delay, \texttt{DemandSupply} and \texttt{MarketCoupling} are published right after the market clearing.} bids datasets considered in this paper are published everyday around 13:00, right after the market clears, so they can realistically be used for day-ahead forecasts.}

\begin{figure*}[h!]
     \centering
     
     \begin{subfigure}[b]{0.32\textwidth}
         \includegraphics[width=\textwidth]{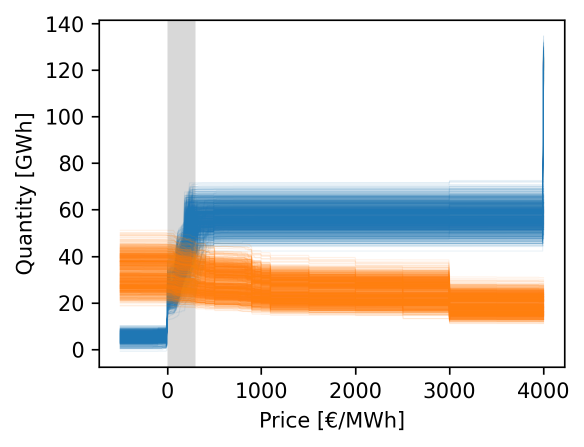}
         \caption{Full (\textbf{GME})}
         \label{fig:curves:full_gme}
     \end{subfigure}
     \begin{subfigure}[b]{0.32\textwidth}
         \includegraphics[width=\textwidth]{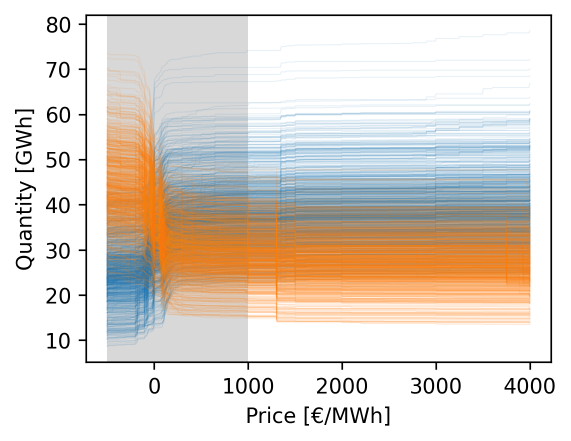}
         \caption{Full (\textbf{EPEX-DE-LU})}
         \label{fig:curves:full_epex-de-lu}
     \end{subfigure}
     \begin{subfigure}[b]{0.32\textwidth}
         \includegraphics[width=\textwidth]{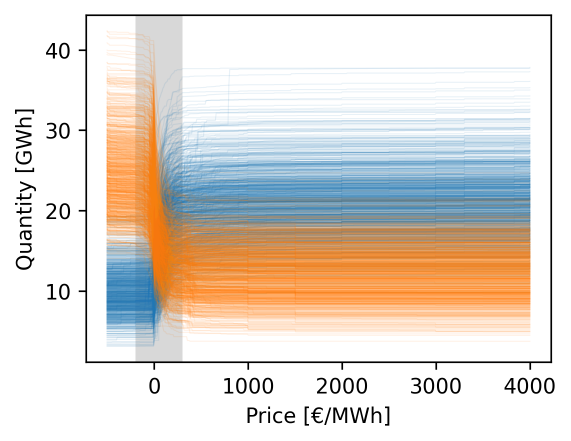}
         \caption{Full (\textbf{EPEX-FR})}
         \label{fig:curves:full_epex-fr}
     \end{subfigure}
     
    \par\vspace{0.5cm}
    
     \begin{subfigure}[b]{0.32\textwidth}
         \includegraphics[width=\textwidth]{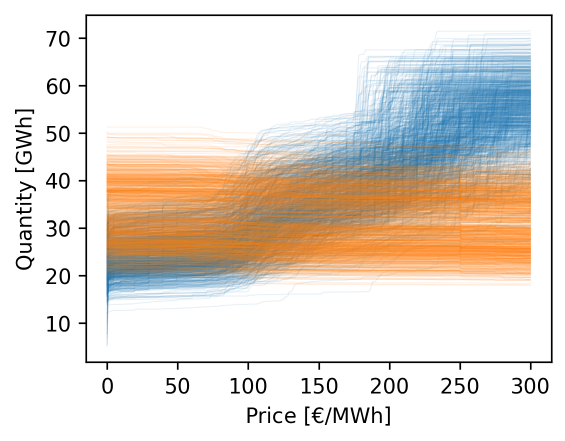}
         \caption{Restricted (\textbf{GME})}
         \label{fig:curves:gme}
     \end{subfigure}
     \begin{subfigure}[b]{0.32\textwidth}
         \includegraphics[width=\textwidth]{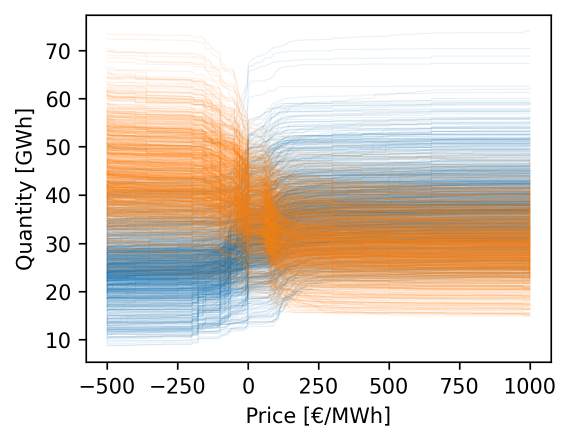}
         \caption{Restricted (\textbf{EPEX-DE-LU})}
         \label{fig:curves:epex-de-lu}
     \end{subfigure}
     \begin{subfigure}[b]{0.32\textwidth}
         \includegraphics[width=\textwidth]{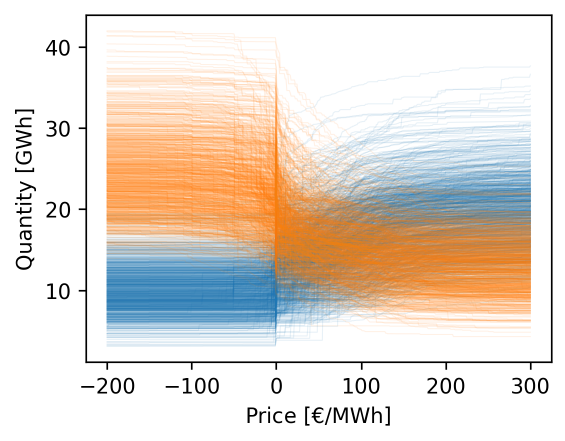}
         \caption{Restricted (\textbf{EPEX-FR})}
         \label{fig:curves:epex-fr}
     \end{subfigure}
        \caption{\rev{(\textit{Color optional}) Sample ($N=1000$) of curves dataset for each market on full and restricted domains. The rectangular grey shaded area identifies the restricted domain on full-domain plots. Supply curves are blue-colored while demand curves are orange-colored.}}
        \label{fig:curves}
\end{figure*}

\rev{The full curves price domain ranges from --500\euro \ to 4000\euro \ but we analyze them on a \textit{restricted} range. From the perspective of clearing price forecasting, portions of the domain where the intersection point has negligible probability of occurring are of lower interest: they may exhibit arbitrary variability with no impact on market outcomes, and could introduce unwanted noise into the model. We therefore restrict the curves to a price domain that contains at least all prices observed during the period of study. Finally, curves are represented by their evaluations on a uniform price grid of 1\euro/MWh resolution and a linear interpolation scheme. The resulting restricted domain and grid size are reported in Table \ref{tab:domain}. A sample of the curves dataset on both full and restricted domains is pictured in Fig. \ref{fig:curves}.}

\rev{To appreciate the impact of this price range restriction, we reproduced the entire analysis with the full [--500, 4000] domain for all three markets, which can be found in \ref{sec:res_unres}.}

\begin{table}
    \centering
    \rev{
    \begin{tabular}{lccc}
    \toprule
         & \textbf{GME} & \textbf{EPEX-DE-LU} & \textbf{EPEX-FR} \\
    \midrule
        \textbf{Domain} & [0, 300] & [-500, 1000] & [-200, 300]  \\
        \textbf{Grid size} & 301 & 1501 & 501 \\
    \bottomrule
    \end{tabular}
    }
    \caption{\rev{Restricted price domain (in \euro/MWh) and discretization grid size considered for the analysis.}}
    \label{tab:domain}
\end{table}

\bigskip

\noindent \rev{Additionally, we consider the following exogenous variables for both curve-based and price-based models, which are similar to those of \citet{mestre_forecasting_2020} and commonly chosen in EPF studies:
\begin{itemize}
    \item For all three markets, the country-level \textit{day-ahead load forecast} (\texttt{Load}), retrieved from the ENTSO-E Transparency Platform
    \item For all three markets, the country-level combined \textit{day-ahead wind and solar generation forecast} (\texttt{RES}), obtained from a private data provider (LSEG Data \& Analytics)
    \item For \textbf{GME} and \textbf{EPEX-FR} only, the \textit{day-ahead net transfer capacity} (NTC) of the France $\rightarrow$ Italy cross-border connection (\texttt{NTC FR > IT}), retrieved from the ENTSO-E Transparency Platform
    \item For \textbf{GME} only, the NTC of the Switzerland $\rightarrow$ Italy cross-border connection (\texttt{NTC CH > IT}), retrieved from the ENTSO-E Transparency Platform
\end{itemize}
The France $\rightarrow$ Italy and Switzerland $\rightarrow$ Italy NTCs are relevant for curves prediction as Italy (nearly) systematically imports cheaper nuclear and hydropower electricity from France and Switzerland, in the limit of this transfer capacity, which frequently saturates. Therefore, NTC variations directly affect market results --- specifically the SDAC imports from France and the regular imports (subject to the merit-order) from Switzerland. Note that no other connection was considered because of smaller NTCs and the absence of a dominant flow direction.}

As for bid data, all predictors are available at least two hour before the day-ahead market closure at 12:00 and can therefore realistically produce market results forecasts informing trading strategies "in time".

\bigskip

\noindent \rev{Consequently, we have a total of
\begin{itemize}
    \item 2 exogenous variables for \textbf{EPEX-DE-LU} (\texttt{Load}, \texttt{RES})
    \item 3 exogenous variables for \textbf{EPEX-FR} (\texttt{Load}, \texttt{RES}, \texttt{NTC FR > IT})
    \item 4 exogenous variables for \textbf{GME} (\texttt{Load}, \texttt{RES}, \texttt{NTC FR > IT}, \texttt{NTC CH > IT})
\end{itemize}}

\subsection{\rev{Curves representation}}

\begin{figure*}[h!]
     \centering
     \begin{subfigure}[b]{0.49\textwidth}
         \includegraphics[width=\textwidth]{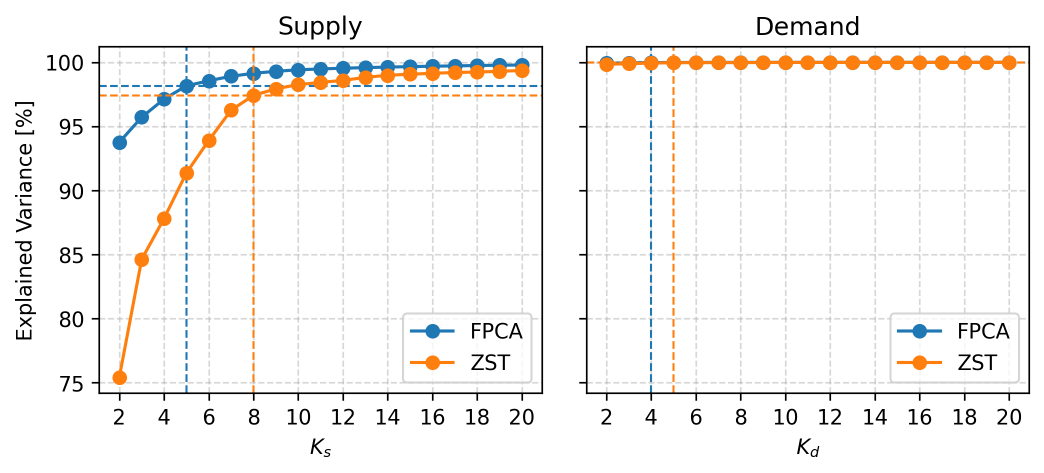}
         \caption{Curves (\textbf{GME})}
         \label{fig:K:curve_gme}
     \end{subfigure}
     \begin{subfigure}[b]{0.49\textwidth}
         \includegraphics[width=\textwidth]{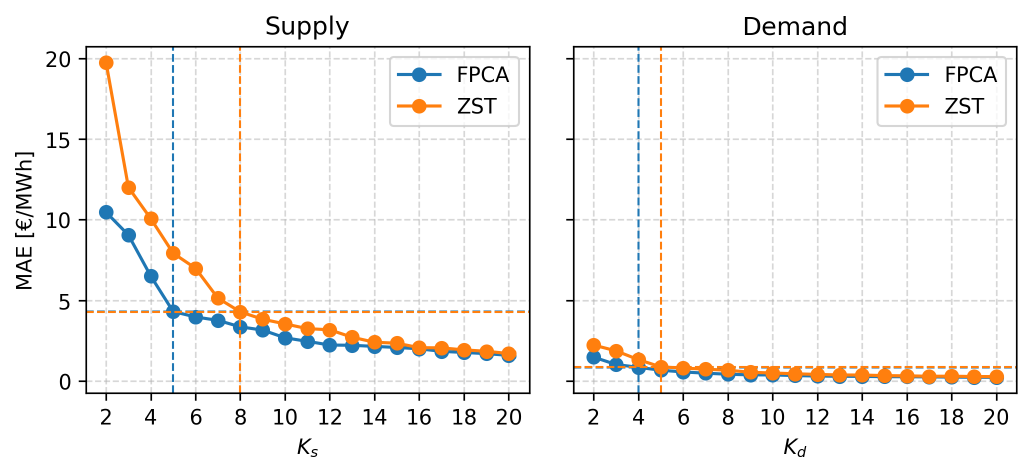}
         \caption{Clearing price (\textbf{GME})}
         \label{fig:K:mcp_gme}
     \end{subfigure}
     \par\vspace{0.5cm}
     \begin{subfigure}[b]{0.49\textwidth}
         \includegraphics[width=\textwidth]{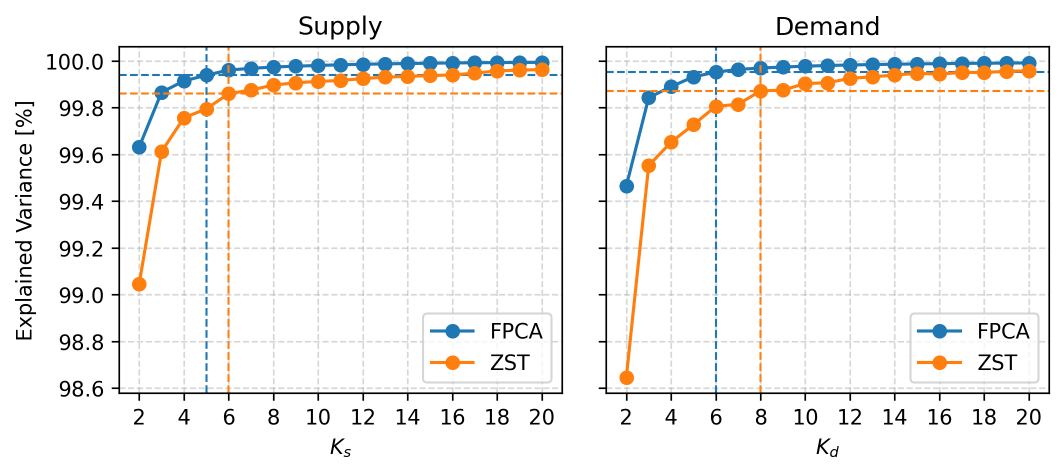}
         \caption{Curves (\textbf{EPEX-DE-LU})}
         \label{fig:K:curve_epex-de-lu}
     \end{subfigure}
     \begin{subfigure}[b]{0.49\textwidth}
         \includegraphics[width=\textwidth]{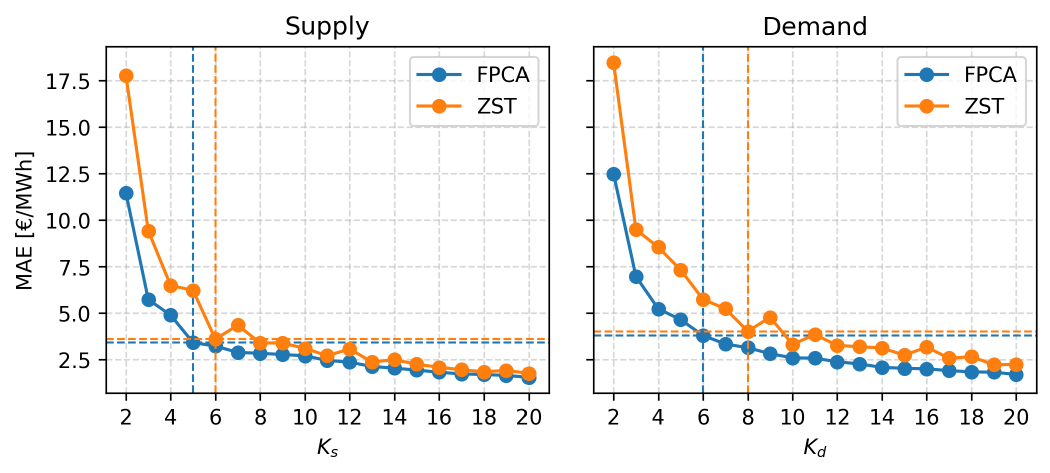}
         \caption{Clearing price (\textbf{EPEX-DE-LU})}
         \label{fig:K:mcp_epex-de-lu}
     \end{subfigure}
     \par\vspace{0.5cm}
     \begin{subfigure}[b]{0.49\textwidth}
         \includegraphics[width=\textwidth]{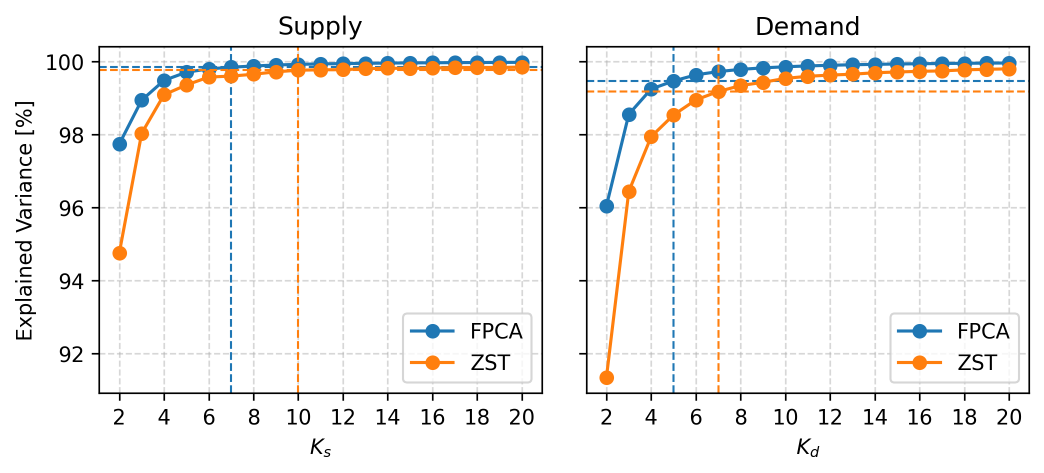}
         \caption{Curves (\textbf{EPEX-FR})}
         \label{fig:K:curve_epex-fr}
     \end{subfigure}
     \begin{subfigure}[b]{0.49\textwidth}
         \includegraphics[width=\textwidth]{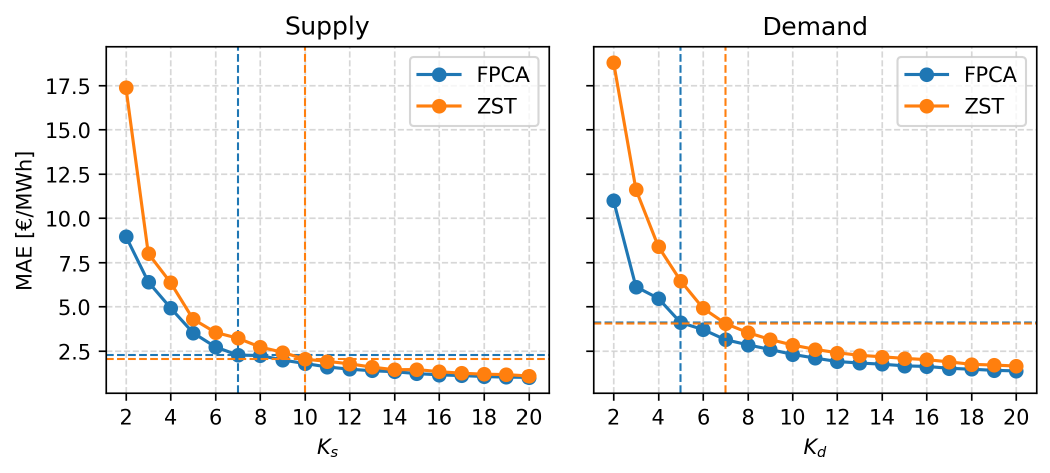}
         \caption{Clearing price (\textbf{EPEX-FR})}
         \label{fig:K:mcp_epex-fr}
     \end{subfigure}
        \caption{\rev{(\textit{Color optional}) Curves (left panel) and clearing price (right panel) \textit{approximation} error for 2023 (inital estimation window) for each representation method with respect to the number of components considered for supply curves ($K_s$) and demand curves ($K_d$). When computing the clearing price for different $K_s$, true demand curves are used and, conversely, when computing the clearing price for different $K_d$ values, true supply curves are used. The vertical and horizontal dashed lines identify the $K_s$ and $K_d$ values selected for each representation and market (which are also reported in Table \ref{tab:K}).}}
        \label{fig:K}
\end{figure*}

\rev{The curves and clearing price approximation error with respect to the number of supply and demand components $K_s$ and $K_d$ is shown in Fig. \ref{fig:K} and the selected $K_s, K_d$ values for each representation and market are reported in Table \ref{tab:K}. Observing Fig. \ref{fig:K}, it should be noted that \textbf{ZST}, although being less parsimonious than \textbf{FPCA} --- as one could expect --- is nevertheless not so far behind and surprisingly efficient considering that it's based on a discrete representation.}

\begin{table}[h!]
\centering
\rev{
\begin{tabular}{lc@{\hspace{0.5em}}cc c@{\hspace{0.5em}}c}
\toprule
& \multicolumn{2}{c}{\textbf{FPCA}}
&& \multicolumn{2}{c}{\textbf{ZST}} \\
\cmidrule(lr){2-3}
\cmidrule(lr){5-6}
\textbf{Market} & $K_s$ & $K_d$ &&  $K_s$ & $K_d$ \\
\midrule
GME         & 5 & 4 && 8 & 5 \\
EPEX-DE-LU  & 5 & 6 && 6 & 8 \\
EPEX-FR     & 7 & 5 && 10 & 7 \\
\bottomrule
\end{tabular}
}
\caption{\rev{Number of components selected for supply curves ($K_s$) and demand curves ($K_d$)}}
\label{tab:K}
\end{table}

\rev{The first four dynamic functional principal components (FPCs) for each side and market are plotted in Figs. \ref{fig:fpc_gme}, \ref{fig:fpc_epex-de-lu} and \ref{fig:fpc_epex-fr} while an interpretation of the static FPCs can be found in Figs. \ref{fig:fpc_effect_supply} and \ref{fig:fpc_effect_demand}, in \ref{sec:fpca-res}.}

\subsection{\rev{Curves prediction}}

\begin{figure*}[h!]
     \centering
     \begin{subfigure}[b]{0.8\textwidth}
         \includegraphics[width=\textwidth]{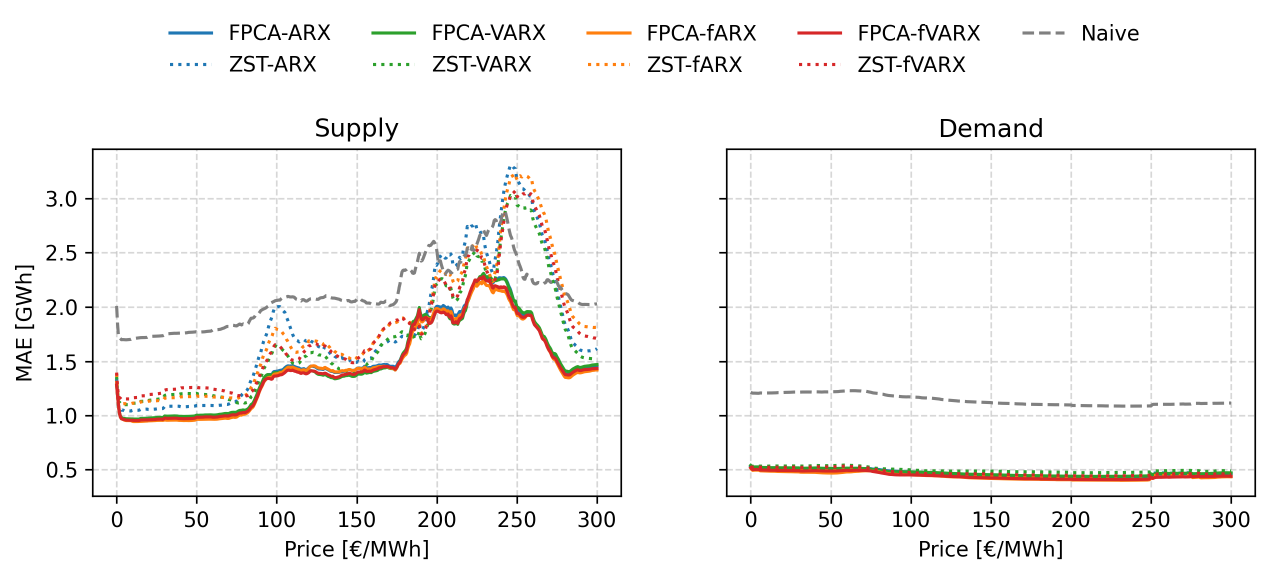}
         \caption{\textbf{GME}}
         \label{fig:curve_mae:gme}
     \end{subfigure}
     \begin{subfigure}[b]{0.8\textwidth}
         \includegraphics[width=\textwidth]{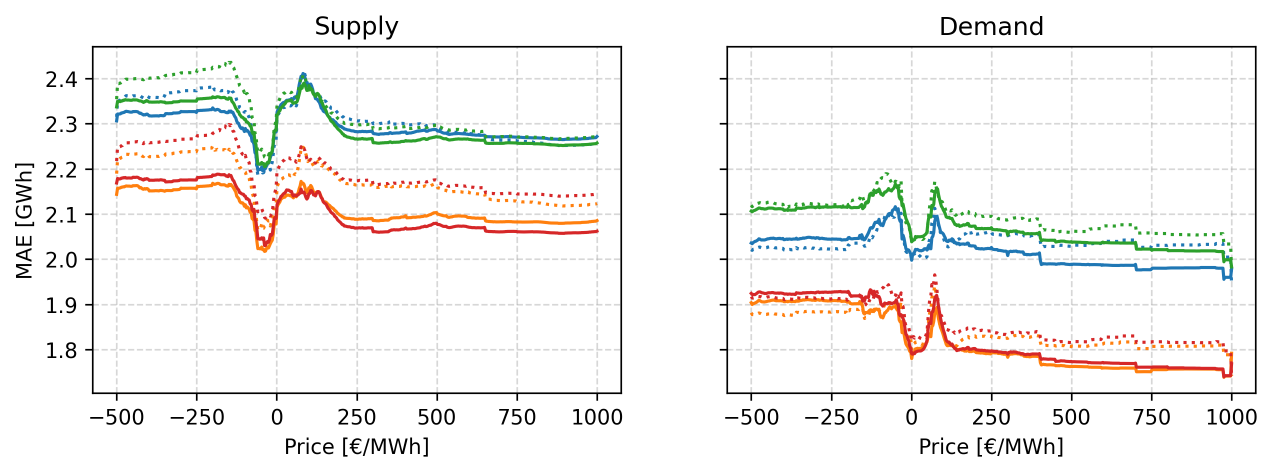}
         \caption{\textbf{EPEX-DE-LU}}
         \label{fig:curve_mae:epex-de-lu}
     \end{subfigure}
     \begin{subfigure}[b]{0.8\textwidth}
         \includegraphics[width=\textwidth]{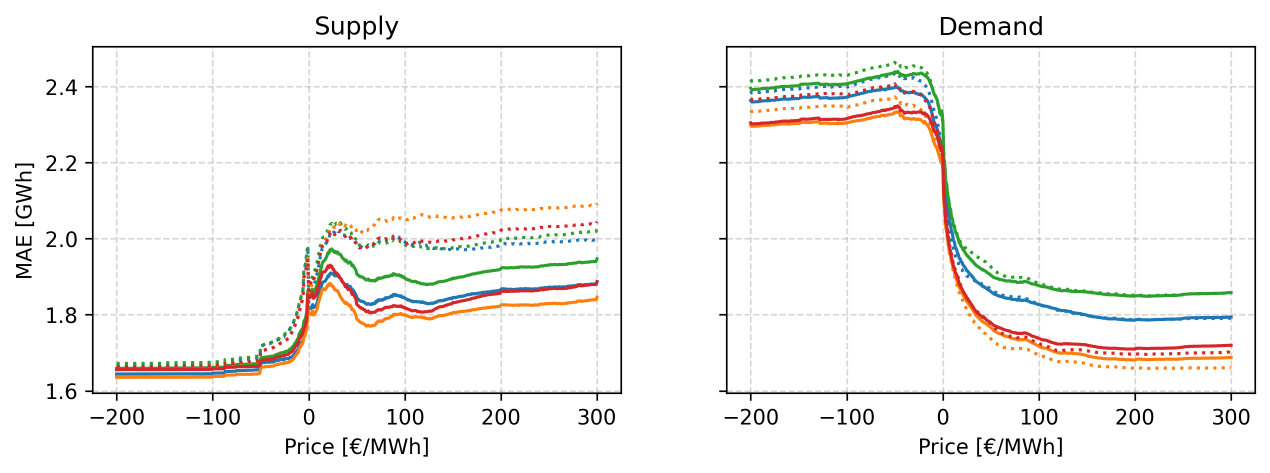}
         \caption{\textbf{EPEX-FR}}
         \label{fig:curve_mae:epex-fr}
     \end{subfigure}
        \caption{\rev{(\textit{Color optional}) Pointwise mean absolute error (MAE) of supply and demand curves forecasts}}
        \label{fig:curve_mae}
\end{figure*}

\rev{The functional error metrics for curves forecasts are presented in Table \ref{tab:gme_supply_demand} (\textbf{GME}), Table \ref{tab:epex-de-lu_supply_demand} (\textbf{EPEX-DE-LU}) and Table \ref{tab:epex_fr_supply_demand} (\textbf{EPEX-FR}) and the pointwise mean absolute error is plotted in Fig. \ref{fig:curve_mae}. The DM tests are shown in Fig. \ref{fig:dm_curve} in \ref{sec:dm_curves}}.

\begin{table}[h!]
\centering
\footnotesize
\begin{tabular}{rcccccccc}
\toprule
&
\multicolumn{4}{c}{\textbf{Supply}} &
\multicolumn{4}{c}{\textbf{Demand}} \\
\cmidrule(lr){2-5}
\cmidrule(lr){6-9}
&
\makecell{\textbf{FMAE}\\{}[GWh]} &
\makecell{\textbf{FRMSE}\\{}[GWh]} &
\makecell{\textbf{FMAPE}\\{}[\%]} &
\textbf{rFMAE} &
\makecell{\textbf{FMAE}\\{}[GWh]} &
\makecell{\textbf{FRMSE}\\{}[GWh]} &
\makecell{\textbf{FMAPE}\\{}[\%]} &
\textbf{rFMAE} \\
\midrule

\textbf{Naive}
& \cellcolor[RGB]{218,134,118}2.11
& \cellcolor[RGB]{218,134,118}2.95
& \cellcolor[RGB]{218,134,118}5.86
& \cellcolor[RGB]{218,134,118}1.000
& \cellcolor[RGB]{218,134,118}1.14
& \cellcolor[RGB]{218,134,118}1.99
& \cellcolor[RGB]{218,134,118}3.78
& \cellcolor[RGB]{218,134,118}1.000 \\

\textbf{ZST-ARX}
& \cellcolor[RGB]{247,214,119}1.78
& \cellcolor[RGB]{246,209,119}2.48
& \cellcolor[RGB]{220,208,124}4.66
& \cellcolor[RGB]{247,214,119}0.846
& \cellcolor[RGB]{123,187,140}0.47
& \cellcolor[RGB]{118,186,141}0.62
& \cellcolor[RGB]{123,187,140}1.46
& \cellcolor[RGB]{123,187,140}0.408 \\

\textbf{ZST-VARX}
& \cellcolor[RGB]{214,207,125}1.70
& \cellcolor[RGB]{216,207,125}2.33
& \cellcolor[RGB]{194,203,128}4.47
& \cellcolor[RGB]{214,207,125}0.806
& \cellcolor[RGB]{135,189,138}0.50
& \cellcolor[RGB]{124,187,140}0.65
& \cellcolor[RGB]{134,189,138}1.56
& \cellcolor[RGB]{135,189,138}0.435 \\

\textbf{ZST-fARX}
& \cellcolor[RGB]{245,207,119}1.81
& \cellcolor[RGB]{245,207,119}2.49
& \cellcolor[RGB]{228,210,123}4.72
& \cellcolor[RGB]{245,207,119}0.859
& \cellcolor[RGB]{117,186,141}0.45
& \cellcolor[RGB]{114,185,141}0.60
& \cellcolor[RGB]{117,186,141}1.40
& \cellcolor[RGB]{117,186,141}0.394 \\

\textbf{ZST-fVARX}
& \cellcolor[RGB]{246,214,120}1.77
& \cellcolor[RGB]{241,213,120}2.42
& \cellcolor[RGB]{220,208,124}4.66
& \cellcolor[RGB]{246,214,120}0.843
& \cellcolor[RGB]{126,188,139}0.47
& \cellcolor[RGB]{119,186,140}0.62
& \cellcolor[RGB]{126,188,139}1.49
& \cellcolor[RGB]{126,188,139}0.414 \\

\textbf{FPCA-ARX}
& \cellcolor[RGB]{121,186,140}1.47
& \cellcolor[RGB]{123,187,140}1.98
& \cellcolor[RGB]{118,186,141}3.90
& \cellcolor[RGB]{121,186,140}0.700
& \cellcolor[RGB]{121,186,140}0.46
& \cellcolor[RGB]{117,185,141}0.61
& \cellcolor[RGB]{121,186,140}1.44
& \cellcolor[RGB]{121,186,140}0.403 \\

\textbf{FPCA-VARX}
& \cellcolor[RGB]{120,186,140}1.47
& \cellcolor[RGB]{118,186,141}1.96
& \cellcolor[RGB]{119,186,140}3.91
& \cellcolor[RGB]{120,186,140}0.698
& \cellcolor[RGB]{125,187,139}0.47
& \cellcolor[RGB]{118,186,141}0.62
& \cellcolor[RGB]{125,187,139}1.48
& \cellcolor[RGB]{125,187,139}0.412 \\

\textbf{FPCA-fARX}
& \cellcolor[RGB]{113,185,142}1.45
& \cellcolor[RGB]{114,185,141}1.95
& \cellcolor[RGB]{113,185,142}3.86
& \cellcolor[RGB]{113,185,142}0.689
& \cellcolor[RGB]{113,185,142}0.44
& \cellcolor[RGB]{113,185,142}0.59
& \cellcolor[RGB]{113,185,142}1.36
& \cellcolor[RGB]{113,185,142}0.383 \\

\textbf{FPCA-fVARX}
& \cellcolor[RGB]{113,185,141}1.45
& \cellcolor[RGB]{113,185,142}1.94
& \cellcolor[RGB]{113,185,141}3.86
& \cellcolor[RGB]{113,185,141}0.690
& \cellcolor[RGB]{115,185,141}0.44
& \cellcolor[RGB]{113,185,141}0.59
& \cellcolor[RGB]{115,185,141}1.39
& \cellcolor[RGB]{115,185,141}0.388 \\

\bottomrule
\end{tabular}

\caption{\rev{Functional error metrics for supply and demand curves prediction (\textbf{GME}). The green-yellow-red colormap is applied column-wise.}}
\label{tab:gme_supply_demand}
\end{table}

\rev{For \textbf{GME}, the \textbf{Naive} model is outperformed by all other models on both sides. On the demand side, all non-naive models perform nearly identically, which is most likely attributable to the low demand elasticity of the Italian day-ahead market: the quantity demanded is approximately constant and equal to the load forecast --- a predictor available to all models --- within the restricted price range considered. On the supply side, \textbf{FPCA} substantially outperforms \textbf{ZST}, while the inclusion of cross-component and cross-hour effects yields no meaningful improvement, particularly for \textbf{FPCA} models. The pointwise MAE (Fig. \ref{fig:curve_mae:gme}) confirms that this pattern holds across the entire price domain but also shows that the functional metrics in Table \ref{tab:gme_supply_demand} are inflated by the large performance drop of \textbf{ZST} models around 250\euro/MWh, probably due to lower resolution and poor interpolation quality in this portion.}

\begin{table}[h!]
\centering
\footnotesize
\begin{tabular}{rcccccccc}
\toprule
&
\multicolumn{4}{c}{\textbf{Supply}} &
\multicolumn{4}{c}{\textbf{Demand}} \\
\cmidrule(lr){2-5}
\cmidrule(lr){6-9}
&
\makecell{\textbf{FMAE}\\{}[GWh]} &
\makecell{\textbf{FRMSE}\\{}[GWh]} &
\makecell{\textbf{FMAPE}\\{}[\%]} &
\textbf{rFMAE} &
\makecell{\textbf{FMAE}\\{}[GWh]} &
\makecell{\textbf{FRMSE}\\{}[GWh]} &
\makecell{\textbf{FMAPE}\\{}[\%]} &
\textbf{rFMAE} \\
\midrule
\textbf{Naive}
& \cellcolor[RGB]{218,134,118}3.56
& \cellcolor[RGB]{218,134,118}5.13
& \cellcolor[RGB]{218,134,118}9.90
& \cellcolor[RGB]{218,134,118}1.000
& \cellcolor[RGB]{218,134,118}3.01
& \cellcolor[RGB]{218,134,118}4.43
& \cellcolor[RGB]{218,134,118}7.93
& \cellcolor[RGB]{218,134,118}1.000 \\
\textbf{ZST-ARX}
& \cellcolor[RGB]{150,193,135}2.31
& \cellcolor[RGB]{134,189,138}2.99
& \cellcolor[RGB]{154,194,135}6.56
& \cellcolor[RGB]{150,193,135}0.649
& \cellcolor[RGB]{163,196,133}2.04
& \cellcolor[RGB]{145,192,136}2.65
& \cellcolor[RGB]{169,197,132}5.50
& \cellcolor[RGB]{163,196,133}0.677 \\
\textbf{ZST-VARX}
& \cellcolor[RGB]{153,194,135}2.33
& \cellcolor[RGB]{138,190,137}3.03
& \cellcolor[RGB]{159,195,134}6.62
& \cellcolor[RGB]{153,194,135}0.653
& \cellcolor[RGB]{175,198,131}2.09
& \cellcolor[RGB]{152,193,135}2.71
& \cellcolor[RGB]{180,199,131}5.62
& \cellcolor[RGB]{175,198,131}0.695 \\
\textbf{ZST-fARX}
& \cellcolor[RGB]{125,187,139}2.17
& \cellcolor[RGB]{117,185,141}2.84
& \cellcolor[RGB]{126,188,139}6.15
& \cellcolor[RGB]{125,187,139}0.611
& \cellcolor[RGB]{118,186,141}1.84
& \cellcolor[RGB]{117,186,141}2.44
& \cellcolor[RGB]{119,186,141}4.92
& \cellcolor[RGB]{118,186,141}0.611 \\
\textbf{ZST-fVARX}
& \cellcolor[RGB]{129,188,139}2.19
& \cellcolor[RGB]{123,187,140}2.89
& \cellcolor[RGB]{130,188,139}6.21
& \cellcolor[RGB]{129,188,139}0.616
& \cellcolor[RGB]{122,187,140}1.86
& \cellcolor[RGB]{118,186,141}2.45
& \cellcolor[RGB]{123,187,140}4.98
& \cellcolor[RGB]{122,187,140}0.617 \\
\textbf{FPCA-ARX}
& \cellcolor[RGB]{148,192,136}2.29
& \cellcolor[RGB]{133,189,138}2.99
& \cellcolor[RGB]{150,193,135}6.49
& \cellcolor[RGB]{148,192,136}0.645
& \cellcolor[RGB]{158,195,134}2.02
& \cellcolor[RGB]{142,191,137}2.63
& \cellcolor[RGB]{164,196,133}5.44
& \cellcolor[RGB]{158,195,134}0.671 \\
\textbf{FPCA-VARX}
& \cellcolor[RGB]{148,192,136}2.29
& \cellcolor[RGB]{135,189,138}3.00
& \cellcolor[RGB]{153,194,135}6.55
& \cellcolor[RGB]{148,192,136}0.645
& \cellcolor[RGB]{170,197,132}2.07
& \cellcolor[RGB]{149,193,136}2.68
& \cellcolor[RGB]{178,199,131}5.60
& \cellcolor[RGB]{170,197,132}0.688 \\
\textbf{FPCA-fARX}
& \cellcolor[RGB]{114,185,141}2.11
& \cellcolor[RGB]{113,185,142}2.81
& \cellcolor[RGB]{113,185,142}5.95
& \cellcolor[RGB]{114,185,141}0.593
& \cellcolor[RGB]{113,185,142}1.82
& \cellcolor[RGB]{113,185,141}2.41
& \cellcolor[RGB]{113,185,142}4.86
& \cellcolor[RGB]{113,185,142}0.603 \\
\textbf{FPCA-fVARX}
& \cellcolor[RGB]{113,185,142}2.10
& \cellcolor[RGB]{113,185,141}2.81
& \cellcolor[RGB]{114,185,141}5.97
& \cellcolor[RGB]{113,185,142}0.591
& \cellcolor[RGB]{115,185,141}1.83
& \cellcolor[RGB]{113,185,142}2.41
& \cellcolor[RGB]{117,185,141}4.90
& \cellcolor[RGB]{115,185,141}0.607 \\
\bottomrule
\end{tabular}
\caption{\rev{Functional error metrics for supply and demand curves prediction (\textbf{EPEX-DE-LU}). The green-yellow-red colormap is applied column-wise.}}
\label{tab:epex-de-lu_supply_demand}
\end{table}

\rev{For \textbf{EPEX-DE-LU}, all models perform substantially better than \textbf{Naive}. Performance is also more stable across the price domain than for \textbf{GME}, on both sides. Interestingly, the relative ranking of models differs from that observed on \textbf{GME}: here, the choice between a full and a concurrent modeling approach --- particularly for demand curves --- appears to matter more than the choice between \textbf{FPCA} and \textbf{ZST} representations, with a clear advantage for the full approach. The results also show that incorporating cross-component effects provides no improvement and may even degrade performance in the concurrent setting (Fig. \ref{fig:curve_mae:epex-de-lu}, right panel). Among full models, pairwise comparisons between \textbf{FPCA} and \textbf{ZST} suggest a slight advantage for \textbf{FPCA} across the entire domain for supply, and over the [200, 1000] \euro/MWh interval for demand.}

\begin{table}[h!]
\centering
\footnotesize
\begin{tabular}{rcccccccc}
\toprule
&
\multicolumn{4}{c}{\textbf{Supply}} &
\multicolumn{4}{c}{\textbf{Demand}} \\
\cmidrule(lr){2-5}
\cmidrule(lr){6-9}
&
\makecell{\textbf{FMAE}\\{}[GWh]} &
\makecell{\textbf{FRMSE}\\{}[GWh]} &
\makecell{\textbf{FMAPE}\\{}[\%]} &
\textbf{rFMAE} &
\makecell{\textbf{FMAE}\\{}[GWh]} &
\makecell{\textbf{FRMSE}\\{}[GWh]} &
\makecell{\textbf{FMAPE}\\{}[\%]} &
\textbf{rFMAE} \\
\midrule
\textbf{Naive}
& \cellcolor[RGB]{218,134,118}2.87
& \cellcolor[RGB]{218,134,118}3.91
& \cellcolor[RGB]{218,134,118}21.23
& \cellcolor[RGB]{218,134,118}1.000
& \cellcolor[RGB]{218,134,118}2.60
& \cellcolor[RGB]{218,134,118}3.46
& \cellcolor[RGB]{218,134,118}14.68
& \cellcolor[RGB]{218,134,118}1.000 \\
\textbf{ZST-ARX}
& \cellcolor[RGB]{141,191,137}1.87
& \cellcolor[RGB]{134,189,138}2.38
& \cellcolor[RGB]{137,190,137}14.04
& \cellcolor[RGB]{141,191,137}0.652
& \cellcolor[RGB]{154,194,135}2.06
& \cellcolor[RGB]{151,193,135}2.64
& \cellcolor[RGB]{155,194,135}11.57
& \cellcolor[RGB]{154,194,135}0.791 \\
\textbf{ZST-VARX}
& \cellcolor[RGB]{143,191,136}1.88
& \cellcolor[RGB]{138,190,137}2.40
& \cellcolor[RGB]{142,191,137}14.18
& \cellcolor[RGB]{143,191,136}0.655
& \cellcolor[RGB]{175,198,131}2.11
& \cellcolor[RGB]{171,198,132}2.72
& \cellcolor[RGB]{183,200,130}11.95
& \cellcolor[RGB]{175,198,131}0.810 \\
\textbf{ZST-fARX}
& \cellcolor[RGB]{151,193,135}1.91
& \cellcolor[RGB]{143,191,136}2.44
& \cellcolor[RGB]{143,191,136}14.21
& \cellcolor[RGB]{151,193,135}0.665
& \cellcolor[RGB]{113,185,141}1.96
& \cellcolor[RGB]{118,186,141}2.53
& \cellcolor[RGB]{113,185,142}10.99
& \cellcolor[RGB]{113,185,141}0.754 \\
\textbf{ZST-fVARX}
& \cellcolor[RGB]{144,191,136}1.88
& \cellcolor[RGB]{138,190,137}2.40
& \cellcolor[RGB]{140,191,137}14.12
& \cellcolor[RGB]{144,191,136}0.656
& \cellcolor[RGB]{127,188,139}1.99
& \cellcolor[RGB]{131,189,138}2.58
& \cellcolor[RGB]{132,189,138}11.25
& \cellcolor[RGB]{127,188,139}0.767 \\
\textbf{FPCA-ARX}
& \cellcolor[RGB]{120,186,140}1.78
& \cellcolor[RGB]{118,186,141}2.28
& \cellcolor[RGB]{118,186,141}13.48
& \cellcolor[RGB]{120,186,140}0.621
& \cellcolor[RGB]{149,193,135}2.04
& \cellcolor[RGB]{147,192,136}2.63
& \cellcolor[RGB]{154,194,135}11.55
& \cellcolor[RGB]{149,193,135}0.787 \\
\textbf{FPCA-VARX}
& \cellcolor[RGB]{129,188,139}1.82
& \cellcolor[RGB]{127,188,139}2.33
& \cellcolor[RGB]{128,188,139}13.76
& \cellcolor[RGB]{129,188,139}0.635
& \cellcolor[RGB]{170,197,132}2.09
& \cellcolor[RGB]{166,196,133}2.70
& \cellcolor[RGB]{180,200,130}11.91
& \cellcolor[RGB]{170,197,132}0.806 \\
\textbf{FPCA-fARX}
& \cellcolor[RGB]{113,185,142}1.75
& \cellcolor[RGB]{113,185,142}2.25
& \cellcolor[RGB]{113,185,142}13.31
& \cellcolor[RGB]{113,185,142}0.610
& \cellcolor[RGB]{113,185,142}1.96
& \cellcolor[RGB]{113,185,142}2.51
& \cellcolor[RGB]{118,186,141}11.06
& \cellcolor[RGB]{113,185,142}0.753 \\
\textbf{FPCA-fVARX}
& \cellcolor[RGB]{119,186,140}1.78
& \cellcolor[RGB]{118,186,141}2.28
& \cellcolor[RGB]{119,186,140}13.51
& \cellcolor[RGB]{119,186,140}0.620
& \cellcolor[RGB]{121,186,140}1.98
& \cellcolor[RGB]{121,186,140}2.54
& \cellcolor[RGB]{130,188,139}11.23
& \cellcolor[RGB]{121,186,140}0.760 \\
\bottomrule
\end{tabular}
\caption{\rev{Functional error metrics for supply and demand curves prediction (\textbf{EPEX-FR}). The green-yellow-red colormap is applied column-wise.}}
\label{tab:epex_fr_supply_demand}
\end{table}

\rev{For \textbf{EPEX-FR}, all models again perform substantially better than \textbf{Naive}. For both sides, the pointwise MAE (Fig. \ref{fig:curve_mae:epex-fr}) reveals two distinct regimes: one for the negative price range and one for the positive price range. In the negative range, all models perform equally (and better than in the positive range) on the supply side, while no clear trend emerges on the demand side (where performance is worse than in the positive range). In the positive range, all \textbf{FPCA} models outperform any \textbf{ZST} model on the supply side, regardless of the vector modeling strategy. On the demand side, the conclusions align with those from \textbf{EPEX-DE-LU}. As before, the inclusion of cross-component effects provides no improvement on either side.}

\bigskip

\rev{Finally, as shown in Tables \ref{tab:stat_vs_dyna_supply} and \ref{tab:stat_vs_dyna_demand} in \ref{sec:stat_vs_dyna}, the choice between a static and a dynamic representation has little impact on global curve prediction performance, with the exception of GME supply curves, where the dynamic approach appears to benefit \textbf{FPCA} models while degrading \textbf{ZST} models.}


\subsection{Clearing price prediction}

\begin{table}[h!]
\centering
\footnotesize
\begin{tabular}{lrccc|ccc|ccc}
\toprule
& &
\multicolumn{3}{c}{GME} &
\multicolumn{3}{c}{EPEX-DE-LU} &
\multicolumn{3}{c}{EPEX-FR} \\
\cmidrule(lr){3-5}   
\cmidrule(lr){6-8}   
\cmidrule(lr){9-11}  
& &
\textbf{MAE} & \textbf{RMSE} & \multicolumn{1}{c}{\textbf{rMAE}} & 
\textbf{MAE} & \textbf{RMSE} & \multicolumn{1}{c}{\textbf{rMAE}} & 
\textbf{MAE} & \textbf{RMSE} & \textbf{rMAE} \\
\midrule
\multirow{4}{*}{\textit{price-based}} & \textbf{Naive}
& \cellcolor[RGB]{218,134,118}11.32
& \cellcolor[RGB]{218,134,118}17.23
& \cellcolor[RGB]{218,134,118}1.000
& \cellcolor[RGB]{218,134,118}27.17
& \cellcolor[RGB]{218,134,118}43.22
& \cellcolor[RGB]{218,134,118}1.000
& \cellcolor[RGB]{218,134,118}21.48
& \cellcolor[RGB]{218,134,118}29.67
& \cellcolor[RGB]{218,134,118}1.000 \\
& \textbf{ARX}
& \cellcolor[RGB]{154,194,135}8.32
& \cellcolor[RGB]{136,190,138}11.87
& \cellcolor[RGB]{154,194,135}0.735
& \cellcolor[RGB]{137,190,138}14.64
& \cellcolor[RGB]{113,185,141}26.65
& \cellcolor[RGB]{137,190,138}0.539
& \cellcolor[RGB]{179,199,131}15.85
& \cellcolor[RGB]{151,193,135}20.22
& \cellcolor[RGB]{179,199,131}0.738 \\
& \textbf{fARX}
& \cellcolor[RGB]{127,188,139}7.98
& \cellcolor[RGB]{114,185,141}11.39
& \cellcolor[RGB]{127,188,139}0.705
& \cellcolor[RGB]{126,187,139}14.07
& \cellcolor[RGB]{118,186,141}26.96
& \cellcolor[RGB]{126,187,139}0.518
& \cellcolor[RGB]{134,189,138}14.58
& \cellcolor[RGB]{120,186,140}18.95
& \cellcolor[RGB]{134,189,138}0.679 \\
& \textbf{LEAR}
& \cellcolor[RGB]{149,193,136}8.26
& \cellcolor[RGB]{135,189,138}11.84
& \cellcolor[RGB]{149,193,136}0.730
& \cellcolor[RGB]{113,185,142}13.40
& \cellcolor[RGB]{113,185,142}26.62
& \cellcolor[RGB]{113,185,142}0.493
& \cellcolor[RGB]{135,189,138}14.62
& \cellcolor[RGB]{127,188,139}19.23
& \cellcolor[RGB]{135,189,138}0.681 \\
\midrule
\multirow{8}{*}{\textit{curve-based}} & \textbf{ZST-ARX}
& \cellcolor[RGB]{246,209,119}9.66
& \cellcolor[RGB]{211,206,125}13.50
& \cellcolor[RGB]{246,209,119}0.854
& \cellcolor[RGB]{177,199,131}16.67
& \cellcolor[RGB]{214,207,125}32.91
& \cellcolor[RGB]{177,199,131}0.614
& \cellcolor[RGB]{181,200,130}15.90
& \cellcolor[RGB]{175,198,131}21.19
& \cellcolor[RGB]{181,200,130}0.740 \\
& \textbf{ZST-VARX}
& \cellcolor[RGB]{178,199,131}8.63
& \cellcolor[RGB]{154,194,135}12.26
& \cellcolor[RGB]{178,199,131}0.763
& \cellcolor[RGB]{170,197,132}16.32
& \cellcolor[RGB]{188,201,129}31.28
& \cellcolor[RGB]{170,197,132}0.601
& \cellcolor[RGB]{180,199,131}15.86
& \cellcolor[RGB]{175,198,131}21.18
& \cellcolor[RGB]{180,199,131}0.738 \\
& \textbf{ZST-fARX}
& \cellcolor[RGB]{235,212,121}9.39
& \cellcolor[RGB]{205,205,126}13.38
& \cellcolor[RGB]{235,212,121}0.830
& \cellcolor[RGB]{247,213,119}20.39
& \cellcolor[RGB]{218,134,118}43.29
& \cellcolor[RGB]{247,213,119}0.750
& \cellcolor[RGB]{241,198,119}18.50
& \cellcolor[RGB]{229,164,118}27.62
& \cellcolor[RGB]{241,198,119}0.861 \\
& \textbf{ZST-fVARX}
& \cellcolor[RGB]{209,206,126}9.05
& \cellcolor[RGB]{187,201,129}12.98
& \cellcolor[RGB]{209,206,126}0.800
& \cellcolor[RGB]{230,211,122}19.37
& \cellcolor[RGB]{225,153,118}41.33
& \cellcolor[RGB]{230,211,122}0.713
& \cellcolor[RGB]{225,210,123}17.12
& \cellcolor[RGB]{245,214,120}24.06
& \cellcolor[RGB]{225,210,123}0.797 \\
& \textbf{FPCA-ARX}
& \cellcolor[RGB]{134,189,138}8.07
& \cellcolor[RGB]{137,190,137}11.90
& \cellcolor[RGB]{134,189,138}0.713
& \cellcolor[RGB]{154,194,135}15.53
& \cellcolor[RGB]{142,191,137}28.43
& \cellcolor[RGB]{154,194,135}0.572
& \cellcolor[RGB]{113,185,142}14.00
& \cellcolor[RGB]{113,185,142}18.63
& \cellcolor[RGB]{113,185,142}0.652 \\
& \textbf{FPCA-VARX}
& \cellcolor[RGB]{113,185,142}7.78
& \cellcolor[RGB]{113,185,142}11.36
& \cellcolor[RGB]{113,185,142}0.688
& \cellcolor[RGB]{157,194,134}15.65
& \cellcolor[RGB]{188,201,129}31.28
& \cellcolor[RGB]{157,194,134}0.576
& \cellcolor[RGB]{132,189,138}14.55
& \cellcolor[RGB]{133,189,138}19.45
& \cellcolor[RGB]{132,189,138}0.677 \\
& \textbf{FPCA-fARX}
& \cellcolor[RGB]{133,189,138}8.05
& \cellcolor[RGB]{135,190,138}11.86
& \cellcolor[RGB]{133,189,138}0.711
& \cellcolor[RGB]{181,200,130}16.90
& \cellcolor[RGB]{206,205,126}32.37
& \cellcolor[RGB]{181,200,130}0.622
& \cellcolor[RGB]{117,185,141}14.12
& \cellcolor[RGB]{124,187,140}19.09
& \cellcolor[RGB]{117,185,141}0.657 \\
& \textbf{FPCA-fVARX}
& \cellcolor[RGB]{117,186,141}7.84
& \cellcolor[RGB]{125,187,139}11.63
& \cellcolor[RGB]{117,186,141}0.693
& \cellcolor[RGB]{170,197,132}16.32
& \cellcolor[RGB]{220,208,124}33.28
& \cellcolor[RGB]{170,197,132}0.601
& \cellcolor[RGB]{121,186,140}14.22
& \cellcolor[RGB]{129,188,139}19.29
& \cellcolor[RGB]{121,186,140}0.662 \\
\bottomrule
\end{tabular}
\caption{\rev{Clearing price prediction performance. MAE and RMSE are expressed in \euro/MWh. The green-yellow-red colormap is applied column-wise.}}
\label{tab:price_all_markets}
\end{table}

\begin{figure*}[h!]
     \centering
     \begin{subfigure}[b]{0.355\textwidth}
         \includegraphics[width=\textwidth]{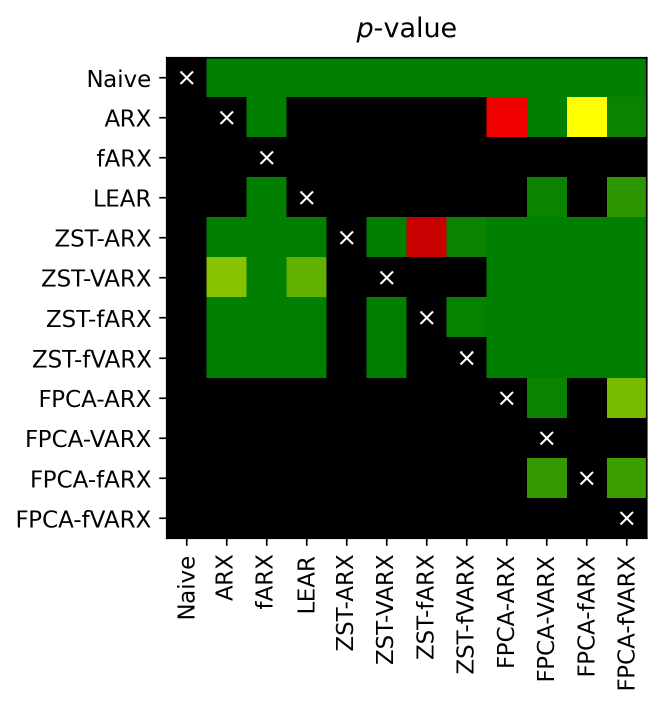}
         \caption{GME}
         \label{fig:dm_price_daily:gme}
     \end{subfigure}
     \begin{subfigure}[b]{0.278\textwidth}
         \includegraphics[width=\textwidth]{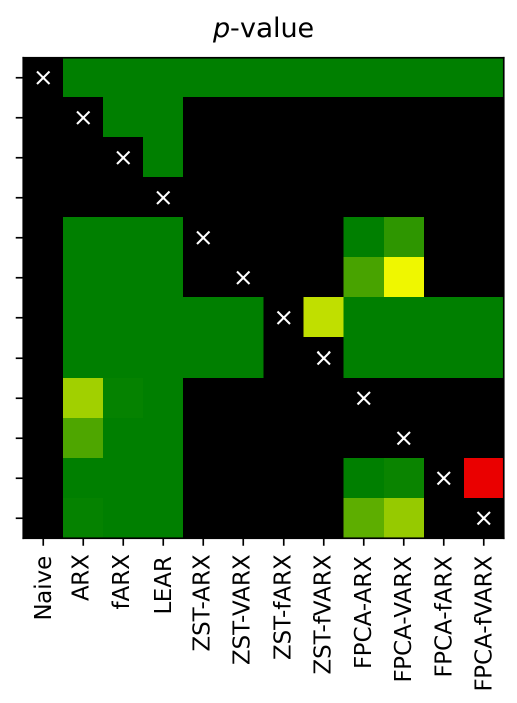}
         \caption{EPEX-DE-LU}
         \label{fig:dm_price_daily:epex-de-lu}
     \end{subfigure}
     \begin{subfigure}[b]{0.351\textwidth}
         \includegraphics[width=\textwidth]{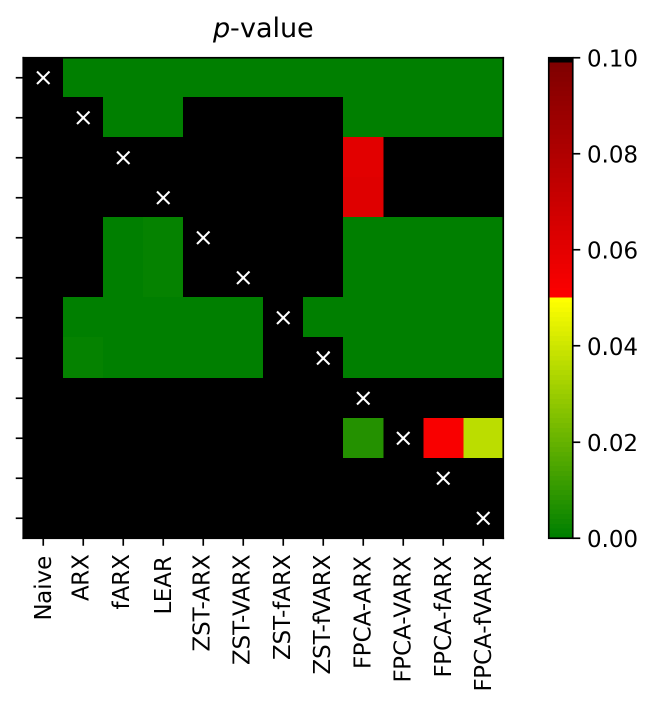}
         \caption{EPEX-FR}
         \label{fig:dm_price_daily:epex-fr}
     \end{subfigure}
        \caption{\rev{(\textit{Color optional}) Results of the Diebold-Mariano test for the difference in clearing price forecasting performance on average \textit{daily} absolute errors. The alternative hypothesis is that models on the $x$-axis outperform those on the $y$-axis (one-sided test).}}
        \label{fig:dm_price_daily}
\end{figure*}

Examples of curves and price forecast for each of the nine curve-based models and each market are pictured in Figs. \ref{fig:preds_gme}, \ref{fig:preds_epex-de-lu} and \ref{fig:preds_epex-fr}, in \ref{sec:preds} and a analysis of the computational efficiency of each model is presented in Table \ref{tab:effpoint} in \ref{sec:compute}. \rev{The global price forecasting performance for the four \textit{price-based} and eight \textit{curve-based} models is reported in Table \ref{tab:price_all_markets}, with the associated DM tests in Fig. \ref{fig:dm_price_daily}.}

\rev{All models significantly outperform \textbf{Naive} in terms of MAE. For the EPEX markets, however, full \textbf{ZST} models yield high RMSE values, indicating the presence of occasional large errors.}

\rev{Regarding the comparison between \textbf{FPCA} and \textbf{ZST}, every \textbf{FPCA} curve-based model significantly outperforms its \textbf{ZST} counterpart at the 0.1\% level for both \textbf{GME} and \textbf{EPEX-FR}. For \textbf{EPEX-DE-LU}, the same holds for all pairings except \textbf{VARX}, where significance is reached at the 5\% level only.}

\rev{A key finding is that strong curve forecasting performance does not necessarily translate into strong clearing price forecasting performance. For instance, on the EPEX markets, the full \textbf{ZST} models rank among the top performers for curve forecasting on both sides, yet perform poorly on price forecasting.}

\rev{Among \textbf{ZST} models, the best vector modeling approaches appear to be the concurrent ones --- without clear distinction between them --- for the two EPEX markets and the multivariate concurrent (\textbf{VARX}) for \textbf{GME}. Among \textbf{FPCA} models, incorporating cross-component effects (multivariate models) yields a significant improvement for \textbf{GME} only (as confirmed by the DM tests), while accounting for cross-hour dependence (full models) provides no consistent benefit across any of the three markets.}

\rev{A further key result is that \textbf{FPCA} curve-based models are competitive with price-based models, unlike their \textbf{ZST} counterparts. For \textbf{EPEX-DE-LU}, \textbf{fARX} and \textbf{LEAR} (price-based) significantly outperform all \textbf{FPCA} models at the 0.1\% level. For \textbf{GME} and \textbf{EPEX-FR}, however, no price-based model significantly outperforms any \textbf{FPCA} model. Moreover, all \textbf{FPCA} models significantly outperform the price-based \textbf{ARX} at the 5\% level for \textbf{GME} (except \textbf{FPCA-ARX}) and at the 0.1\% level for \textbf{EPEX-FR}. On \textbf{GME}, \textbf{FPCA-VARX} and \textbf{FPCA-fVARX} additionally outperform \textbf{LEAR} at the 1\% level.}

\begin{figure*}[h!]
     \centering
     \begin{subfigure}[b]{0.335\textwidth}
         \includegraphics[width=\textwidth]{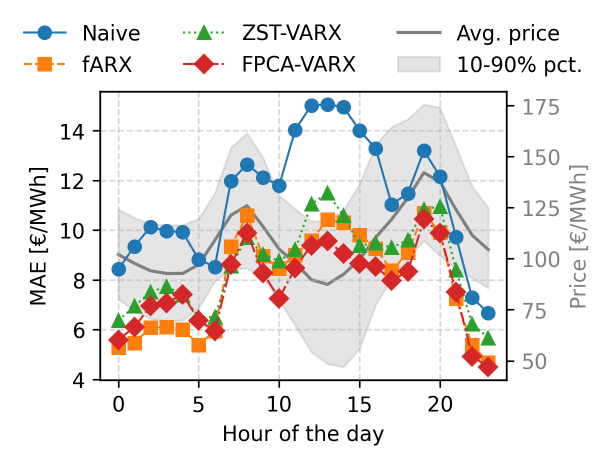}
         \caption{GME}
         \label{fig:mae_price_hourly:gme}
     \end{subfigure}
     \hspace{-3mm}
     \begin{subfigure}[b]{0.335\textwidth}
         \includegraphics[width=\textwidth]{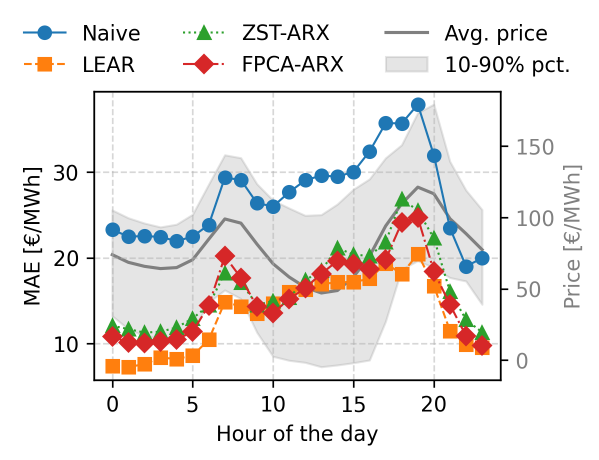}
         \caption{EPEX-DE-LU}
         \label{fig:mae_price_hourly:epex-de-lu}
     \end{subfigure}
     \hspace{-3mm}
     \begin{subfigure}[b]{0.335\textwidth}
         \includegraphics[width=\textwidth]{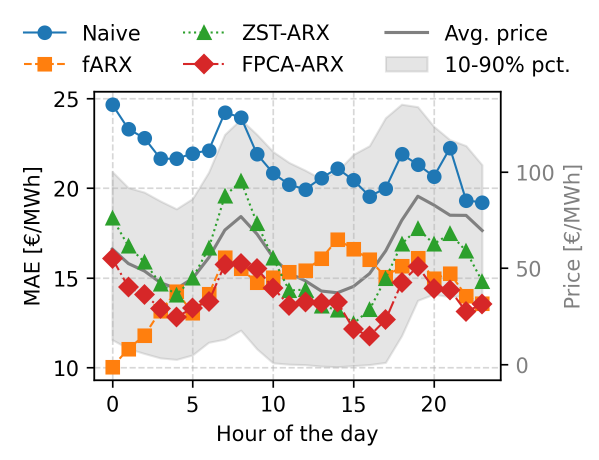}
         \caption{EPEX-FR}
         \label{fig:mae_price_hourly:epex-fr}
     \end{subfigure}
        \caption{\rev{(\textit{Color optional}) Mean absolute error of price forecasts vs price distribution observed during the test period for each hour. For legibility, only the best price-based, ZST-curve-based and FPCA-curve-based are represented.}}
        \label{fig:mae_price_hourly}
\end{figure*}

\begin{figure*}[h!]
     \centering
     \begin{subfigure}[b]{0.362\textwidth}
         \includegraphics[width=\textwidth]{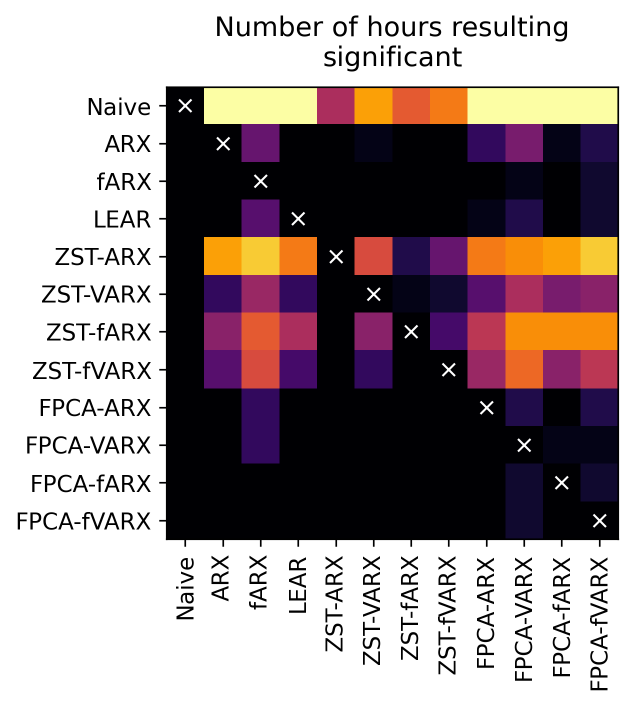}
         \caption{GME}
         \label{fig:dm_price_hourly:gme}
     \end{subfigure}
     \begin{subfigure}[b]{0.280\textwidth}
         \includegraphics[width=\textwidth]{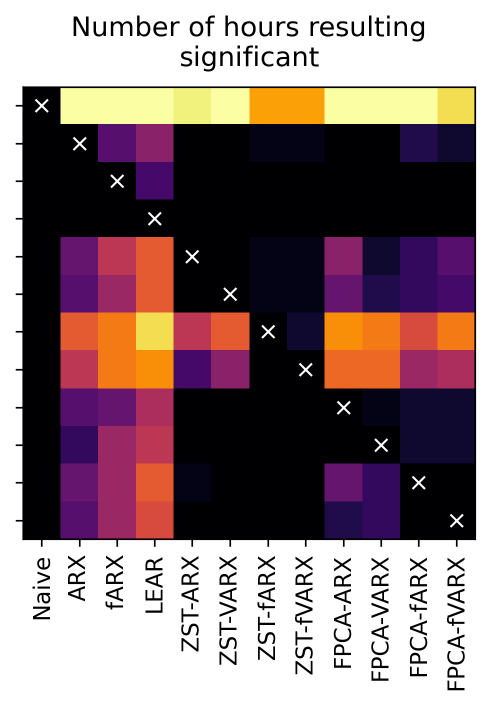}
         \caption{EPEX-DE-LU}
         \label{fig:dm_price_hourly:epex-de-lu}
     \end{subfigure}
     \begin{subfigure}[b]{0.344\textwidth}
         \includegraphics[width=\textwidth]{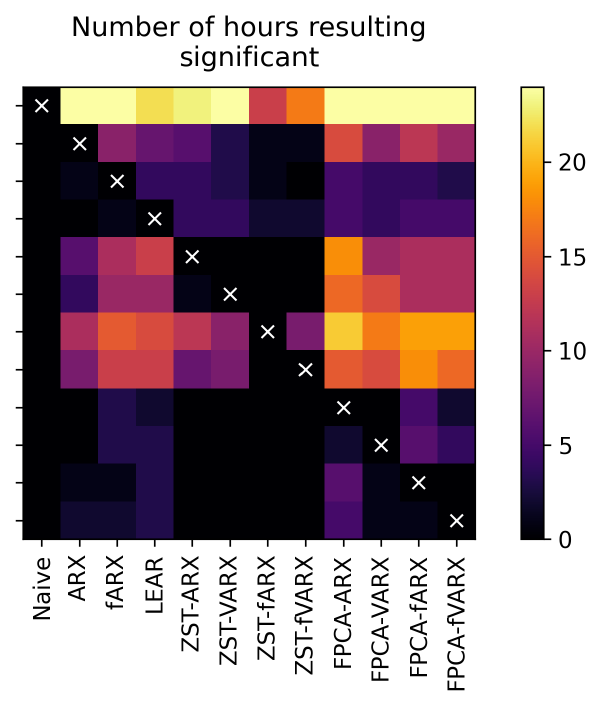}
         \caption{EPEX-FR}
         \label{fig:dm_price_hourly:epex-fr}
     \end{subfigure}
        \caption{\rev{(\textit{Color optional}) Results of the Diebold-Mariano tests for the difference in clearing price forecasting performance on absolute errors, at the hour level. The alternative hypothesis is that models on the $x$-axis outperform those on the $y$-axis (one-sided test). The significance threshold is set to 0.5\% \citep{benjamin_redefine_2017}.}}
        \label{fig:dm_price_hourly}
\end{figure*}

\medskip

\rev{Fig. \ref{fig:mae_price_hourly} shows the hourly mean absolute error of price forecasts, with associated DM tests in Fig. \ref{fig:dm_price_hourly}. Across all three markets, price-based models tend to perform better at night, coinciding with low-demand periods. For \textbf{EPEX-DE-LU} (Figs. \ref{fig:mae_price_hourly:epex-de-lu} and \ref{fig:dm_price_hourly:epex-de-lu}), \textbf{LEAR} dominates, significantly outperforming all curve-based models at the 0.5\% level for at least half of the 24 hours.  For \textbf{GME}, Fig. \ref{fig:mae_price_hourly:gme} suggests that \textbf{FPCA-VARX} performs marginally better than the best price-based model (\textbf{fARX}) during midday hours (10:00–17:00), when prices occasionally drop due to high renewable generation, though this difference is statistically significant at the 0.5\% level for only one hour (Fig. \ref{fig:dm_price_hourly:gme}). For \textbf{EPEX-FR}, both the best \textbf{ZST} and \textbf{FPCA} curve-based models clearly outperform the best price-based model (\textbf{fARX}) during the 10:00–17:00 window (Fig. \ref{fig:mae_price_hourly:epex-fr}), when null or negative prices are most frequently observed, with this advantage reaching significance at the 0.5\% level for several hours --- five hours for \textbf{FPCA-ARX} and four for \textbf{ZST-ARX} (Fig. \ref{fig:dm_price_hourly:epex-fr}). These differences can be substantial: for the 15:00–16:00 interval, \textbf{fARX} yields an MAE of 16.61\euro/MWh, compared to 12.14\euro/MWh for \textbf{FPCA-ARX} and 12.45\euro/MWh for \textbf{ZST-ARX}.}

\bigskip

\rev{Finally, Table \ref{tab:stat_vs_dyna_price} in \ref{sec:stat_vs_dyna} suggests that the dynamic representation strategy slightly improves clearing price prediction performance, most evidently for \textbf{EPEX-DE-LU}, although the effect is marginal and occasionally negative for some models.}

\subsection{\rev{Sensitivity to number of supply and demand components ($K_s$ and $K_d)$}}

\begin{figure*}[h!]
     \centering
     \begin{subfigure}[b]{0.49\textwidth}
         \includegraphics[width=\textwidth]{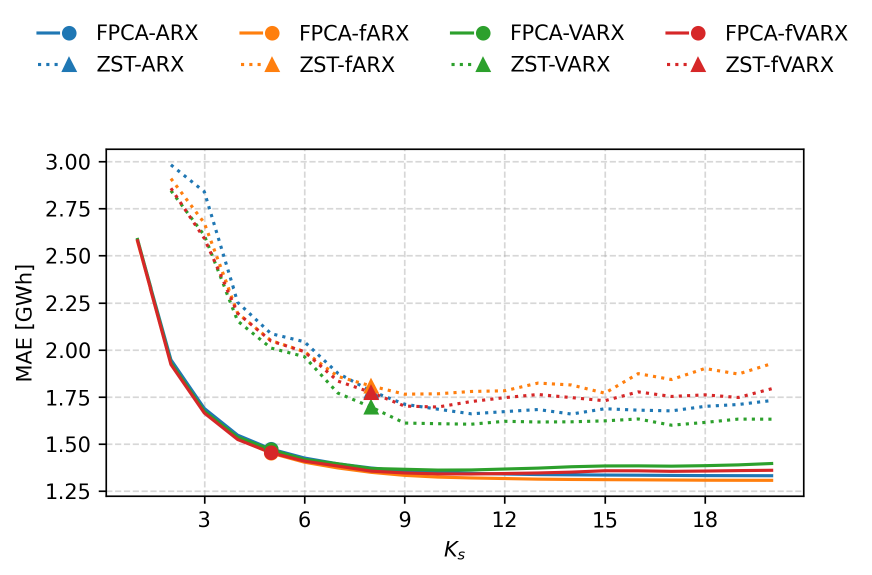}
         \caption{Supply (\textbf{GME})}
         \label{fig:sens_curve_mae:gme_supply}
     \end{subfigure}
     \begin{subfigure}[b]{0.49\textwidth}
         \includegraphics[width=\textwidth]{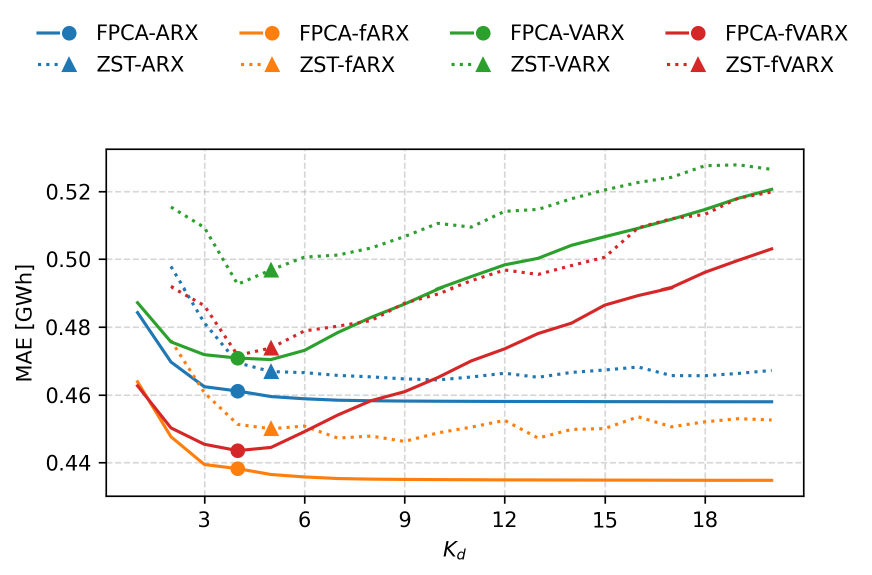}
         \caption{Demand (\textbf{GME})}
         \label{fig:sens_curve_mae:gme_demand}
     \end{subfigure}
     \begin{subfigure}[b]{0.49\textwidth}
         \includegraphics[width=\textwidth]{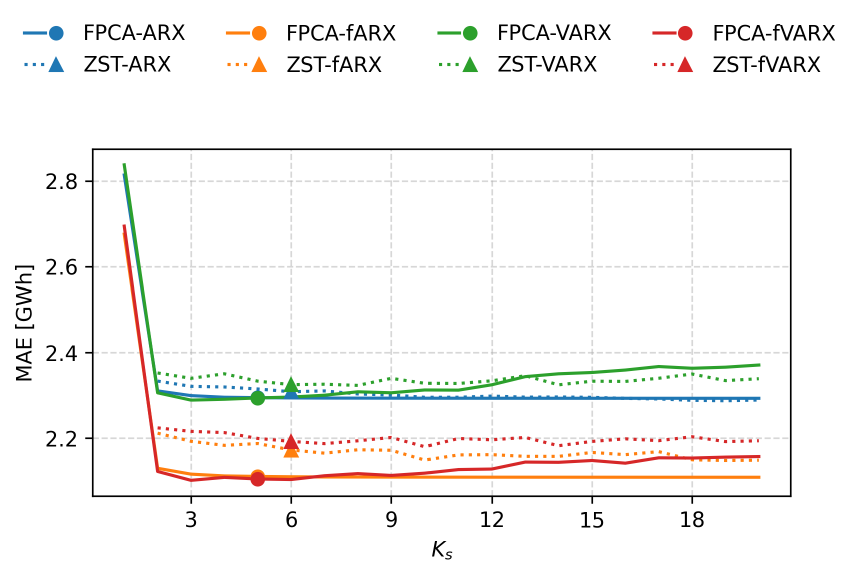}
         \caption{Supply (\textbf{EPEX-DE-LU})}
         \label{fig:sens_curve_mae:epex-de-lu_supply}
     \end{subfigure}
     \begin{subfigure}[b]{0.49\textwidth}
         \includegraphics[width=\textwidth]{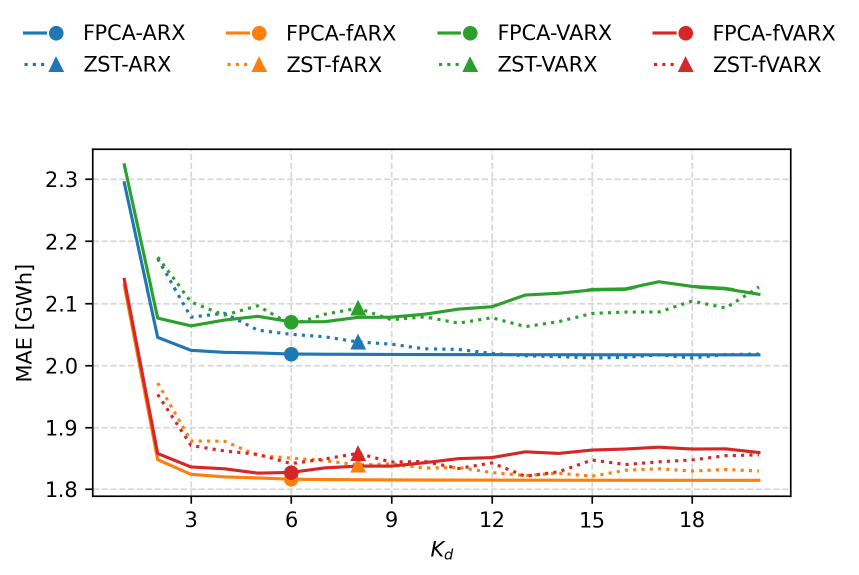}
         \caption{Demand (\textbf{EPEX-DE-LU})}
         \label{fig:sens_curve_mae:epex-de-lu_demand}
     \end{subfigure}
     \begin{subfigure}[b]{0.49\textwidth}
         \includegraphics[width=\textwidth]{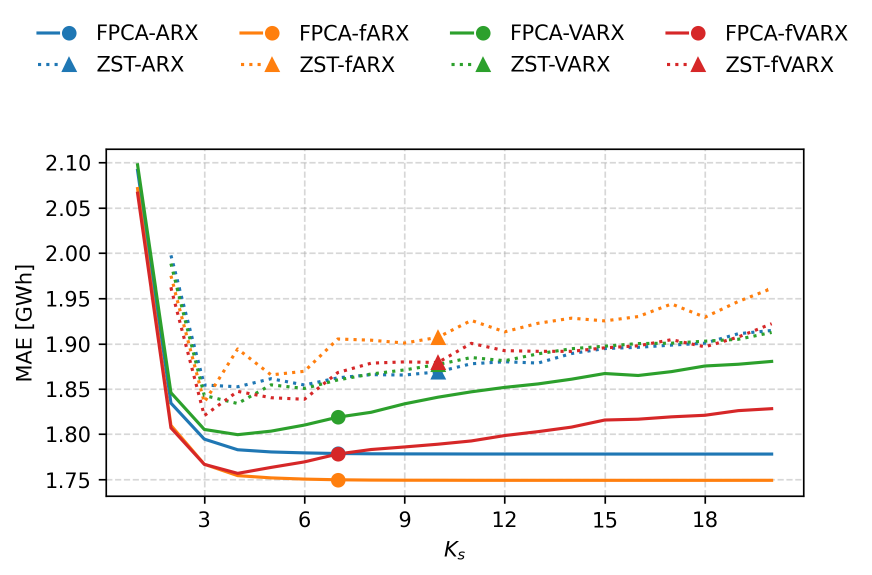}
         \caption{Supply (\textbf{EPEX-FR})}
         \label{fig:sens_curve_mae:epex-fr_supply}
     \end{subfigure}
     \begin{subfigure}[b]{0.49\textwidth}
         \includegraphics[width=\textwidth]{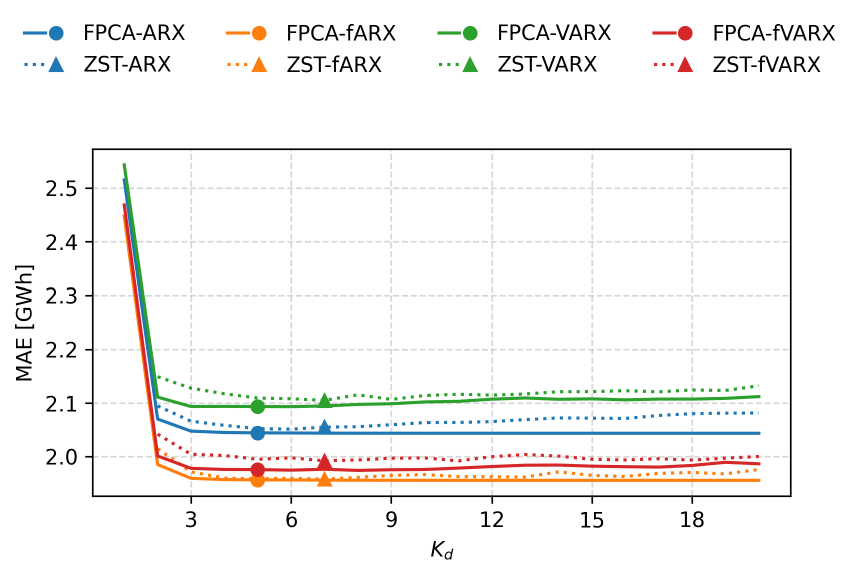}
         \caption{Demand (\textbf{EPEX-FR})}
         \label{fig:sens_curve_mae:epex-fr_demand}
     \end{subfigure}
        \caption{\rev{(\textit{Color optional}) Sensitivity of \textit{supply} (left panel) and \textit{demand} (right panel) curve forecasting performance, as measured by the mean absolute error, with respect to number of vector components considered for each curve type  ($K_s$ for supply, $K_d$ for demand). The markers show the values that were chosen for the analysis.}}
        \label{fig:sens_curve_mae}
\end{figure*}

\rev{The effect of the number of components ($K_s$ for supply and $K_d$ for demand) on the functional mean absolute error of curves forecasts is plotted in Fig. \ref{fig:sens_curve_mae}.}

\rev{For \textbf{GME}, the conclusions are largely insensitive to $K_s$ on the supply side, and the ex-ante selection procedure appears effective, as increasing $K_s$ beyond the selected value yields little improvement. On the demand side, increasing $K_d$ has negligible effect for curve models based on univariate vector models, but clearly degrades performance for those based on multivariate vector models. The explanation is straightforward: a larger $K_d$ increases the number of predictors (which may not be relevant in the case of \textbf{GME} demand curves, given their inelasticity) and, despite LASSO regularization, raises the risk of overfitting. Nonetheless, the selected values of $K_d$ remain optimal for both \textbf{FPCA} and \textbf{ZST}.}

\rev{For \textbf{EPEX-DE-LU}, lower values of $K_s$ and $K_d$ would have sufficed without sacrificing performance: $K_s=2$ for both \textbf{FPCA} and \textbf{ZST} (versus 5 and 6 selected), and $K_d=3$ (versus 6 for \textbf{FPCA} and 8 for \textbf{ZST}). The conclusions remain robust throughout, and the same deterioration with increasing $K_d$ observed for multivariate models on \textbf{GME} is replicated here, though for \textbf{FPCA} models only --- a finding that favors univariate models in terms of robustness.}

\rev{For \textbf{EPEX-FR}, slightly better supply curve performance could have been achieved with a lower $K_s$, such as $K_s=3$ for \textbf{FPCA} (versus 7 selected) or $K_s=4$ for \textbf{ZST} (versus 10 selected). Notably, this does not apply to univariate \textbf{FPCA} models, further confirming their robustness to overestimation of the optimal dimension. On the demand side, the results are highly stable and comparable performance could have been achieved with $K_d=2$ or 3 (versus 5 for \textbf{FPCA} and 7 for \textbf{ZST}).}

\begin{figure*}[h!]
     \centering
     \begin{subfigure}[b]{0.49\textwidth}
         \includegraphics[width=\textwidth]{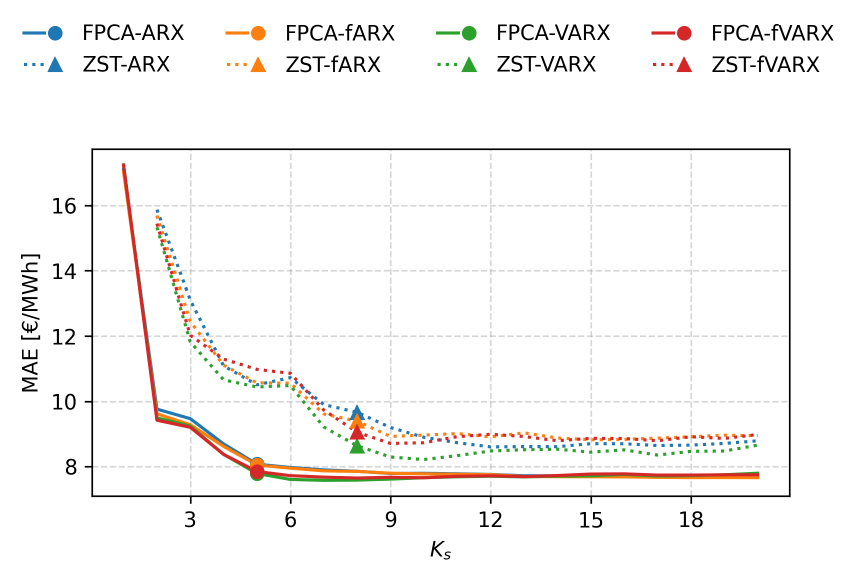}
         \caption{$K_s$ effect (\textbf{GME})}
         \label{fig:sens_price_mae:gme_supply}
     \end{subfigure}
     \begin{subfigure}[b]{0.49\textwidth}
         \includegraphics[width=\textwidth]{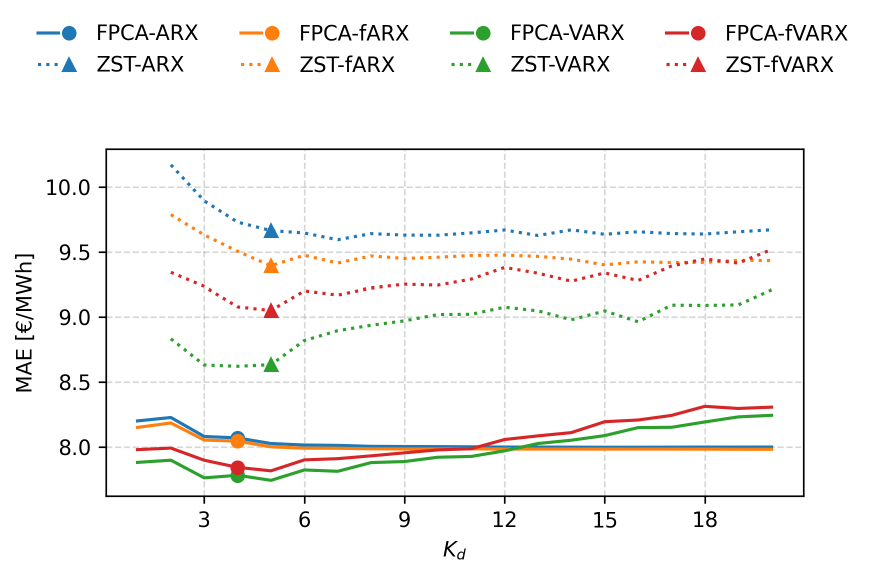}
         \caption{$K_d$ effect (\textbf{GME})}
         \label{fig:sens_price_mae:gme_demand}
     \end{subfigure}
     \begin{subfigure}[b]{0.49\textwidth}
         \includegraphics[width=\textwidth]{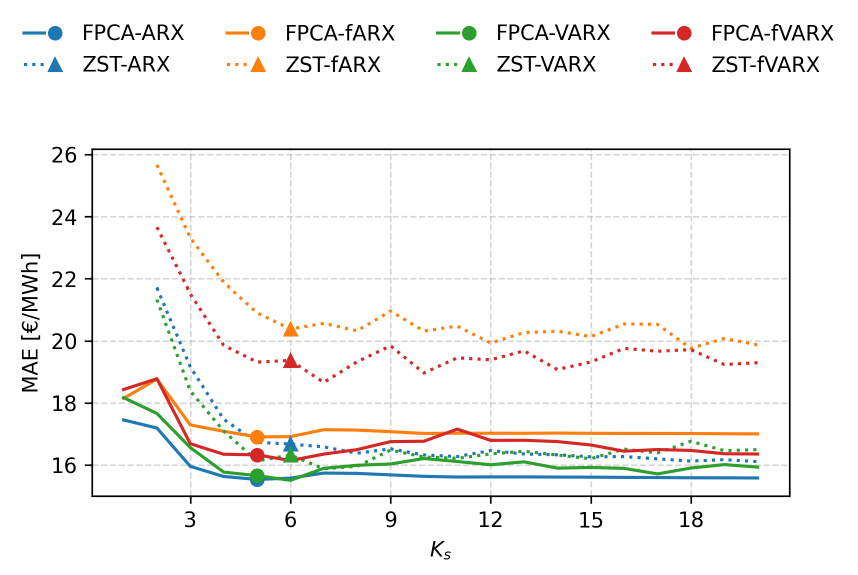}
         \caption{$K_s$ effect (\textbf{EPEX-DE-LU})}
         \label{fig:sens_price_mae:epex-de-lu_supply}
     \end{subfigure}
     \begin{subfigure}[b]{0.49\textwidth}
         \includegraphics[width=\textwidth]{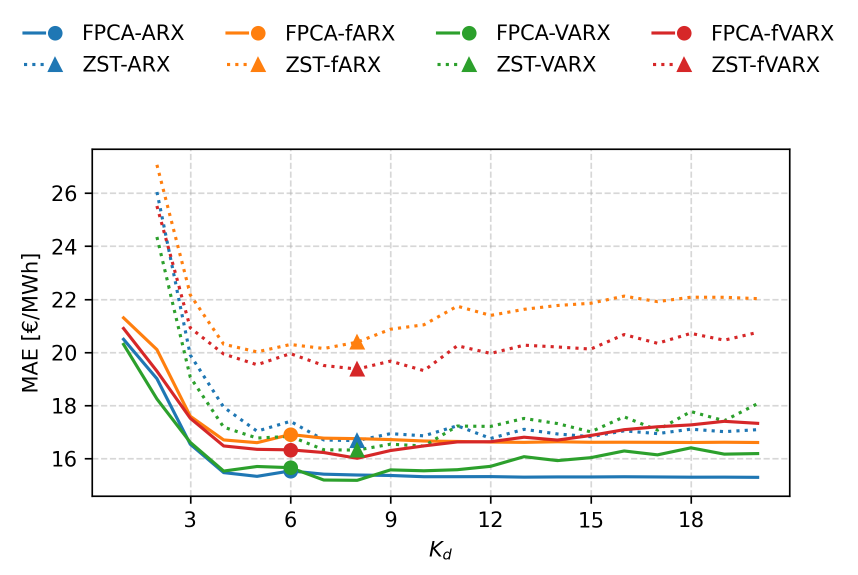}
         \caption{$K_d$ effect (\textbf{EPEX-DE-LU})}
         \label{fig:sens_price_mae:epex-de-lu_demand}
     \end{subfigure}
     \begin{subfigure}[b]{0.49\textwidth}
         \includegraphics[width=\textwidth]{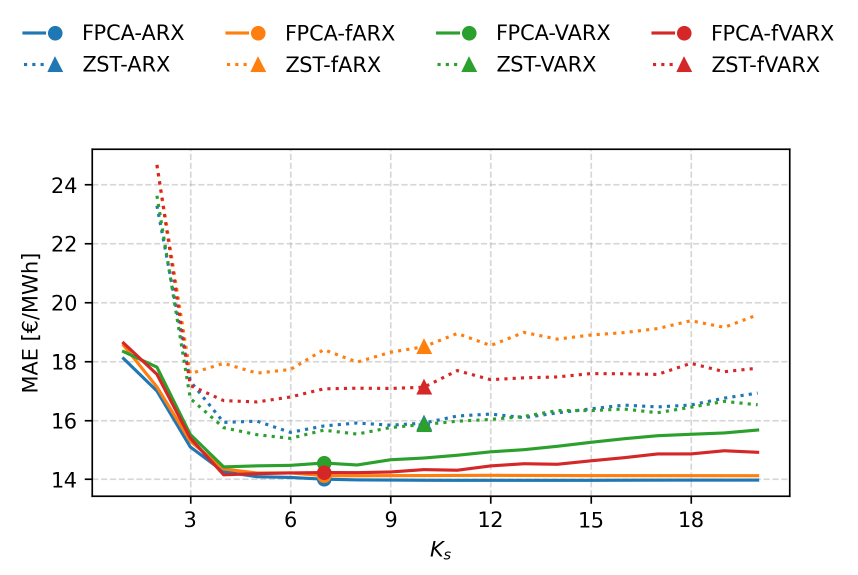}
         \caption{$K_s$ effect (\textbf{EPEX-FR})}
         \label{fig:sens_price_mae:epex-fr_supply}
     \end{subfigure}
     \begin{subfigure}[b]{0.49\textwidth}
         \includegraphics[width=\textwidth]{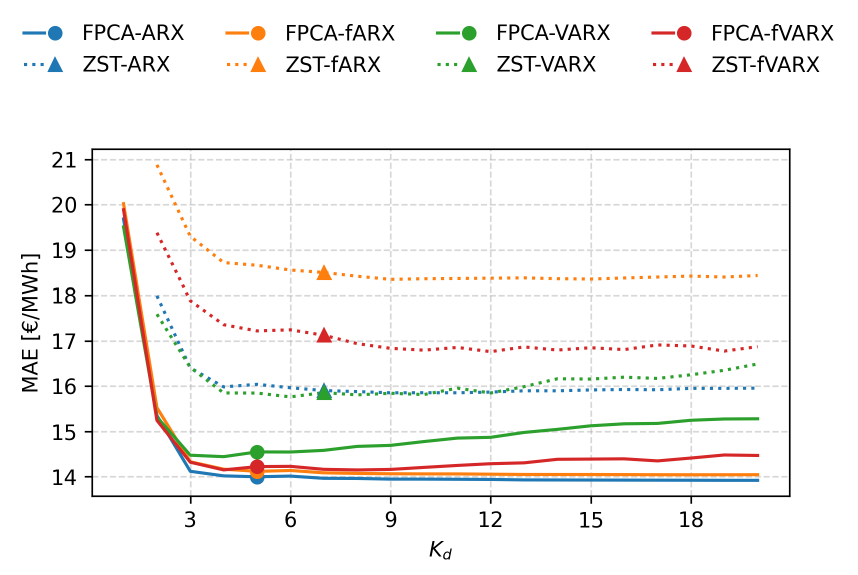}
         \caption{$K_d$ effect (\textbf{EPEX-FR})}
         \label{fig:sens_price_mae:epex-fr_demand}
     \end{subfigure}
        \caption{\rev{(\textit{Color optional}) Sensitivity of \textit{clearing price} forecasting performance, as measured by the mean absolute error, with respect to number of vector components considered for supply curves ($K_s$, left panel) and demand curves ($K_d$, right panel). The markers show the values that were chosen for the analysis.}}
        \label{fig:sens_price_mae}
\end{figure*}

\bigskip

\rev{The effect of $K_s$ and $K_d$ on price predictions' MAE is plotted in Fig. \ref{fig:sens_price_mae}.}
\rev{For \textbf{GME}, the selected values ($K_s = 5$) for \textbf{FPCA}, $K_s = 8$ for \textbf{ZST}) offer a good trade-off between accuracy and model complexity. A marginal improvement (on the order of a few cents per MWh) could be obtained with one additional component for each, but gains plateau quickly beyond that. Notably, for large $K_s$, the performance of all \textbf{FPCA} models converge. The selected $K_d$ turns out to be optimal, and the characteristic performance degradation of multivariate models under increasing $K_d$ is again evident. Overall, conclusions are robust to small variations in both $K_s$ and $K_d$.}

\rev{For \textbf{EPEX-DE-LU}, the selected values are near-optimal, with comparable performance achievable using $K_s=4$ (versus 5 for \textbf{FPCA} and 6 for \textbf{ZST}) and $K_d = 4$ (versus 6 for \textbf{FPCA} and 5 for \textbf{ZST}). \textbf{FPCA-ARX} stands out as the most stable model across component choices and, given its simplicity, should be preferred in practice. The comparative conclusions between models remain unchanged across component choices.}

\rev{For \textbf{EPEX-FR}, the selected values appear conservative. Equivalent performance is achievable with $K_s=4$ (versus 7 for \textbf{FPCA} and 10 for \textbf{ZST}) and $K_d=3$ (versus 5 for \textbf{FPCA} and 4 for \textbf{ZST}). Despite this, the conclusion that \textbf{FPCA} outperforms \textbf{ZST} in clearing price forecasting holds robustly across all component configurations.}

\bigskip

\rev{We conclude this section with a brief discussion of results obtained without restricting the curves (\ref{sec:res_unres}). Curve prediction conclusions are unchanged: functional error metrics are globally shifted as they now integrate pointwise errors over a larger domain, but relative model rankings are preserved. One exception concerns \textbf{GME} demand curves, where \textbf{FPCA} models appear to struggle in small portions of the domain above 800\euro/MWh, which inflates their functional metrics. For price prediction, overall performance is lower and a higher number of curve components is required, further justifying the choice to restrict the price domain when price forecasting is the primary objective. Comparative conclusions between curve-based models remain unaffected.}

\section{\rev{Discussion and conclusion}}
\label{sec:conclusion}

This work introduced a general functional data analysis framework for forecasting day-ahead merit-order curves \rev{with a focus on (clearing) price prediction}, leveraging functional principal component analysis to represent a pair of supply and demand curves in a low-dimensional vector space and employing regularized vector autoregressive models for their prediction. The application to the Italian (GME), \rev{German (EPEX-DE-LU) and French (EPEX-FR)} day-ahead markets during the 2023-2024 period not only demonstrated the method's effectiveness in forecasting merit-order curves but also in producing competitive electricity price forecasts. We tested four variations of the model, each treating the hourly time series as 24 independent daily series and differing in whether they accounted for (i) cross-dependence between hours and (ii) cross-dependence between components of the vector representation.

\rev{We found that accounting for cross-component dependence does not lead to performance gains in curve prediction, but may improve price forecasts --- as observed for GME --- though this effect vanishes when a sufficient number of curve components is included. This implies, in particular, that supply and demand curves can be modeled independently without meaningful loss of performance.  The inclusion of cross-hour dependence was found to improve curve predictions in most cases, but never improved price predictions. On this basis, we recommend the simplest ARX form for forecasting the FPCA vector representation --- one that incorporates neither cross-hour nor cross-component effects --- as it provides strong curve and price forecasting performance at very low computational cost and high level of interpretability. Additionally, the number of components can be reliably selected by identifying the elbow point of the clearing price approximation error on an initial training set, without resorting to a costly cross-validation procedure. Furthermore, if price forecasting is the central objective (the perspective we adopt in this paper), restricting the curves domain to a realistic clearing price range leads to more accurate forecasts with fewer curve components. Generally, we observed that choosing 4 components for each side (supply and demand) is a safe choice for both curve and price prediction, regardless of whether a full or restricted price range is considered.}

\medskip

\rev{The proposed functional data approach was rigorously compared with the original discretization-based approach of \citet{ziel_electricity_2016}, which shares the same high-dimensional vector autoregression framework but differs in its curve representation strategy (based on curve discretization and denoted as ZST). Although the two representation methods offer comparable curve approximation quality, the functional representation consistently outperforms its discrete counterpart on supply curves and price forecasting, across all three markets. This observation does not hold for demand curves where the two approaches were found to perform equally.}
\rev{A further notable finding is that strong curve forecasting performance does not necessarily translate into strong clearing price forecasting performance. This disconnect, observed most strikingly for ZST models on the EPEX markets, has important practical implications: if price forecasting is one of the objectives, curve prediction models must be evaluated and selected with price forecasting performance as the criterion, rather than on the basis of curve prediction accuracy alone.} 

\medskip

\rev{The proposed curve-based price forecasting approach was also compared with state-of-the-art price-based models (i.e., directly modeling the clearing price) based on the same regularized high-dimensional autoregression framework. The results reveal a market-dependent picture: For GME and EPEX-FR, functional curve-based and price-based models exhibit comparable performance though with a small advantage for the curve-based approach. Moreover, curve-based models consistently outperform price-based ones during midday hours (10:00–17:00), when prices frequently drop due to the combined effect of high renewable generation and low demand — with MAE reductions reaching up to 27\% within this window. For EPEX-DE-LU, by contrast, price-based models retain a clear and statistically significant advantage over all curve-based models.
One possible explanation is the higher volatility of German day-ahead prices compared to Italian or French ones, and the fact that the regularized high-dimensional autoregression framework works particularly well for German prices, as supported by the low rMAE achieved by LEAR on EPEX-DE-LU (0.49, close to values to reported by \citet{lago_forecasting_2021} for the 2016-2017 period), compared to 0.68 on EPEX-FR and 0.73 on GME.}

\medskip

\rev{The present study focused exclusively on point forecasting. A natural extension is to examine whether modeling uncertainty at the curve level --- rather than at the price level --- can also improve probabilistic price forecasts.}
Additional analyses could also examine the effect of alternative curves representations, the use of a single versus hour-specific representation and the calibration window size, on both curve and price forecasting performance. \rev{A detailed analysis on model interpretation, identifying the relative effects of lagged endogenous, exogenous and day-type dummy variables would definitely be valuable.} Besides, more complex non-linear models such as feedforward neural networks could be considered to jointly predict the 24 vector representations within a single model. Finally, the recent switch to 15-min intervals in European day-ahead markets, making no longer 24 but 96 curves pairs to predict, could justify a functional data rather than multivariate perspective to capture intraday dependence. These aspects will be explored in future research.




\section*{Declaration of generative AI and AI-assisted technologies in the writing process}
\noindent During the preparation of this work the authors used OpenAI ChatGPT and Google Gemini in order to obtain suggestions for improving writing fluency and clarity. After using this tool/service, the authors reviewed and edited the content as needed and take full responsibility for the content of the published article.




\bibliographystyle{model5-names}
\bibliography{references}



\clearpage

\appendix

\section{\rev{Reformulation of the FPCA-based models as FAR models}}
\label{sec:far}

\rev{In the following section, we show that the FPCA-based models can be reformulated as functional autoregressive (FAR) models. We focus on the concurrent ARX (\ref{subsec:arx_far}) and VARX (\ref{subsec:varx_far}) models only but the proof can be easily extended to the full models.}

\bigskip

\subsection{\rev{Concurrent ARX}}
\label{subsec:arx_far}

\rev{The truncated KL decomposition for either supply ($K=K_s$) or demand curves ($K=K_d$) reads:
$$Q_{d,h}(p) \approx \overline{Q}(p) + \sum_{k=1}^K y_{d, h, k}. \xi_k(p)$$
Let $\tilde{Q}_{d,h}(p)=Q_{d,h}(p)-\overline{Q}(p)$ be the centered functions. The concurrent ARX model (\textbf{ARX}), as specified in section \ref{sec:methods}, writes:
\begin{equation}
\begin{split}
y_{d,h,k} &= c_{h, k} + \phi_{1,h,k}y_{d-1,h,k} + \phi_{2,h,k}y_{d-2,h,k} + \phi_{3,h,k}y_{d-3,h,k} + \phi_{7,h,k}y_{d-7,h,k} \\
&\quad + \boldsymbol{\beta}^{\top}_{0,h,k} \mathbf{x}_{d,h} + \boldsymbol{\beta}_{1,h,k}^{\top} \mathbf{x}_{d-1,h} +  \boldsymbol{\beta}_{7,h,k}^{\top} \mathbf{x}_{d-7,h} + \boldsymbol{\theta}^{\top}_{h,k} \mathbf{z}_{d} + \varepsilon_{d,h,k}
\end{split}
\end{equation}}

\rev{By the orthornormal property of the FPCs, we have:
$$y_{d,h,k}=\int_{\mathcal{D}}\tilde{Q}_{d,h}(p)\xi_k(p)dp$$
where $\mathcal{D}$ is the price domain over which supply and demand curves are defined.}

\bigskip

\noindent \rev{Replacing in the above equation:
\begin{equation}
\begin{split}
y_{d,h,k} &= c_{h,k} +
\phi_{1,h,k}\int_{\mathcal{D}}\tilde{Q}_{d-1,h}(p)\xi_k(p)dp +
\phi_{2,h,k}\int_{\mathcal{D}}\tilde{Q}_{d-2,h}(p)\xi_k(p)dp +
\phi_{3,h,k}\int_{\mathcal{D}}\tilde{Q}_{d-3,h}(p)\xi_k(p)dp \\
&\quad +
\phi_{7,h,k}\int_{\mathcal{D}}\tilde{Q}_{d-7,h}(p)\xi_k(p)dp +
\boldsymbol{\beta}^{\top}_{0,h,k} \mathbf{x}_{d,h} +
\boldsymbol{\beta}_{1,h,k}^{\top} \mathbf{x}_{d-1,h} +
\boldsymbol{\beta}_{7,h,k}^{\top} \mathbf{x}_{d-7,h} +
\boldsymbol{\theta}^{\top}_{h,k} \mathbf{z}_{d} +
\varepsilon_{d,h,k}
\end{split}
\end{equation}
simplified:
\begin{equation}
\begin{split}
y_{d,h,k} &= c_{h,k} +
\int_{\mathcal{D}} \left[\phi_{1,h,k}\tilde{Q}_{d-1,h}(p)+\phi_{2,h,k}\tilde{Q}_{d-2,h}(p) + \phi_{3,h,k}\tilde{Q}_{d-3,h}(p) + \phi_{7,h,k}\tilde{Q}_{d-7,h}(p) \right]\xi_k(p)dp \\
&\quad +
\boldsymbol{\beta}^{\top}_{0,h,k} \mathbf{x}_{d,h} +
\boldsymbol{\beta}_{1,h,k}^{\top} \mathbf{x}_{d-1,h} +
\boldsymbol{\beta}_{7,h,k}^{\top} \mathbf{x}_{d-7,h} +
\boldsymbol{\theta}^{\top}_{h,k} \mathbf{z}_{d} +
\varepsilon_{d,h,k}
\end{split}
\end{equation}
}

\noindent \rev{Now plugging the expression above in the KL expansion:}

\rev{
\begin{equation*}
\begin{split}
    \tilde{Q}_{d,h}(p) &= \sum_{k=1}^K c_{h,k} \xi_k(p) +
    \int_{\mathcal{D}} \left[\sum_{k=1}^K\phi_{1,h,k}\xi_k(p)\xi_k(u)\right]\tilde{Q}_{d-1,h}(u)du 
    + \int_{\mathcal{D}} \left[\sum_{k=1}^K\phi_{2,h,k}\xi_k(p)\xi_k(u)\right]\tilde{Q}_{d-2,h}(u)du    \\ &\quad
    + \int_{\mathcal{D}} \left[\sum_{k=1}^K\phi_{3,h,k}\xi_k(p)\xi_k(u)\right]\tilde{Q}_{d-3,h}(u)du 
    + \int_{\mathcal{D}} \left[\sum_{k=1}^K\phi_{7,h,k}\xi_k(p)\xi_k(u)\right]\tilde{Q}_{d-7,h}(u)du \\ &\quad
    +\left[\sum_{k=1}^K\boldsymbol{\beta}_{0,h,k} \xi_k(p)\right]^{\top} \mathbf{x}_{d,h} +
    \left[\sum_{k=1}^K\boldsymbol{\beta}_{1,h,k} \xi_k(p)\right]^{\top} \mathbf{x}_{d-1,h} +
    \left[\sum_{k=1}^K\boldsymbol{\beta}_{7,h,k} \xi_k(p)\right]^{\top} \mathbf{x}_{d-7,h} \\ &\quad +
    \left[\sum_{k=1}^K\boldsymbol{\theta}_{h,k} \xi_k(p)\right]^{\top} \mathbf{z}_{d} + \sum_{k=1}^K\varepsilon_{d,h,k} \xi_k(p)
\end{split}
\end{equation*}
Which can be expressed as:
\begin{equation*}
\begin{split}
    \tilde{Q}_{d,h}(p) &= c_h(p)+
    \int_{\mathcal{D}} \psi_{1,h}(p, u)\tilde{Q}_{d-1,h}(u)du 
    + \int_{\mathcal{D}} \psi_{2,h}(p, u)\tilde{Q}_{d-2,h}(u)du  
    + \int_{\mathcal{D}} \psi_{3,h}(p, u)\tilde{Q}_{d-3,h}(u)du  \\ &\quad
    + \int_{\mathcal{D}}  \psi_{7,h}(p, u)\tilde{Q}_{d-7,h}(u)du
    +\boldsymbol{\beta}_{0,h}(p)^{\top} \mathbf{x}_{d,h} +
    \boldsymbol{\beta}_{1,h}(p)^{\top} \mathbf{x}_{d-1,h} +
    \boldsymbol{\beta}_{7,h}(p)^{\top} \mathbf{x}_{d-7,h} \\ &\quad +
    \boldsymbol{\theta}_{h}(p)^{\top} \mathbf{z}_{d} + \varepsilon_{d,h}(p)
\end{split}
\end{equation*}
}

\rev{This is in fact the expression of a \textit{functional autoregressive model with exogenous covariates} (FARX, not to be confused with fARX which stands for \textit{full} ARX) model where the autoregressive kernels $\psi_{\cdot,h}$ are constrained to be symmetric.}

\subsection{\rev{Concurrent VARX}}
\label{subsec:varx_far}

\rev{The KL expansions for centered supply quantity curves $\tilde{Q}_{d,h}^{(s)}(p)$ and centered demand quantity curves $\tilde{Q}_{d,h}^{(d)}(p)$ read:
\begin{equation}
\tilde{Q}_{d,h}^{(s)}(p) \approx \sum_{k=1}^{K_s} y_{d, h, k}^{(s)} \xi_k^{(s)}(p), \quad \text{and} \quad \tilde{Q}_{d,h}^{(d)}(p) \approx \sum_{m=1}^{K_d} y_{d, h, m}^{(d)} \xi_m^{(d)}(p)
\end{equation}
where $\xi_k^{(s)}(p)$ and $\xi_m^{(d)}(p)$ are the orthonormal functional principal components (FPCs), and $y_{d,h,k}^{(s)}$ and $y_{d,h,m}^{(d)}$ are their corresponding time-dependent scores.}

\medskip

\rev{As defined in section \ref{sec:methods}, concatenating the $K_s$ supply scores and $K_d$ demand scores, we obtain a joint $K$-dimensional vector representation 
($K = K_s + K_d$):
\begin{equation}
\mathbf{y}_{d,h} = \left[ y_{d,h,1}^{(s)}, \dots, y_{d,h,K_s}^{(s)}, y_{d,h,1}^{(d)}, \dots, y_{d,h,K_d}^{(d)} \right]^{\top}
\end{equation}}

\rev{In the concurrent VARX model (\textbf{VARX}), any generic score component $y_{d,h,k}$ (for $k =1,...,K$) depends on its own past lags as well as the past lags of all other components across both curve types. Splitting the autoregressive summation into supply and demand components yields:
\begin{equation}
\begin{split}
y_{d,h,k} &= c_{h, k} + \sum_{j \in \{1,2,3,7\}} \left( \sum_{l=1}^{K_s}\phi_{j,h,k,l}y_{d-j,h,l}^{(s)} + \sum_{m=1}^{K_d}\phi_{j,h,k,m+K_s}y_{d-j,h,m}^{(d)} \right) \\
&\quad + \bm{\beta}^{\top}_{0,h,k} \mathbf{x}_{d,h} + \bm{\beta}_{1,h,k}^{\top} \mathbf{x}_{d-1,h} +  \bm{\beta}_{7,h,k}^{\top} \mathbf{x}_{d-7,h} + \bm{\theta}^{\top}_{h,k} \mathbf{z}_{d} + \varepsilon_{d,h,k}
\end{split}
\end{equation}}

\rev{By the orthonormal property of the FPCs we have:
\begin{equation}
y_{d,h,l}^{(s)} = \int_{\mathcal{D}}\tilde{Q}_{d,h}^{(s)}(u)\xi_l^{(s)}(u)du, \quad \text{and} \quad y_{d,h,m}^{(d)} = \int_{\mathcal{D}}\tilde{Q}_{d,h}^{(d)}(u)\xi_m^{(d)}(u)du
\end{equation}}

\rev{Replacing in the above equation:
\begin{equation}
\begin{split}
y_{d,h,k} &= c_{h,k} + \sum_{j \in \{1,2,3,7\}} \left( \int_{\mathcal{D}} \left[ \sum_{l=1}^{K_s} \phi_{j,h,k,l} \xi_l^{(s)}(u) \right] \tilde{Q}_{d-j,h}^{(s)}(u) du + \int_{\mathcal{D}} \left[ \sum_{m=1}^{K_d} \phi_{j,h,k,m+K_s} \xi_m^{(d)}(u) \right] \tilde{Q}_{d-j,h}^{(d)}(u) du \right) \\
&\quad + \bm{\beta}^{\top}_{0,h,k} \mathbf{x}_{d,h} + \bm{\beta}_{1,h,k}^{\top} \mathbf{x}_{d-1,h} + \bm{\beta}_{7,h,k}^{\top} \mathbf{x}_{d-7,h} + \bm{\theta}^{\top}_{h,k} \mathbf{z}_{d} + \varepsilon_{d,h,k}
\end{split}
\end{equation}}

\rev{Now plugging this expression back into the Karhunen-Lo\`eve expansions transforms the score vector autoregression into a \textit{bivariate} FARX model, which we write component-wise for simplicity:}

\bigskip

\rev{\noindent \textbf{Supply curves} (plugging $y_{d.h,k}$ for $k \in \{1, \dots, K_s\}$):
\begin{equation*}
\begin{split}
    \tilde{Q}_{d,h}^{(s)}(p) &= c_h^{(s)}(p) + \sum_{j \in \{1,2,3,7\}} \left( \int_{\mathcal{D}} \psi_{j,h}^{(ss)}(p, u)\tilde{Q}_{d-j,h}^{(s)}(u)du + \int_{\mathcal{D}} \psi_{j,h}^{(sd)}(p, u)\tilde{Q}_{d-j,h}^{(d)}(u)du \right) \\
    &\quad + \bm{\beta}_{0,h}^{(s)}(p)^{\top} \mathbf{x}_{d,h} + \bm{\beta}_{1,h}^{(s)}(p)^{\top} \mathbf{x}_{d-1,h} + \bm{\beta}_{7,h}^{(s)}(p)^{\top} \mathbf{x}_{d-7,h} + \bm{\theta}_{h}^{(s)}(p)^{\top} \mathbf{z}_{d} + \varepsilon_{d,h}^{(s)}(p)
\end{split}
\end{equation*}}

\rev{\noindent \textbf{Demand curves} (plugging $y_{d.h,k}$ for $k \in \{K_s+1, \dots, K_s+K_d\}$):
\begin{equation*}
\begin{split}
    \tilde{Q}_{d,h}^{(d)}(p) &= c_h^{(d)}(p) + \sum_{j \in \{1,2,3,7\}} \left( \int_{\mathcal{D}} \psi_{j,h}^{(ds)}(p, u)\tilde{Q}_{d-j,h}^{(s)}(u)du + \int_{\mathcal{D}} \psi_{j,h}^{(dd)}(p, u)\tilde{Q}_{d-j,h}^{(d)}(u)du \right) \\
    &\quad + \bm{\beta}_{0,h}^{(d)}(p)^{\top} \mathbf{x}_{d,h} + \bm{\beta}_{1,h}^{(d)}(p)^{\top} \mathbf{x}_{d-1,h} + \bm{\beta}_{7,h}^{(d)}(p)^{\top} \mathbf{x}_{d-7,h} + \bm{\theta}_{h}^{(d)}(p)^{\top} \mathbf{z}_{d} + \varepsilon_{d,h}^{(d)}(p)
\end{split}
\end{equation*}}

\rev{The Hilbert-Schmidt autoregressive operators with kernels $\psi_{j,h}^{(\cdot\cdot)}$ applying on the bivariate functional observations $[\tilde{Q}_{d,h}^{(s)}, \tilde{Q}_{d,h}^{(d)}]^\top$ can be written component-wise:
\begin{align*}
    \psi_{j,h}^{(ss)}(p, u) &= \sum_{k=1}^{K_s} \sum_{l=1}^{K_s} \phi_{j,h,k,l} \xi_k^{(s)}(p) \xi_l^{(s)}(u) \quad \text{(Supply $\rightarrow$ Supply)} \\
    \psi_{j,h}^{(sd)}(p, u) &= \sum_{k=1}^{K_s} \sum_{m=1}^{K_d} \phi_{j,h,k,m+K_s} \xi_k^{(s)}(p) \xi_m^{(d)}(u) \quad \text{(Demand $\rightarrow$ Supply)} \\
    \psi_{j,h}^{(ds)}(p, u) &= \sum_{m=1}^{K_d} \sum_{l=1}^{K_s} \phi_{j,h,m+K_s,l} \xi_m^{(d)}(p) \xi_l^{(s)}(u) \quad \text{(Supply $\rightarrow$ Demand)} \\
    \psi_{j,h}^{(dd)}(p, u) &= \sum_{m=1}^{K_d} \sum_{n=1}^{K_d} \phi_{j,h,m+K_s,n+K_s} \xi_m^{(d)}(p) \xi_n^{(d)}(u) \quad \text{(Demand $\rightarrow$ Demand)}
\end{align*}}

\rev{\textbf{Important note}: As $K_s, K_d \to \infty$, the set $\{\xi_i^{(\cdot)}\}_{i=1}^{\infty}$ forms an orthonormal basis of $\mathcal{D}$. Therefore, by the tensor product theorem for Hilbert spaces, $\{\xi_i^{(\cdot)}\xi_j^{(\cdot)}\}_{i,j=1}^{\infty}$ forms an orthonormal basis of $\mathcal{D}^2$. Consequently, in the asymptotic case, the score-based method imposes no restrictions on the class of autoregressive kernels that can be represented and estimated.}

\section{\rev{Copper plate model for Italy}}\label{sec:copper}

\noindent \rev{In Italy,} around 35\% of the time \rev{--- frequency observed in the period of study ---} the country-level economic optimum is not feasible due to transmission capacity constraints \rev{among the seven bidding zones composing Italy}, which means that the market cannot be cleared from the country-level supply and demand curves. In this case, \rev{GME uses a \textit{zonal pricing} mechanism}, which splits the entire country-level market pool in a minimum number of "subpools" of market zones such that, in each subpool, the economic optimum is feasible. As a result, there can be several pairs of supply and demand curves --- hence several market clearing prices --- for a single hour, one for each subpool. To keep the analysis simple, we did as if the country-level economic optimum was always feasible and built the country-level supply and demand curves for any hour, ignoring the problem of transmission congestion, \rev{i.e., considering a \textit{copper plate} model}. As a result, the constructed curves reflect the true market clearing 65\% of the time, the remaining 35\% corresponding to a "virtual" market clearing only. Note the same curves construction procedure was considered in \citet{mestre_forecasting_2020}. \rev{The associated market clearing price is what GME calls the "National" or "Italy (unconstrained)" price and reports together with the zonal prices (see \url{https://www.mercatoelettrico.org/en-us/Home/Results/Electricity/MGP/Results/ZonalPrices})}.

\section{Computational efficiency} \label{sec:compute}

The computational time requirements for clearing price forecasting models (Table \ref{tab:effpoint}) were evaluated on an Apple M2 Pro chip with a 12-core CPU and 16GB of RAM. We note that the \textbf{ARX} model could be made more efficient by parallelizing model fits across hours or vector components. The remaining models benefit from NumPy’s built-in multithreading for large matrix operations and therefore execute in a multicore fashion.

\begin{table}[h!]
    \centering
    \small
    \rev{
    \begin{tabular}{ccccccccccc}
    \toprule
    \multicolumn{3}{c}{\textit{}} &
    \multicolumn{8}{c}{\textit{curve-based}} \\
    \cmidrule(lr){4-11}
    \multicolumn{3}{c}{\textit{price-based}}
    & \multicolumn{2}{c}{\textbf{ARX}}
    & \multicolumn{2}{c}{\textbf{VARX}}
    & \multicolumn{2}{c}{\textbf{fARX}}
    & \multicolumn{2}{c}{\textbf{fVARX}} \\
    \cmidrule(lr){1-3}
    \cmidrule(lr){4-5}
    \cmidrule(lr){6-7}
    \cmidrule(lr){8-9}
    \cmidrule(lr){10-11}
    \textbf{ARX} & \textbf{fARX} & \textbf{LEAR}
    & \textbf{ZST} & \textbf{FPCA}
    & \textbf{ZST} & \textbf{FPCA}
    & \textbf{ZST} & \textbf{FPCA}
    & \textbf{ZST} & \textbf{FPCA} \\
    \midrule
    0.2s
    & 0.6s
    & 3.0s
    & 1.5s
    & 1.8s
    & 3.1s
    & 3.4s
    & 6.5s
    & 6.0s
    & 9.2s
    & 8.6s \\
    \bottomrule
    \end{tabular}
    }
    \caption{\rev{Average computational (wall) time required for one daily iteration of clearing price forecasting on EPEX-FR with $K_s=K_d=5$. A daily iteration consists of (i) scaling/transforming the data \& fitting the model to the last 364 days (ii) predicting the 24 prices of the next day.}}
    \label{tab:effpoint}
\end{table}

\clearpage
\section{\rev{Comparison of static and dynamic approaches for curves representation}}
\label{sec:stat_vs_dyna}

\begin{table}[h!]
\centering
\footnotesize
\begin{tabular}{rcc|cc|cc}
\toprule
&
\multicolumn{2}{c}{GME} &
\multicolumn{2}{c}{EPEX-DE-LU} &
\multicolumn{2}{c}{EPEX-FR} \\
\cmidrule(lr){2-3}   
\cmidrule(lr){4-5}   
\cmidrule(lr){6-7}  
&
\textbf{Static} & \multicolumn{1}{c}{\textbf{Dynamic}} & 
\textbf{Static} & \multicolumn{1}{c}{\textbf{Dynamic}} & 
\textbf{Static} & \textbf{Dynamic} \\
\midrule
\textbf{Naive}
& \cellcolor[RGB]{218,134,118}2.11
& \cellcolor[RGB]{218,134,118}2.11
& \cellcolor[RGB]{218,134,118}3.56
& \cellcolor[RGB]{218,134,118}3.56
& \cellcolor[RGB]{218,134,118}2.87
& \cellcolor[RGB]{218,134,118}2.87 \\

\textbf{ZST-ARX}
& \cellcolor[RGB]{211,206,125}1.69
& \cellcolor[RGB]{247,214,119}1.78
& \cellcolor[RGB]{152,193,135}2.32
& \cellcolor[RGB]{150,193,135}2.31
& \cellcolor[RGB]{138,190,137}1.86
& \cellcolor[RGB]{141,191,137}1.87 \\

\textbf{ZST-VARX}
& \cellcolor[RGB]{194,203,128}1.65
& \cellcolor[RGB]{214,207,125}1.70
& \cellcolor[RGB]{154,194,135}2.33
& \cellcolor[RGB]{153,194,135}2.33
& \cellcolor[RGB]{142,191,137}1.87
& \cellcolor[RGB]{143,191,136}1.88 \\

\textbf{ZST-fARX}
& \cellcolor[RGB]{231,211,122}1.74
& \cellcolor[RGB]{245,207,119}1.81
& \cellcolor[RGB]{126,188,139}2.18
& \cellcolor[RGB]{125,187,139}2.17
& \cellcolor[RGB]{148,192,136}1.90
& \cellcolor[RGB]{151,193,135}1.91 \\

\textbf{ZST-fVARX}
& \cellcolor[RGB]{216,207,125}1.70
& \cellcolor[RGB]{246,214,120}1.77
& \cellcolor[RGB]{128,188,139}2.19
& \cellcolor[RGB]{129,188,139}2.19
& \cellcolor[RGB]{140,191,137}1.86
& \cellcolor[RGB]{144,191,136}1.88 \\

\textbf{FPCA-ARX}
& \cellcolor[RGB]{148,192,136}1.54
& \cellcolor[RGB]{121,186,140}1.47
& \cellcolor[RGB]{149,193,136}2.30
& \cellcolor[RGB]{148,192,136}2.29
& \cellcolor[RGB]{121,186,140}1.79
& \cellcolor[RGB]{120,186,140}1.78 \\

\textbf{FPCA-VARX}
& \cellcolor[RGB]{147,192,136}1.53
& \cellcolor[RGB]{120,186,140}1.47
& \cellcolor[RGB]{146,192,136}2.29
& \cellcolor[RGB]{148,192,136}2.29
& \cellcolor[RGB]{130,188,139}1.82
& \cellcolor[RGB]{129,188,139}1.82 \\

\textbf{FPCA-fARX}
& \cellcolor[RGB]{141,191,137}1.52
& \cellcolor[RGB]{113,185,142}1.45
& \cellcolor[RGB]{115,185,141}2.12
& \cellcolor[RGB]{114,185,141}2.11
& \cellcolor[RGB]{115,185,141}1.76
& \cellcolor[RGB]{113,185,142}1.75 \\

\textbf{FPCA-fVARX}
& \cellcolor[RGB]{141,191,137}1.52
& \cellcolor[RGB]{113,185,141}1.45
& \cellcolor[RGB]{113,185,141}2.11
& \cellcolor[RGB]{113,185,142}2.10
& \cellcolor[RGB]{118,186,141}1.77
& \cellcolor[RGB]{119,186,140}1.78 \\

\bottomrule
\end{tabular}
\caption{\rev{Functional mean absolute error (FMAE, expressed in GWh) of \textit{supply} curves forecasts in the static and dynamic curves representation approaches. The green-yellow-red colormap is applied market-wise.}}
\label{tab:stat_vs_dyna_supply}
\end{table}

\begin{table}[h!]
\centering
\footnotesize
\begin{tabular}{rcc|cc|cc}
\toprule
&
\multicolumn{2}{c}{GME} &
\multicolumn{2}{c}{EPEX-DE-LU} &
\multicolumn{2}{c}{EPEX-FR} \\
\cmidrule(lr){2-3}   
\cmidrule(lr){4-5}   
\cmidrule(lr){6-7}  
&
\textbf{Static} & \multicolumn{1}{c}{\textbf{Dynamic}} & 
\textbf{Static} & \multicolumn{1}{c}{\textbf{Dynamic}} & 
\textbf{Static} & \textbf{Dynamic} \\
\midrule
\textbf{Naive}
& \cellcolor[RGB]{218,134,118}1.14
& \cellcolor[RGB]{218,134,118}1.14
& \cellcolor[RGB]{218,134,118}3.01
& \cellcolor[RGB]{218,134,118}3.01
& \cellcolor[RGB]{218,134,118}2.60
& \cellcolor[RGB]{218,134,118}2.60 \\

\textbf{ZST-ARX}
& \cellcolor[RGB]{124,187,140}0.47
& \cellcolor[RGB]{123,187,140}0.47
& \cellcolor[RGB]{172,198,132}2.08
& \cellcolor[RGB]{163,196,133}2.04
& \cellcolor[RGB]{151,193,135}2.05
& \cellcolor[RGB]{154,194,135}2.06 \\

\textbf{ZST-VARX}
& \cellcolor[RGB]{137,190,137}0.50
& \cellcolor[RGB]{135,189,138}0.50
& \cellcolor[RGB]{185,201,130}2.13
& \cellcolor[RGB]{175,198,131}2.09
& \cellcolor[RGB]{174,198,131}2.10
& \cellcolor[RGB]{175,198,131}2.11 \\

\textbf{ZST-fARX}
& \cellcolor[RGB]{117,185,141}0.45
& \cellcolor[RGB]{117,186,141}0.45
& \cellcolor[RGB]{126,188,139}1.88
& \cellcolor[RGB]{118,186,141}1.84
& \cellcolor[RGB]{114,185,141}1.96
& \cellcolor[RGB]{113,185,141}1.96 \\

\textbf{ZST-fVARX}
& \cellcolor[RGB]{129,188,139}0.48
& \cellcolor[RGB]{126,188,139}0.47
& \cellcolor[RGB]{130,188,139}1.89
& \cellcolor[RGB]{122,187,140}1.86
& \cellcolor[RGB]{127,188,139}1.99
& \cellcolor[RGB]{127,188,139}1.99 \\

\textbf{FPCA-ARX}
& \cellcolor[RGB]{123,187,140}0.46
& \cellcolor[RGB]{121,186,140}0.46
& \cellcolor[RGB]{159,195,134}2.02
& \cellcolor[RGB]{158,195,134}2.02
& \cellcolor[RGB]{149,193,136}2.04
& \cellcolor[RGB]{149,193,135}2.04 \\

\textbf{FPCA-VARX}
& \cellcolor[RGB]{126,188,139}0.47
& \cellcolor[RGB]{125,187,139}0.47
& \cellcolor[RGB]{166,196,133}2.05
& \cellcolor[RGB]{170,197,132}2.07
& \cellcolor[RGB]{169,197,132}2.09
& \cellcolor[RGB]{170,197,132}2.09 \\

\textbf{FPCA-fARX}
& \cellcolor[RGB]{114,185,141}0.44
& \cellcolor[RGB]{113,185,142}0.44
& \cellcolor[RGB]{114,185,141}1.82
& \cellcolor[RGB]{113,185,142}1.82
& \cellcolor[RGB]{113,185,141}1.96
& \cellcolor[RGB]{113,185,142}1.96 \\

\textbf{FPCA-fVARX}
& \cellcolor[RGB]{116,185,141}0.45
& \cellcolor[RGB]{115,185,141}0.44
& \cellcolor[RGB]{114,185,141}1.82
& \cellcolor[RGB]{115,185,141}1.83
& \cellcolor[RGB]{117,186,141}1.97
& \cellcolor[RGB]{121,186,140}1.98 \\

\bottomrule
\end{tabular}
\caption{\rev{Functional mean absolute error (FMAE, expressed in GWh) of \textit{demand} curves forecasts in the static and dynamic curves representation approaches. The green-yellow-red colormap is applied market-wise.}}
\label{tab:stat_vs_dyna_demand}
\end{table}

\begin{table}[h!]
\centering
\footnotesize
\begin{tabular}{rrcc|cc|cc}
\toprule
& &
\multicolumn{2}{c}{GME} &
\multicolumn{2}{c}{EPEX-DE-LU} &
\multicolumn{2}{c}{EPEX-FR} \\
\cmidrule(lr){3-4}   
\cmidrule(lr){5-6}   
\cmidrule(lr){7-8}  
& &
\textbf{Static} & \multicolumn{1}{c}{\textbf{Dynamic}} & 
\textbf{Static} & \multicolumn{1}{c}{\textbf{Dynamic}} & 
\textbf{Static} & \textbf{Dynamic} \\

\midrule
\multirow{4}{*}{\textit{price-based}}

& \textbf{Naive}
& \cellcolor[RGB]{218,134,118}11.32
& \cellcolor[RGB]{218,134,118}11.32
& \cellcolor[RGB]{218,134,118}27.17
& \cellcolor[RGB]{218,134,118}27.17
& \cellcolor[RGB]{218,134,118}21.48
& \cellcolor[RGB]{218,134,118}21.48 \\

& \textbf{ARX}
& \cellcolor[RGB]{158,195,134}8.32
& \cellcolor[RGB]{158,195,134}8.32
& \cellcolor[RGB]{137,190,138}14.64
& \cellcolor[RGB]{137,190,138}14.64
& \cellcolor[RGB]{179,199,131}15.85
& \cellcolor[RGB]{179,199,131}15.85 \\

& \textbf{fARX}
& \cellcolor[RGB]{132,189,138}7.98
& \cellcolor[RGB]{132,189,138}7.98
& \cellcolor[RGB]{126,187,139}14.07
& \cellcolor[RGB]{126,187,139}14.07
& \cellcolor[RGB]{134,189,138}14.58
& \cellcolor[RGB]{134,189,138}14.58 \\

& \textbf{LEAR}
& \cellcolor[RGB]{153,194,135}8.26
& \cellcolor[RGB]{153,194,135}8.26
& \cellcolor[RGB]{113,185,142}13.40
& \cellcolor[RGB]{113,185,142}13.40
& \cellcolor[RGB]{135,189,138}14.62
& \cellcolor[RGB]{135,189,138}14.62 \\

\midrule
\multirow{8}{*}{\textit{curve-based}}

& \textbf{ZST-ARX}
& \cellcolor[RGB]{243,203,119}9.78
& \cellcolor[RGB]{245,208,119}9.66
& \cellcolor[RGB]{198,203,128}17.75
& \cellcolor[RGB]{177,199,131}16.67
& \cellcolor[RGB]{174,198,131}15.70
& \cellcolor[RGB]{181,200,130}15.90 \\

& \textbf{ZST-VARX}
& \cellcolor[RGB]{202,204,127}8.91
& \cellcolor[RGB]{181,200,130}8.63
& \cellcolor[RGB]{195,203,128}17.60
& \cellcolor[RGB]{170,197,132}16.32
& \cellcolor[RGB]{170,197,132}15.59
& \cellcolor[RGB]{180,199,131}15.86 \\

& \textbf{ZST-fARX}
& \cellcolor[RGB]{243,203,119}9.78
& \cellcolor[RGB]{238,212,121}9.39
& \cellcolor[RGB]{245,207,119}20.88
& \cellcolor[RGB]{247,213,119}20.39
& \cellcolor[RGB]{245,207,119}18.09
& \cellcolor[RGB]{241,198,119}18.50 \\

& \textbf{ZST-fVARX}
& \cellcolor[RGB]{247,214,119}9.52
& \cellcolor[RGB]{213,207,125}9.05
& \cellcolor[RGB]{240,213,121}19.88
& \cellcolor[RGB]{230,211,122}19.37
& \cellcolor[RGB]{213,207,125}16.77
& \cellcolor[RGB]{225,210,123}17.12 \\

& \textbf{FPCA-ARX}
& \cellcolor[RGB]{140,191,137}8.09
& \cellcolor[RGB]{139,190,137}8.07
& \cellcolor[RGB]{170,197,132}16.31
& \cellcolor[RGB]{154,194,135}15.53
& \cellcolor[RGB]{118,186,141}14.14
& \cellcolor[RGB]{113,185,142}14.00 \\

& \textbf{FPCA-VARX}
& \cellcolor[RGB]{113,185,142}7.72
& \cellcolor[RGB]{117,186,141}7.78
& \cellcolor[RGB]{170,197,132}16.35
& \cellcolor[RGB]{157,194,134}15.65
& \cellcolor[RGB]{133,189,138}14.58
& \cellcolor[RGB]{132,189,138}14.55 \\

& \textbf{FPCA-fARX}
& \cellcolor[RGB]{153,193,135}8.25
& \cellcolor[RGB]{137,190,137}8.05
& \cellcolor[RGB]{196,203,128}17.67
& \cellcolor[RGB]{181,200,130}16.90
& \cellcolor[RGB]{130,188,139}14.49
& \cellcolor[RGB]{117,185,141}14.12 \\

& \textbf{FPCA-fVARX}
& \cellcolor[RGB]{121,186,140}7.83
& \cellcolor[RGB]{122,187,140}7.84
& \cellcolor[RGB]{182,200,130}16.96
& \cellcolor[RGB]{170,197,132}16.32
& \cellcolor[RGB]{131,189,138}14.51
& \cellcolor[RGB]{121,186,140}14.22 \\

\bottomrule
\end{tabular}
\caption{\rev{Mean absolute error (MAE, expressed in \euro/MWh) of \textit{clearing price} forecasts in the static and dynamic curves representation approaches. The green-yellow-red colormap is applied market-wise.}}
\label{tab:stat_vs_dyna_price}
\end{table}

\clearpage
\section{\rev{Results with unrestricted price domain}} \label{sec:res_unres}

\rev{For computational efficiency, unrestricted curves were discretized on an optimized \textit{non-uniform} price grid, constructed as follows:}
\rev{
\begin{enumerate}
    \item Build the curves with a uniform price grid of 4501 values (1\euro/MWh resolution).
    \item Compute the mean supply and demand curves and invert them to get the price functions
    \item Build two non-uniform price grids of size 4501 by evaluating the mean supply and demand price functions on a uniform quantity grid of the same size (as in the Ziel-Steinert transformation)
    \item Merge the two grids by taking their union
    \item Round each price to the nearest integer (\euro/MWh) and remove duplicates
\end{enumerate}
}

\noindent \rev{The resulting price grid sizes are reported in Table \ref{tab:domain_unres}.}

\begin{table}
    \centering
    \rev{
    \begin{tabular}{lccc}
    \toprule
         & \textbf{GME} & \textbf{EPEX-DE-LU} & \textbf{EPEX-FR} \\
    \midrule
        \textbf{Grid size} & 924 & 965 & 828 \\
    \bottomrule
    \end{tabular}
    }
    \caption{\rev{Optimized \textit{non-uniform} discretization grid sizes for unrestricted curves.}}
    \label{tab:domain_unres}
\end{table}

\begin{figure*}[h!]
     \centering
     \begin{subfigure}[b]{0.49\textwidth}
         \includegraphics[width=\textwidth]{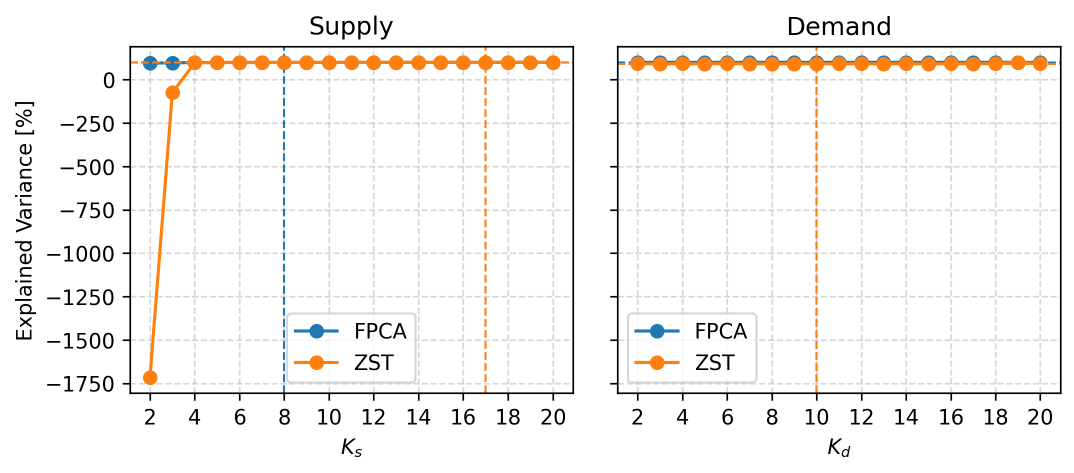}
         \caption{Curves (\textbf{GME})}
         \label{fig:K_unres:curve_gme}
     \end{subfigure}
     \begin{subfigure}[b]{0.49\textwidth}
         \includegraphics[width=\textwidth]{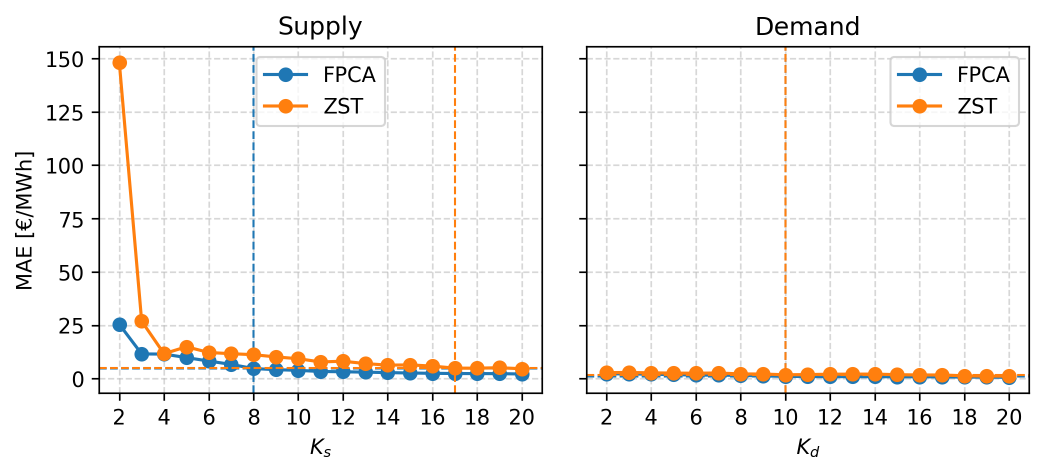}
         \caption{Clearing price (\textbf{GME})}
         \label{fig:K_unres:mcp_gme}
     \end{subfigure}
     \par\vspace{0.5cm}
     \begin{subfigure}[b]{0.49\textwidth}
         \includegraphics[width=\textwidth]{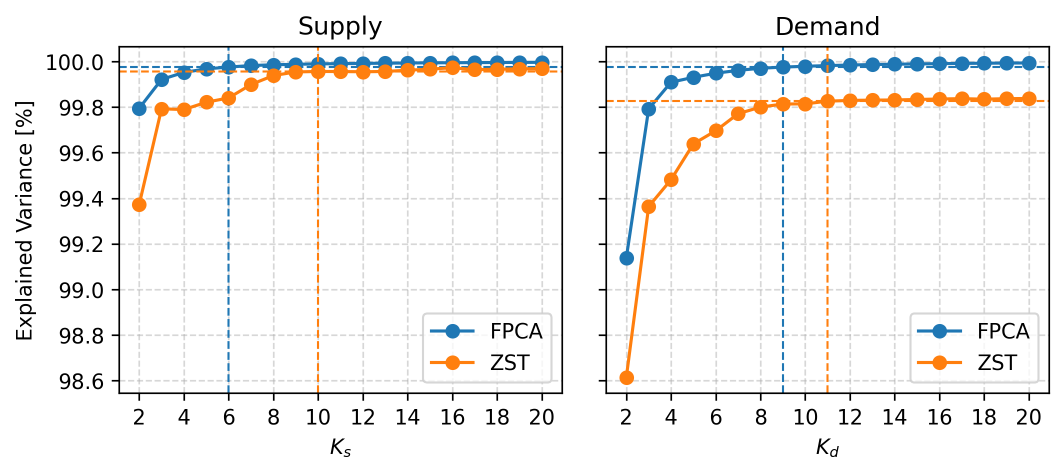}
         \caption{Curves (\textbf{EPEX-DE-LU})}
         \label{fig:K_unres:curve_epex-de-lu}
     \end{subfigure}
     \begin{subfigure}[b]{0.49\textwidth}
         \includegraphics[width=\textwidth]{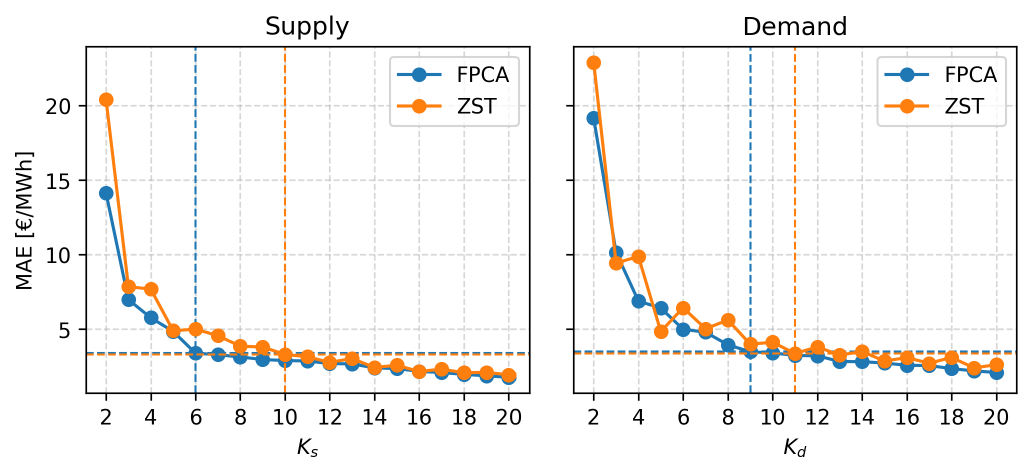}
         \caption{Clearing price (\textbf{EPEX-DE-LU})}
         \label{fig:K_unres:mcp_epex-de-lu}
     \end{subfigure}
     \par\vspace{0.5cm}
     \begin{subfigure}[b]{0.49\textwidth}
         \includegraphics[width=\textwidth]{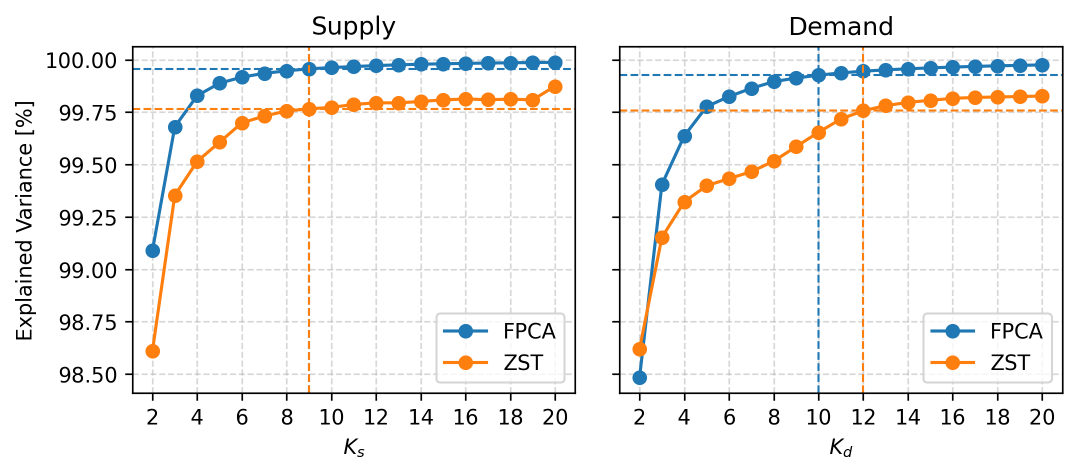}
         \caption{Curves (\textbf{EPEX-FR})}
         \label{fig:K_unres:curve_epex-fr}
     \end{subfigure}
     \begin{subfigure}[b]{0.49\textwidth}
         \includegraphics[width=\textwidth]{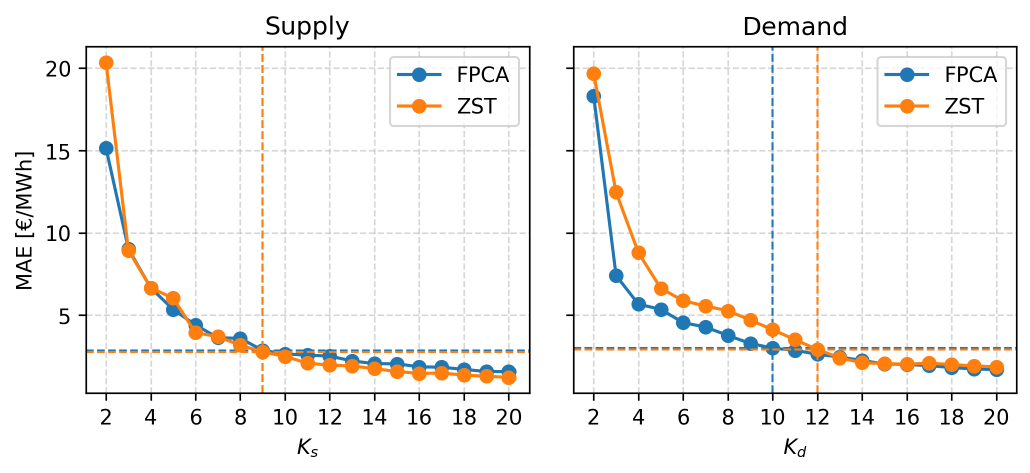}
         \caption{Clearing price (\textbf{EPEX-FR})}
         \label{fig:K_unres:mcp_epex-fr}
     \end{subfigure}
        \caption{\rev{(\textit{Color optional}) Curves (left panel) and clearing price (right panel) \textit{approximation} error for 2023 (inital estimation window) for each representation method with respect to the number of components considered for supply curves ($K_s$) and demand curves ($K_d$) (\textbf{unrestricted domain}). When computing the clearing price for different $K_s$, true demand curves are used and, conversely, when computing the clearing price for different $K_d$ values, true supply curves are used. The vertical and horizontal dashed lines identify the $K_s$ and $K_d$ values selected for each representation and market (which are also reported in Table \ref{tab:K_unres}).}}
        \label{fig:K_unres}
\end{figure*}

\begin{figure*}[h]
     \centering
     \begin{subfigure}[b]{0.8\textwidth}
         \includegraphics[width=\textwidth]{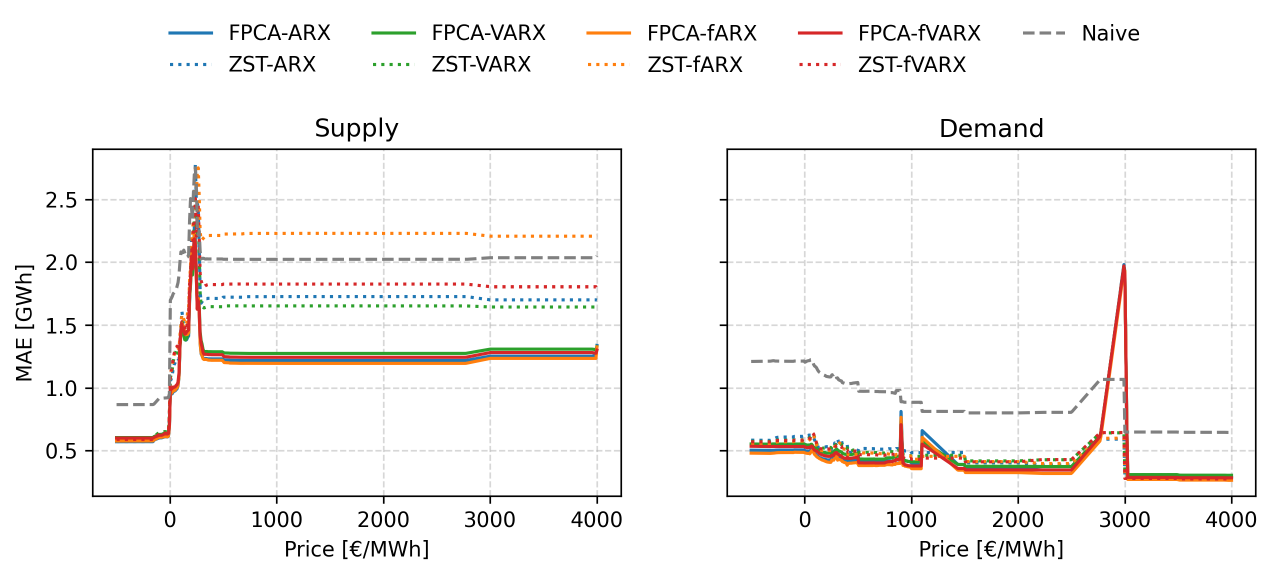}
         \caption{\textbf{GME}}
         \label{fig:curve_mae_unres:gme}
     \end{subfigure}
     \begin{subfigure}[b]{0.8\textwidth}
         \includegraphics[width=\textwidth]{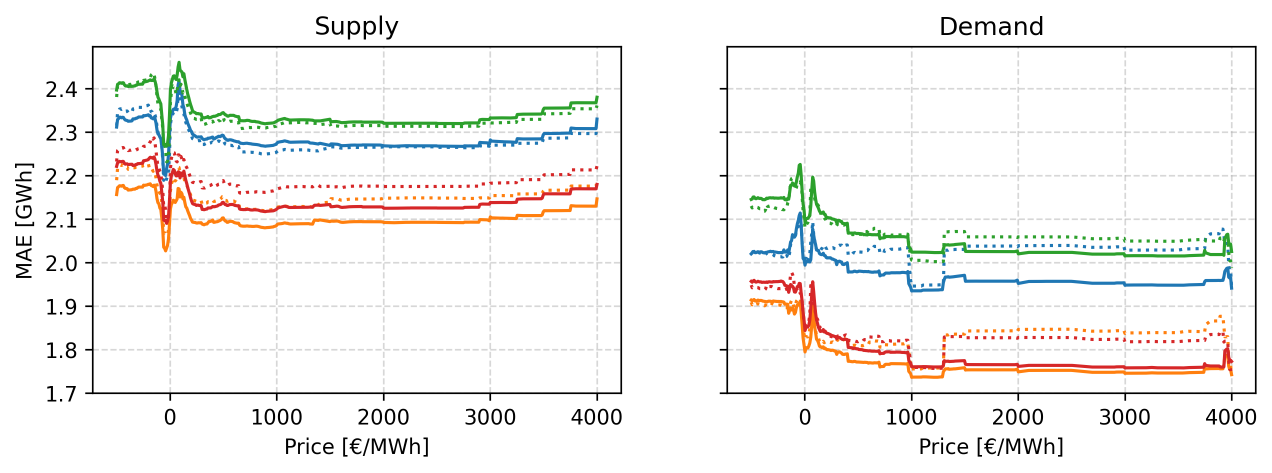}
         \caption{\textbf{EPEX-DE-LU}}
         \label{fig:curve_mae_unres:epex-de-lu}
     \end{subfigure}
     \begin{subfigure}[b]{0.8\textwidth}
         \includegraphics[width=\textwidth]{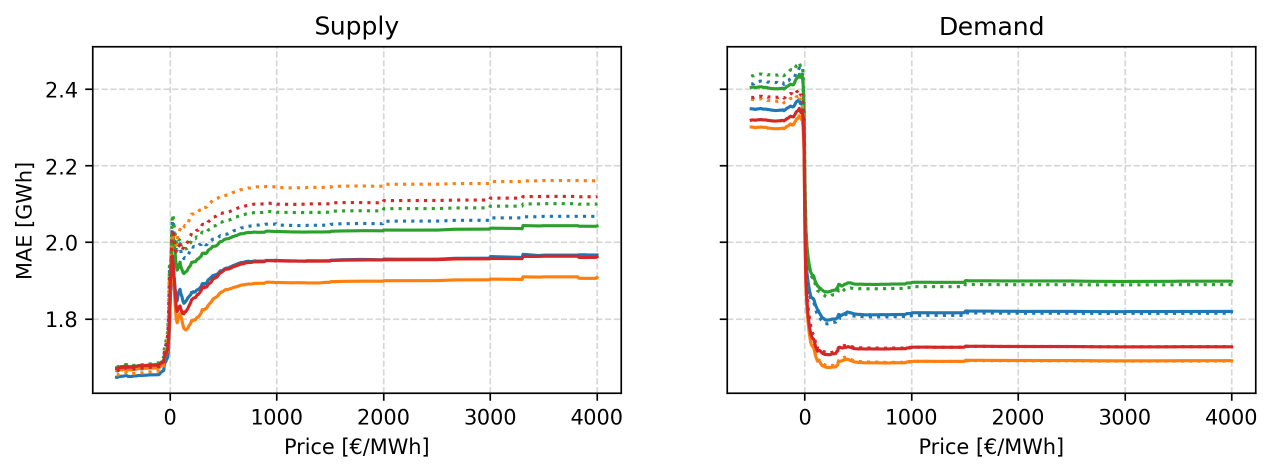}
         \caption{\textbf{EPEX-FR}}
         \label{fig:curve_mae_unres:epex-fr}
     \end{subfigure}
        \caption{\rev{(\textit{Color optional}) Pointwise mean absolute error (MAE) of supply and demand curves forecasts (\textbf{unrestricted domain})}}
        \label{fig:curve_mae_unres}
\end{figure*}

\begin{figure*}[h!]
     \centering
     \begin{subfigure}[b]{0.355\textwidth}
         \includegraphics[width=\textwidth]{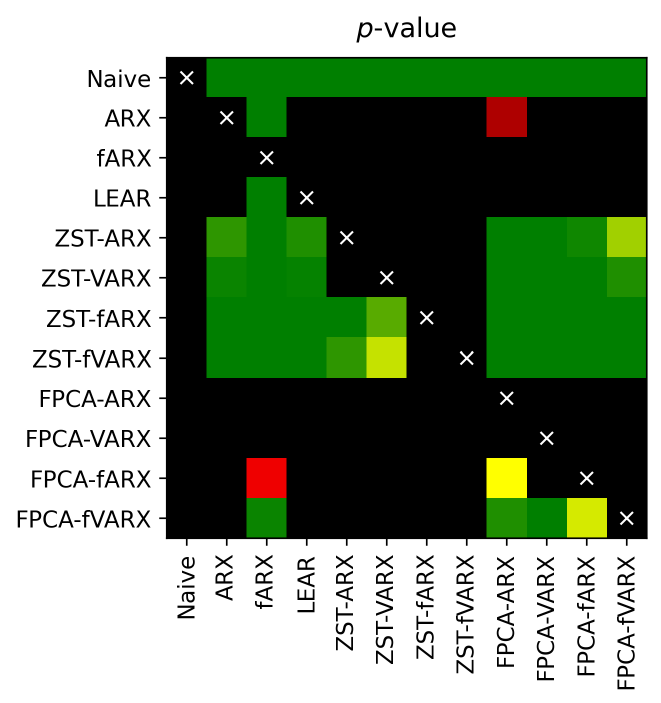}
         \caption{GME}
         \label{fig:dm_price_daily_unres:gme}
     \end{subfigure}
     \begin{subfigure}[b]{0.278\textwidth}
         \includegraphics[width=\textwidth]{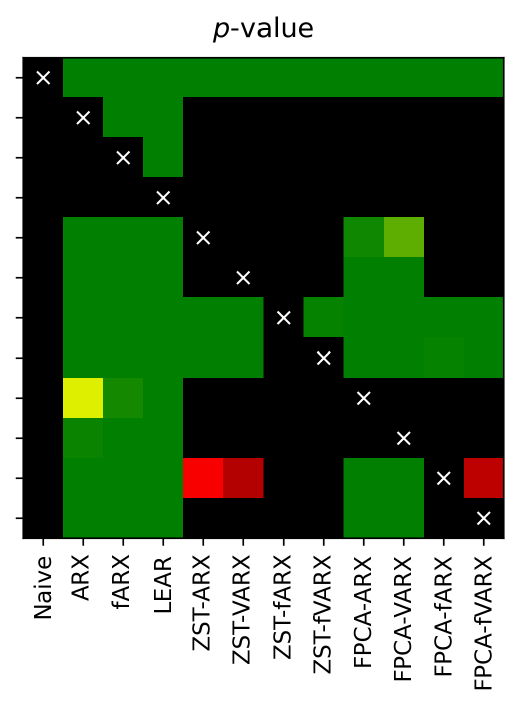}
         \caption{EPEX-DE-LU}
         \label{fig:dm_price_daily_unres:epex-de-lu}
     \end{subfigure}
     \begin{subfigure}[b]{0.351\textwidth}
         \includegraphics[width=\textwidth]{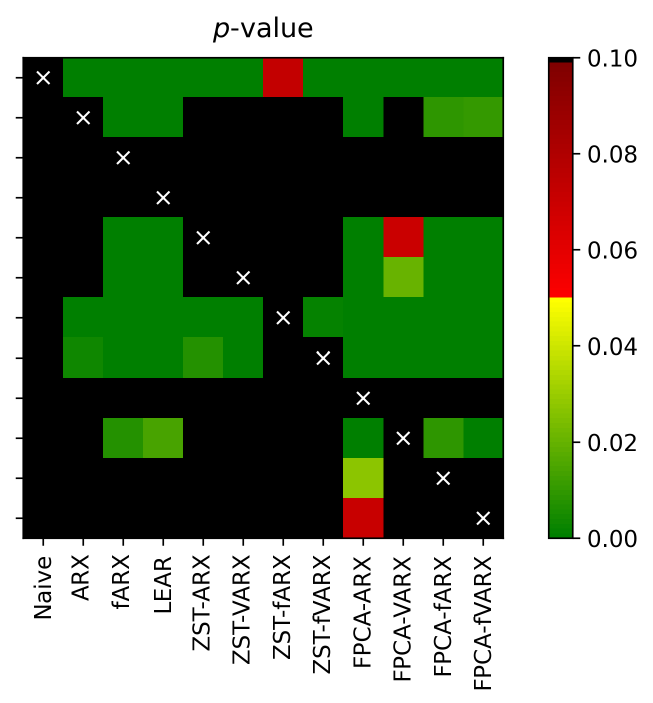}
         \caption{EPEX-FR}
         \label{fig:dm_price_daily_unres:epex-fr}
     \end{subfigure}
        \caption{\rev{(\textit{Color optional}) Results of the Diebold-Mariano test for the difference in clearing price forecasting performance on average \textit{daily} absolute errors (\textbf{unrestricted domain}). The alternative hypothesis is that models on the $x$-axis outperform those on the $y$-axis (one-sided test).}}
        \label{fig:dm_price_daily_unres}
\end{figure*}

\begin{figure*}[h!]
     \centering
     \begin{subfigure}[b]{0.315\textwidth}
         \includegraphics[width=\textwidth]{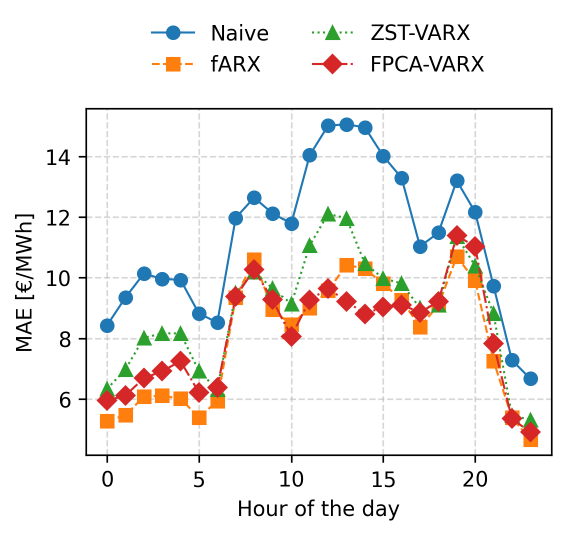}
         \caption{GME}
         \label{fig:mae_price_hourly_unres:gme}
     \end{subfigure}
     \begin{subfigure}[b]{0.315\textwidth}
         \includegraphics[width=\textwidth]{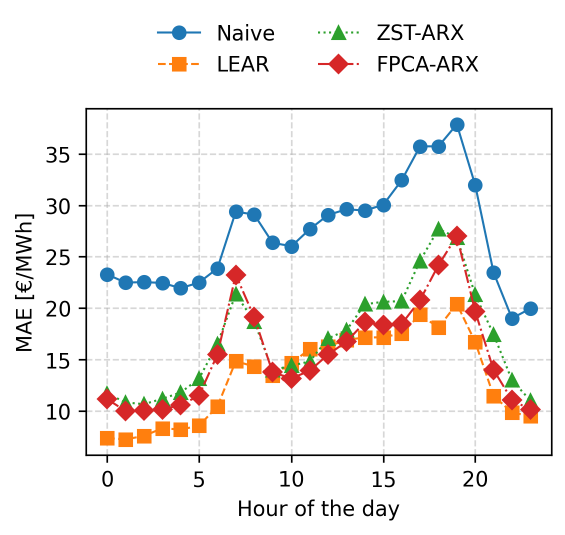}
         \caption{EPEX-DE-LU}
         \label{fig:mae_price_hourly_unres:epex-de-lu}
     \end{subfigure}
     \begin{subfigure}[b]{0.327\textwidth}
         \includegraphics[width=\textwidth]{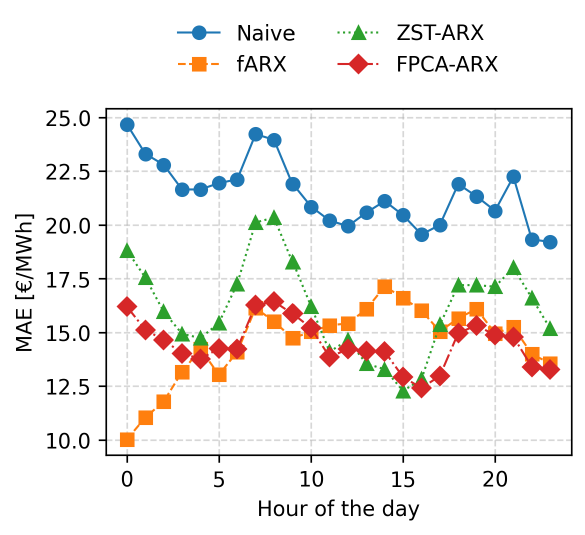}
         \caption{EPEX-FR}
         \label{fig:mae_price_hourly_unres:epex-fr}
     \end{subfigure}
        \caption{\rev{(\textit{Color optional}) Mean absolute error of clearing price forecasts for each hour (\textbf{unrestricted domain}). For legibility, only the best price-based, ZST-curve-based and FPCA-curve-based are represented.}}
        \label{fig:mae_price_hourly_unres}
\end{figure*}

\begin{figure*}[h!]
     \centering
     \begin{subfigure}[b]{0.362\textwidth}
         \includegraphics[width=\textwidth]{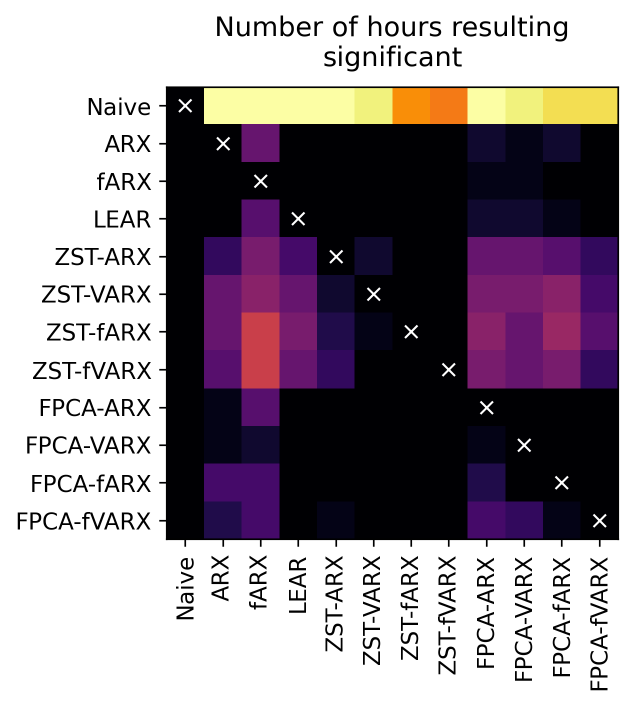}
         \caption{GME}
         \label{fig:dm_price_hourly_unres:gme}
     \end{subfigure}
     \begin{subfigure}[b]{0.280\textwidth}
         \includegraphics[width=\textwidth]{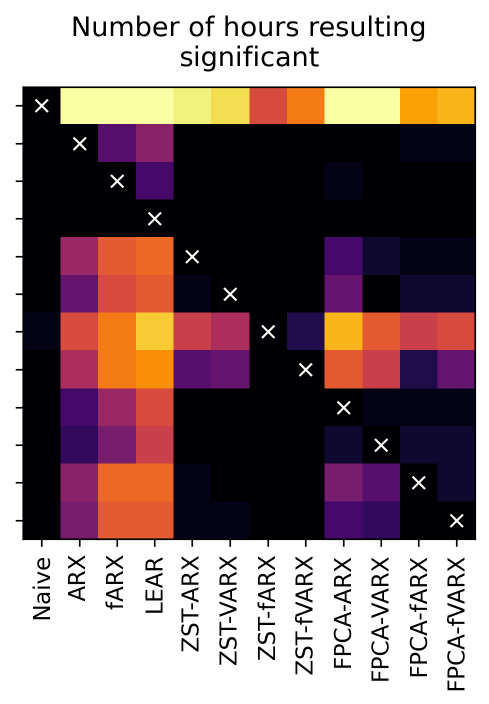}
         \caption{EPEX-DE-LU}
         \label{fig:dm_price_hourly_unres:epex-de-lu}
     \end{subfigure}
     \begin{subfigure}[b]{0.344\textwidth}
         \includegraphics[width=\textwidth]{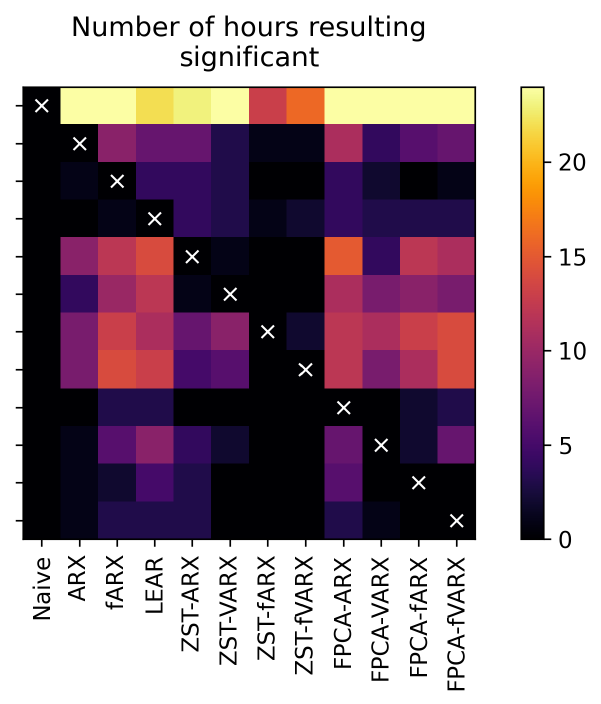}
         \caption{EPEX-FR}
         \label{fig:dm_price_hourly_unres:epex-fr}
     \end{subfigure}
        \caption{\rev{(\textit{Color optional}) Results of the Diebold-Mariano tests for the difference in clearing price forecasting performance on absolute errors, at the hour level (\textbf{unrestricted domain}). The alternative hypothesis is that models on the $x$-axis outperform those on the $y$-axis (one-sided test). The significance threshold is set to 0.005. }}
        \label{fig:dm_price_hourly_unres}
\end{figure*}

\begin{figure*}[h!]
     \centering
     \begin{subfigure}[b]{0.49\textwidth}
         \includegraphics[width=\textwidth]{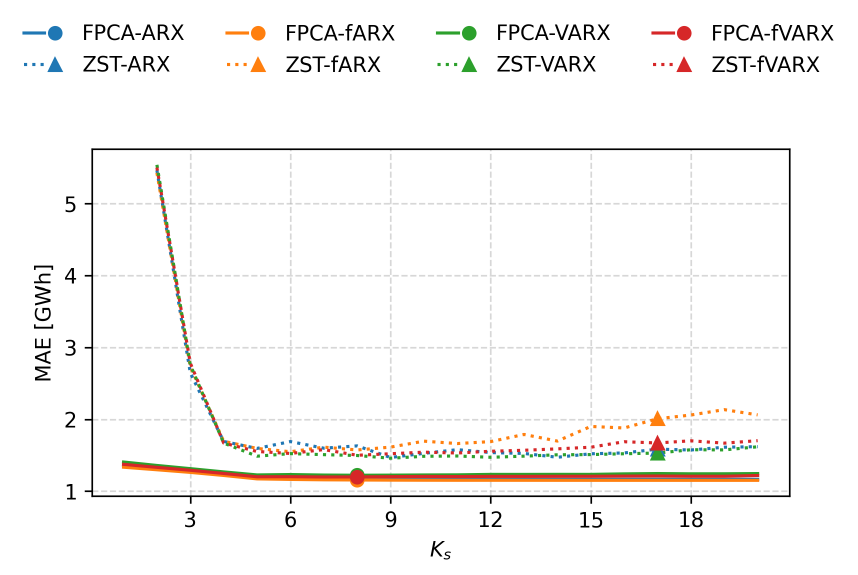}
         \caption{Supply (\textbf{GME})}
         \label{fig:sens_curve_mae_unres:gme_supply}
     \end{subfigure}
     \begin{subfigure}[b]{0.49\textwidth}
         \includegraphics[width=\textwidth]{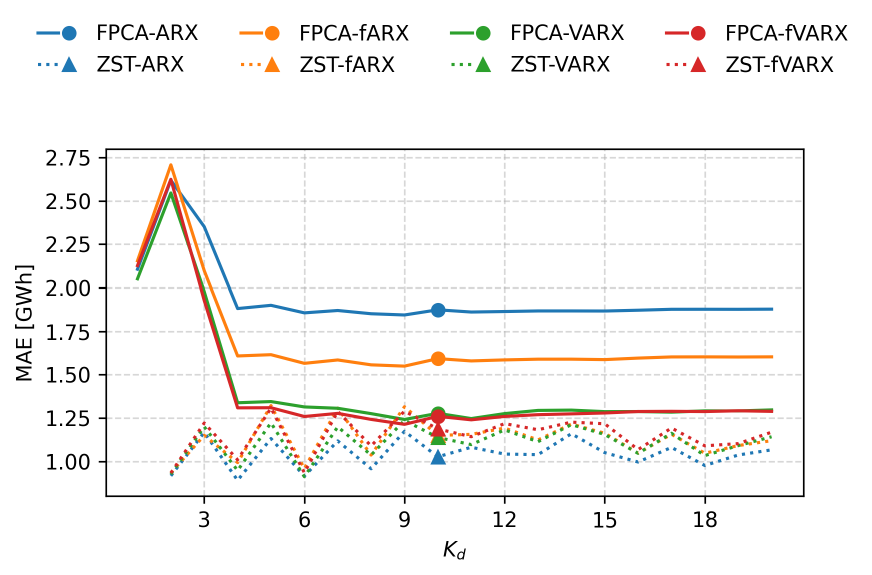}
         \caption{Demand (\textbf{GME})}
         \label{fig:sens_curve_mae_unres:gme_demand}
     \end{subfigure}
     \begin{subfigure}[b]{0.49\textwidth}
         \includegraphics[width=\textwidth]{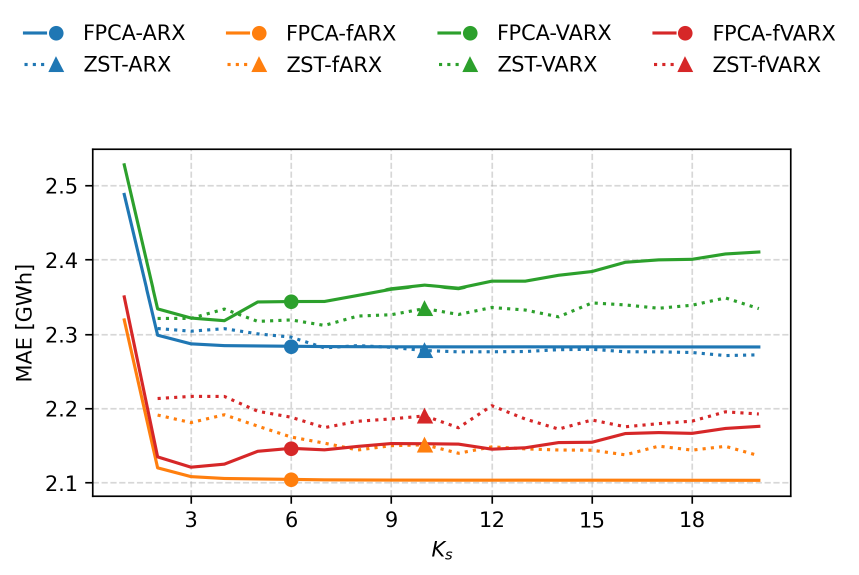}
         \caption{Supply (\textbf{EPEX-DE-LU})}
         \label{fig:sens_curve_mae_unres:epex-de-lu_supply}
     \end{subfigure}
     \begin{subfigure}[b]{0.49\textwidth}
         \includegraphics[width=\textwidth]{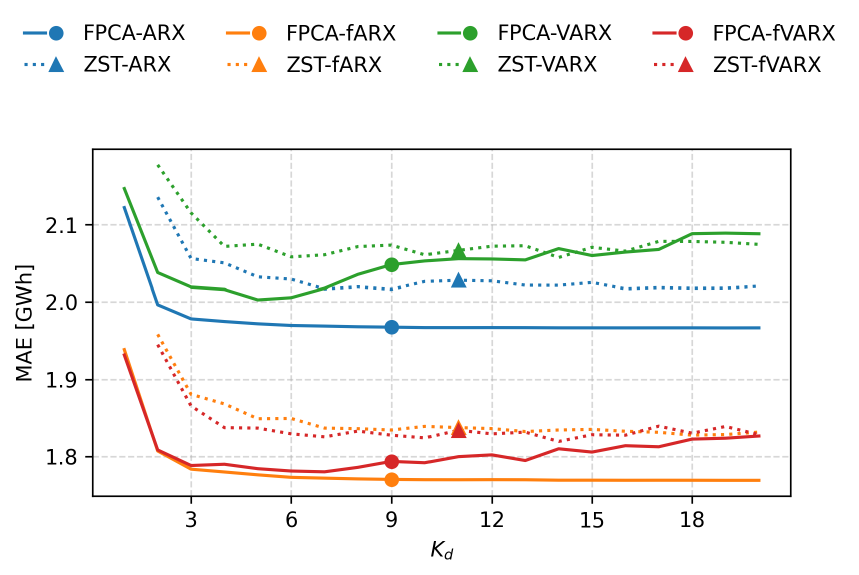}
         \caption{Demand (\textbf{EPEX-DE-LU})}
         \label{fig:sens_curve_mae_unres:epex-de-lu_demand}
     \end{subfigure}
     \begin{subfigure}[b]{0.49\textwidth}
         \includegraphics[width=\textwidth]{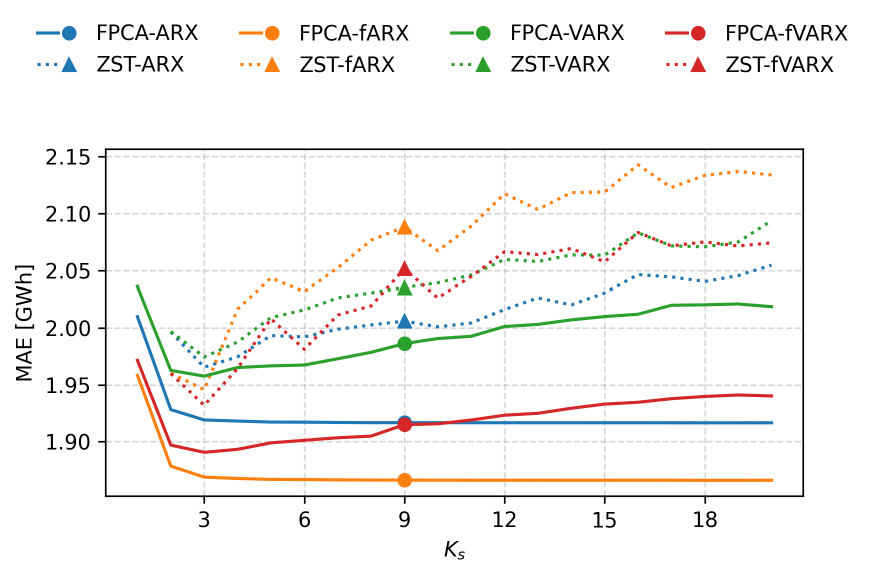}
         \caption{Supply (\textbf{EPEX-FR})}
         \label{fig:sens_curve_mae_unres:epex-fr_supply}
     \end{subfigure}
     \begin{subfigure}[b]{0.49\textwidth}
         \includegraphics[width=\textwidth]{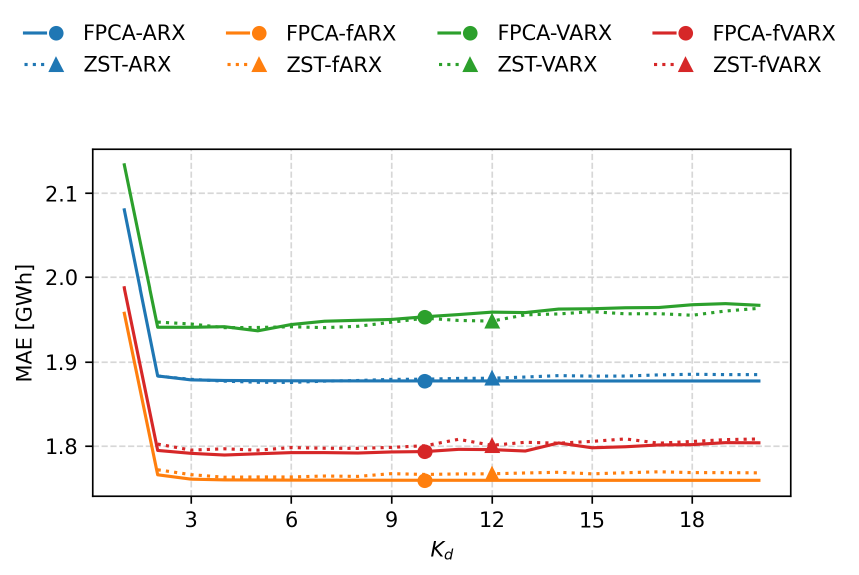}
         \caption{Demand (\textbf{EPEX-FR})}
         \label{fig:sens_curve_mae_unres:epex-fr_demand}
     \end{subfigure}
        \caption{\rev{(\textit{Color optional}) Sensitivity of \textit{supply} (left panel) and \textit{demand} (right panel) curve forecasting performance (\textbf{unrestricted domain}), as measured by the mean absolute error, with respect to number of vector components considered for each curve type  ($K_s$ for supply, $K_d$ for demand). The markers show the values that were chosen for the analysis.}}
        \label{fig:sens_curve_mae_unres}
\end{figure*}

\begin{figure*}[h!]
     \centering
     \begin{subfigure}[b]{0.49\textwidth}
         \includegraphics[width=\textwidth]{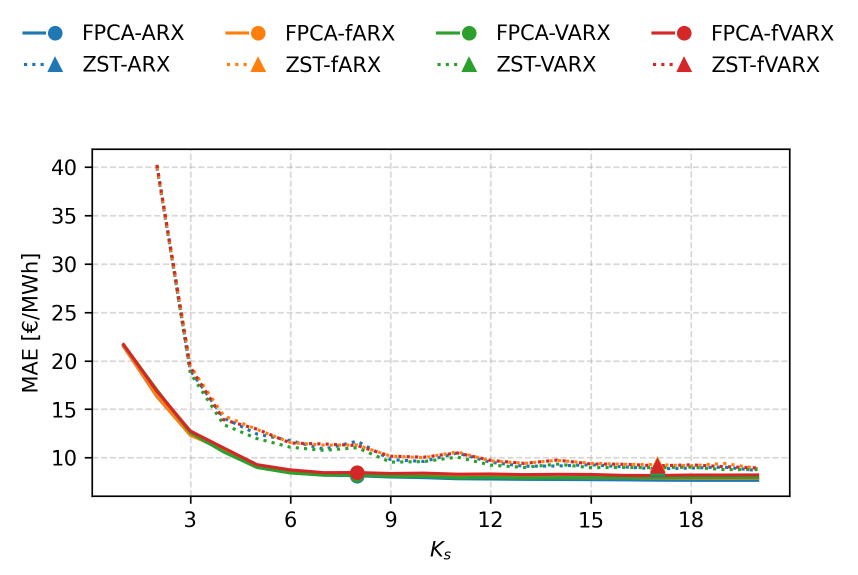}
         \caption{$K_s$ effect (\textbf{GME})}
         \label{fig:sens_price_mae_unres:gme_supply}
     \end{subfigure}
     \begin{subfigure}[b]{0.49\textwidth}
         \includegraphics[width=\textwidth]{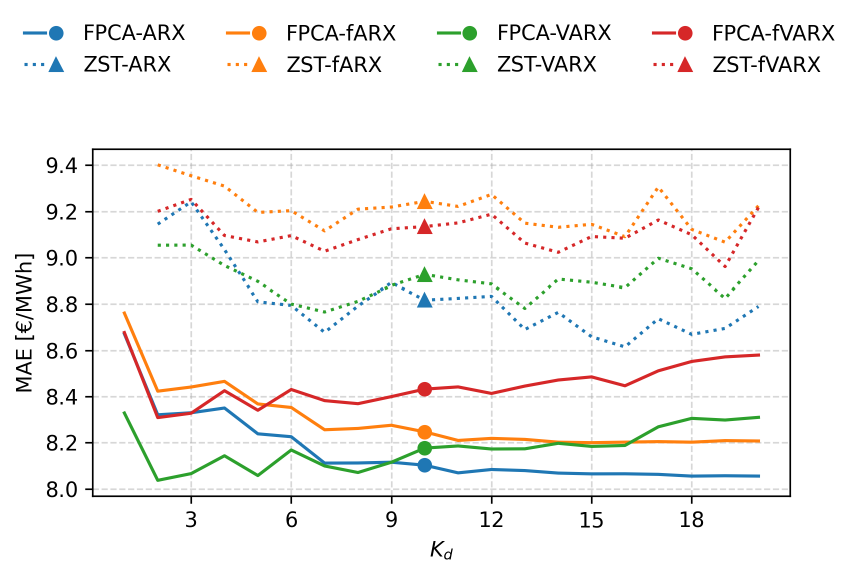}
         \caption{$K_d$ effect (\textbf{GME})}
         \label{fig:sens_price_mae_unres:gme_demand}
     \end{subfigure}
     \begin{subfigure}[b]{0.49\textwidth}
         \includegraphics[width=\textwidth]{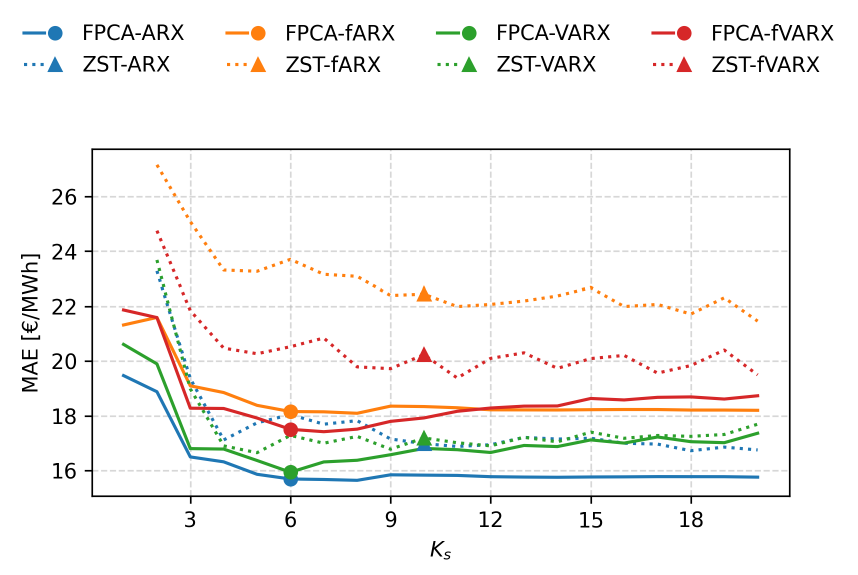}
         \caption{$K_s$ effect (\textbf{EPEX-DE-LU})}
         \label{fig:sens_price_mae_unres:epex-de-lu_supply}
     \end{subfigure}
     \begin{subfigure}[b]{0.49\textwidth}
         \includegraphics[width=\textwidth]{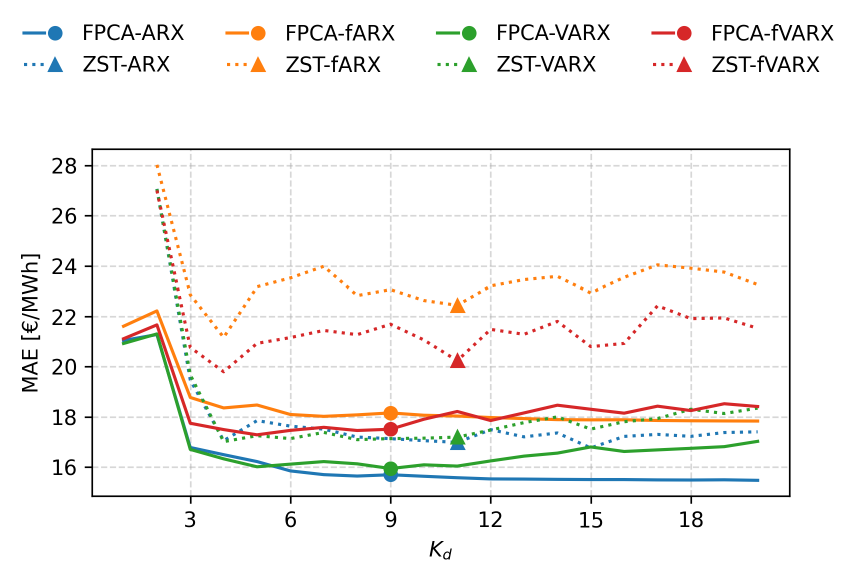}
         \caption{$K_d$ effect (\textbf{EPEX-DE-LU})}
         \label{fig:sens_price_mae_unres:epex-de-lu_demand}
     \end{subfigure}
     \begin{subfigure}[b]{0.49\textwidth}
         \includegraphics[width=\textwidth]{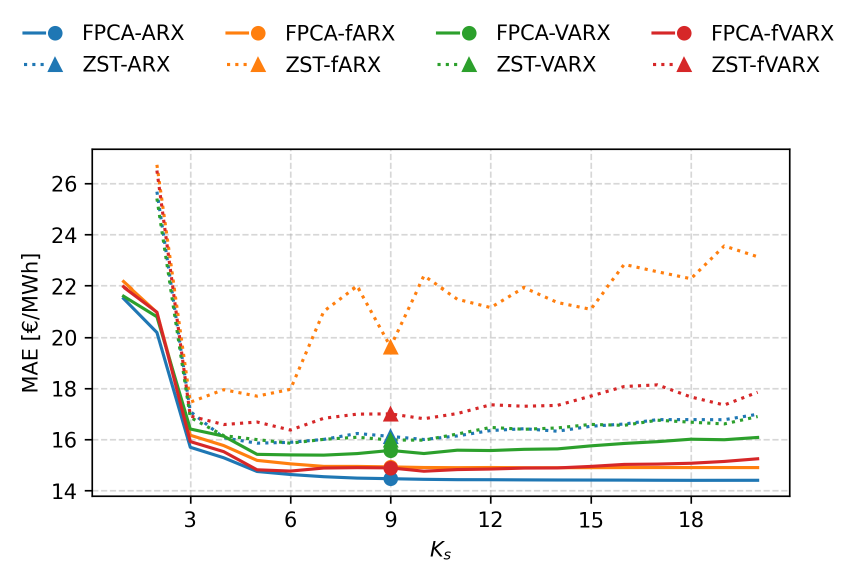}
         \caption{$K_s$ effect (\textbf{EPEX-FR})}
         \label{fig:sens_price_mae_unres:epex-fr_supply}
     \end{subfigure}
     \begin{subfigure}[b]{0.49\textwidth}
         \includegraphics[width=\textwidth]{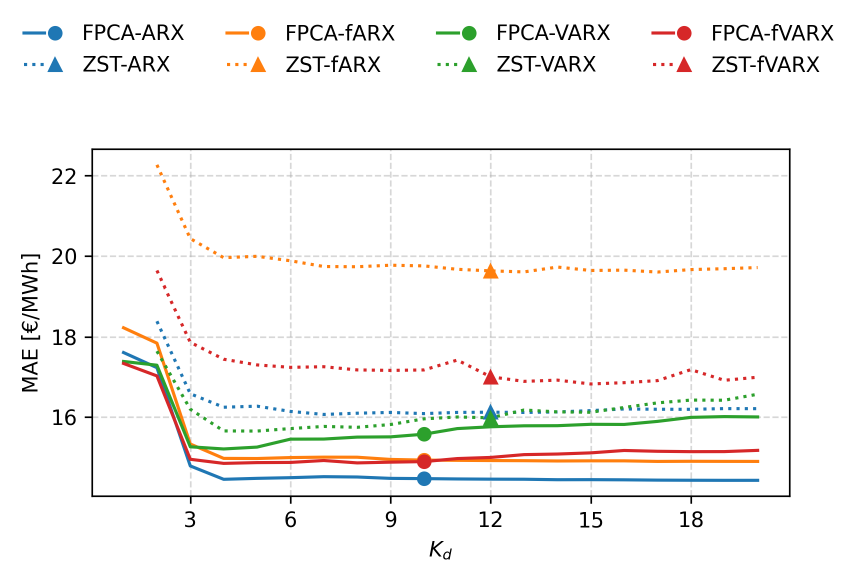}
         \caption{$K_d$ effect (\textbf{EPEX-FR})}
         \label{fig:sens_price_mae_unres:epex-fr_demand}
     \end{subfigure}
        \caption{\rev{(\textit{Color optional}) Sensitivity of \textit{clearing price} forecasting performance, as measured by the mean absolute error (\textbf{unrestricted domain}), with respect to number of vector components considered for supply curves ($K_s$, left panel) and demand curves ($K_d$, right panel). The markers show the values that were chosen for the analysis.}}
        \label{fig:sens_price_mae_unres}
\end{figure*}

\rev{The approximation quality with respect to the number of supply and demand components is shown in Fig. \ref{fig:K_unres}, the selected values being reported in Table \ref{tab:K_unres}. Fig. \ref{fig:curve_mae_unres} plots the pointwise mean absolute error of curves forecasts, while the functional error metrics are presented in Tables \ref{tab:gme_supply_demand_unres}, \ref{tab:epex-de-lu_supply_demand_unres} and \ref{tab:epex_fr_supply_demand_unres}. Clearing price forecast evaluation is provided in Table \ref{tab:price_all_markets_unres} and Fig. \ref{fig:mae_price_hourly_unres}, with associated DM tests in Figs. \ref{fig:dm_price_daily_unres} and \ref{fig:dm_price_hourly_unres}. Finally the sensitivity analysis with respect to the number of curve components can be found in Figs. \ref{fig:sens_curve_mae_unres} and \ref{fig:sens_price_mae_unres}.}


\begin{table}[h]
\centering
\rev{
\begin{tabular}{lc@{\hspace{0.5em}}cc c@{\hspace{0.5em}}c}
\toprule
& \multicolumn{2}{c}{\textbf{FPCA}}
&& \multicolumn{2}{c}{\textbf{ZST}} \\
\cmidrule(lr){2-3}
\cmidrule(lr){5-6}
\textbf{Market} & $K_s$ & $K_d$ &&  $K_s$ & $K_d$ \\
\midrule
GME         & 8 & 10 && 17 & 10 \\
EPEX-DE-LU  & 6 & 9 && 10 & 11 \\
EPEX-FR     & 9 & 10 && 9 & 12 \\
\bottomrule
\end{tabular}
}
\caption{\rev{Number of components selected for supply curves ($K_s$) and demand curves ($K_d$) (\textbf{unrestricted domain})}}
\label{tab:K_unres}
\end{table}


\begin{table}[h!]
\centering
\footnotesize
\begin{tabular}{rcccccccc}
\toprule
&
\multicolumn{4}{c}{\textbf{Supply}} &
\multicolumn{4}{c}{\textbf{Demand}} \\
\cmidrule(lr){2-5}
\cmidrule(lr){6-9}
&
\makecell{\textbf{FMAE}\\{}[GWh]} &
\makecell{\textbf{FRMSE}\\{}[GWh]} &
\makecell{\textbf{FMAPE}\\{}[\%]} &
\textbf{rFMAE} &
\makecell{\textbf{FMAE}\\{}[GWh]} &
\makecell{\textbf{FRMSE}\\{}[GWh]} &
\makecell{\textbf{FMAPE}\\{}[\%]} &
\textbf{rFMAE} \\
\midrule

\textbf{Naive} & \cellcolor[RGB]{225,153,118}1.90 & \cellcolor[RGB]{220,141,118}2.61 & \cellcolor[RGB]{218,134,118}6.70 & \cellcolor[RGB]{225,153,118}1.000 & \cellcolor[RGB]{229,210,122}1.39 & \cellcolor[RGB]{218,134,118}2.33 & \cellcolor[RGB]{218,134,118}5.24 & \cellcolor[RGB]{229,210,122}1.000 \\

\textbf{ZST-ARX} & \cellcolor[RGB]{247,214,119}1.58 & \cellcolor[RGB]{242,213,120}2.07 & \cellcolor[RGB]{176,199,131}4.91 & \cellcolor[RGB]{247,214,119}0.832 & \cellcolor[RGB]{113,185,142}1.03 & \cellcolor[RGB]{113,185,142}1.43 & \cellcolor[RGB]{233,174,119}4.29 & \cellcolor[RGB]{113,185,142}0.737 \\

\textbf{ZST-VARX} & \cellcolor[RGB]{231,211,122}1.53 & \cellcolor[RGB]{223,209,123}1.98 & \cellcolor[RGB]{180,200,130}4.95 & \cellcolor[RGB]{231,211,122}0.804 & \cellcolor[RGB]{148,192,136}1.14 & \cellcolor[RGB]{131,189,139}1.49 & \cellcolor[RGB]{226,157,118}4.68 & \cellcolor[RGB]{148,192,136}0.817 \\

\textbf{ZST-fARX} & \cellcolor[RGB]{218,134,118}2.01 & \cellcolor[RGB]{218,134,118}2.66 & \cellcolor[RGB]{243,203,119}5.69 & \cellcolor[RGB]{218,134,118}1.055 & \cellcolor[RGB]{152,193,135}1.15 & \cellcolor[RGB]{135,190,138}1.50 & \cellcolor[RGB]{224,152,118}4.81 & \cellcolor[RGB]{152,193,135}0.826 \\

\textbf{ZST-fVARX} & \cellcolor[RGB]{241,198,119}1.67 & \cellcolor[RGB]{240,195,119}2.23 & \cellcolor[RGB]{205,205,126}5.16 & \cellcolor[RGB]{241,198,119}0.878 & \cellcolor[RGB]{164,196,133}1.19 & \cellcolor[RGB]{134,189,138}1.50 & \cellcolor[RGB]{222,147,118}4.93 & \cellcolor[RGB]{164,196,133}0.852 \\

\textbf{FPCA-ARX} & \cellcolor[RGB]{117,186,141}1.17 & \cellcolor[RGB]{119,186,140}1.54 & \cellcolor[RGB]{113,185,142}4.36 & \cellcolor[RGB]{117,186,141}0.615 & \cellcolor[RGB]{218,134,118}1.87 & \cellcolor[RGB]{212,207,125}1.76 & \cellcolor[RGB]{239,192,119}3.88 & \cellcolor[RGB]{218,134,118}1.346 \\

\textbf{FPCA-VARX} & \cellcolor[RGB]{133,189,138}1.22 & \cellcolor[RGB]{133,189,138}1.60 & \cellcolor[RGB]{144,191,136}4.63 & \cellcolor[RGB]{133,189,138}0.641 & \cellcolor[RGB]{192,202,128}1.28 & \cellcolor[RGB]{145,192,136}1.54 & \cellcolor[RGB]{116,185,141}1.52 & \cellcolor[RGB]{192,202,128}0.917 \\

\textbf{FPCA-fARX} & \cellcolor[RGB]{113,185,142}1.16 & \cellcolor[RGB]{113,185,142}1.51 & \cellcolor[RGB]{118,186,141}4.41 & \cellcolor[RGB]{113,185,142}0.607 & \cellcolor[RGB]{237,187,119}1.59 & \cellcolor[RGB]{144,192,136}1.54 & \cellcolor[RGB]{197,203,128}2.65 & \cellcolor[RGB]{237,187,119}1.144 \\

\textbf{FPCA-fVARX} & \cellcolor[RGB]{126,188,139}1.20 & \cellcolor[RGB]{124,187,140}1.56 & \cellcolor[RGB]{136,190,138}4.57 & \cellcolor[RGB]{126,188,139}0.629 & \cellcolor[RGB]{187,201,129}1.26 & \cellcolor[RGB]{147,192,136}1.55 & \cellcolor[RGB]{113,185,142}1.47 & \cellcolor[RGB]{187,201,129}0.905 \\

\bottomrule
\end{tabular}
\caption{\rev{Functional error metrics for supply and demand curves prediction (\textbf{GME, unrestricted domain}). The green-yellow-red colormap is applied column-wise.}}
\label{tab:gme_supply_demand_unres}
\end{table}

\begin{table}[h!]
\centering
\footnotesize
\begin{tabular}{rcccccccc}
\toprule
&
\multicolumn{4}{c}{\textbf{Supply}} &
\multicolumn{4}{c}{\textbf{Demand}} \\
\cmidrule(lr){2-5}
\cmidrule(lr){6-9}
&
\makecell{\textbf{FMAE}\\{}[GWh]} &
\makecell{\textbf{FRMSE}\\{}[GWh]} &
\makecell{\textbf{FMAPE}\\{}[\%]} &
\textbf{rFMAE} &
\makecell{\textbf{FMAE}\\{}[GWh]} &
\makecell{\textbf{FRMSE}\\{}[GWh]} &
\makecell{\textbf{FMAPE}\\{}[\%]} &
\textbf{rFMAE} \\
\midrule
\textbf{Naive} & \cellcolor[RGB]{218,134,118}3.81 & \cellcolor[RGB]{218,134,118}5.59 & \cellcolor[RGB]{218,134,118}9.50 & \cellcolor[RGB]{218,134,118}1.000 & \cellcolor[RGB]{218,134,118}2.73 & \cellcolor[RGB]{218,134,118}4.00 & \cellcolor[RGB]{218,134,118}8.22 & \cellcolor[RGB]{218,134,118}1.000 \\

\textbf{ZST-ARX} & \cellcolor[RGB]{140,191,137}2.28 & \cellcolor[RGB]{128,188,139}2.97 & \cellcolor[RGB]{144,191,136}5.82 & \cellcolor[RGB]{140,191,137}0.598 & \cellcolor[RGB]{185,201,130}2.03 & \cellcolor[RGB]{158,195,134}2.63 & \cellcolor[RGB]{189,202,129}6.18 & \cellcolor[RGB]{185,201,130}0.742 \\

\textbf{ZST-VARX} & \cellcolor[RGB]{149,193,136}2.33 & \cellcolor[RGB]{135,190,138}3.05 & \cellcolor[RGB]{155,194,135}5.99 & \cellcolor[RGB]{149,193,136}0.613 & \cellcolor[RGB]{196,203,128}2.07 & \cellcolor[RGB]{167,197,133}2.68 & \cellcolor[RGB]{200,204,127}6.29 & \cellcolor[RGB]{196,203,128}0.756 \\

\textbf{ZST-fARX} & \cellcolor[RGB]{120,186,140}2.15 & \cellcolor[RGB]{115,185,141}2.84 & \cellcolor[RGB]{122,187,140}5.48 & \cellcolor[RGB]{120,186,140}0.565 & \cellcolor[RGB]{131,189,138}1.84 & \cellcolor[RGB]{126,187,139}2.43 & \cellcolor[RGB]{131,189,138}5.56 & \cellcolor[RGB]{131,189,138}0.672 \\

\textbf{ZST-fVARX} & \cellcolor[RGB]{126,188,139}2.19 & \cellcolor[RGB]{122,187,140}2.91 & \cellcolor[RGB]{130,188,139}5.60 & \cellcolor[RGB]{126,188,139}0.575 & \cellcolor[RGB]{130,188,139}1.83 & \cellcolor[RGB]{124,187,140}2.42 & \cellcolor[RGB]{131,189,139}5.56 & \cellcolor[RGB]{130,188,139}0.671 \\

\textbf{FPCA-ARX} & \cellcolor[RGB]{141,191,137}2.28 & \cellcolor[RGB]{129,188,139}2.98 & \cellcolor[RGB]{143,191,136}5.81 & \cellcolor[RGB]{141,191,137}0.599 & \cellcolor[RGB]{168,197,132}1.97 & \cellcolor[RGB]{146,192,136}2.55 & \cellcolor[RGB]{173,198,132}6.00 & \cellcolor[RGB]{168,197,132}0.720 \\

\textbf{FPCA-VARX} & \cellcolor[RGB]{150,193,135}2.34 & \cellcolor[RGB]{136,190,138}3.06 & \cellcolor[RGB]{157,194,134}6.02 & \cellcolor[RGB]{150,193,135}0.615 & \cellcolor[RGB]{190,202,129}2.05 & \cellcolor[RGB]{160,195,134}2.64 & \cellcolor[RGB]{196,203,128}6.25 & \cellcolor[RGB]{190,202,129}0.749 \\

\textbf{FPCA-fARX} & \cellcolor[RGB]{113,185,142}2.10 & \cellcolor[RGB]{113,185,142}2.81 & \cellcolor[RGB]{113,185,142}5.34 & \cellcolor[RGB]{113,185,142}0.552 & \cellcolor[RGB]{113,185,142}1.77 & \cellcolor[RGB]{113,185,142}2.35 & \cellcolor[RGB]{113,185,142}5.36 & \cellcolor[RGB]{113,185,142}0.648 \\

\textbf{FPCA-fVARX} & \cellcolor[RGB]{119,186,140}2.15 & \cellcolor[RGB]{117,185,141}2.86 & \cellcolor[RGB]{122,187,140}5.48 & \cellcolor[RGB]{119,186,140}0.563 & \cellcolor[RGB]{119,186,140}1.79 & \cellcolor[RGB]{115,185,141}2.36 & \cellcolor[RGB]{121,186,140}5.45 & \cellcolor[RGB]{119,186,140}0.656 \\
\bottomrule
\end{tabular}
\caption{\rev{Functional error metrics for supply and demand curves prediction (\textbf{EPEX-DE-LU, unrestricted domain}). The green-yellow-red colormap is applied market-wise.}}
\label{tab:epex-de-lu_supply_demand_unres}
\end{table}

\begin{table}[h!]
\centering
\footnotesize
\begin{tabular}{rcccccccc}
\toprule
&
\multicolumn{4}{c}{\textbf{Supply}} &
\multicolumn{4}{c}{\textbf{Demand}} \\
\cmidrule(lr){2-5}
\cmidrule(lr){6-9}
&
\makecell{\textbf{FMAE}\\{}[GWh]} &
\makecell{\textbf{FRMSE}\\{}[GWh]} &
\makecell{\textbf{FMAPE}\\{}[\%]} &
\textbf{rFMAE} &
\makecell{\textbf{FMAE}\\{}[GWh]} &
\makecell{\textbf{FRMSE}\\{}[GWh]} &
\makecell{\textbf{FMAPE}\\{}[\%]} &
\textbf{rFMAE} \\
\midrule
\textbf{Naive} & \cellcolor[RGB]{218,134,118}3.10 & \cellcolor[RGB]{218,134,118}4.23 & \cellcolor[RGB]{218,134,118}15.68 & \cellcolor[RGB]{218,134,118}1.000 & \cellcolor[RGB]{218,134,118}2.39 & \cellcolor[RGB]{218,134,118}3.19 & \cellcolor[RGB]{218,134,118}16.73 & \cellcolor[RGB]{218,134,118}1.000 \\

\textbf{ZST-ARX} & \cellcolor[RGB]{143,191,137}2.01 & \cellcolor[RGB]{137,190,137}2.57 & \cellcolor[RGB]{140,191,137}10.27 & \cellcolor[RGB]{143,191,137}0.648 & \cellcolor[RGB]{164,196,133}1.88 & \cellcolor[RGB]{158,195,134}2.41 & \cellcolor[RGB]{160,195,134}12.99 & \cellcolor[RGB]{164,196,133}0.787 \\

\textbf{ZST-VARX} & \cellcolor[RGB]{150,193,135}2.04 & \cellcolor[RGB]{144,192,136}2.62 & \cellcolor[RGB]{148,192,136}10.46 & \cellcolor[RGB]{150,193,135}0.657 & \cellcolor[RGB]{193,202,128}1.95 & \cellcolor[RGB]{183,200,130}2.50 & \cellcolor[RGB]{200,204,127}13.66 & \cellcolor[RGB]{193,202,128}0.815 \\

\textbf{ZST-fARX} & \cellcolor[RGB]{161,195,134}2.09 & \cellcolor[RGB]{152,193,135}2.67 & \cellcolor[RGB]{157,194,134}10.66 & \cellcolor[RGB]{161,195,134}0.674 & \cellcolor[RGB]{116,185,141}1.77 & \cellcolor[RGB]{116,185,141}2.27 & \cellcolor[RGB]{116,185,141}12.25 & \cellcolor[RGB]{116,185,141}0.739 \\

\textbf{ZST-fVARX} & \cellcolor[RGB]{153,194,135}2.05 & \cellcolor[RGB]{146,192,136}2.63 & \cellcolor[RGB]{152,193,135}10.55 & \cellcolor[RGB]{153,194,135}0.662 & \cellcolor[RGB]{130,188,139}1.80 & \cellcolor[RGB]{130,188,139}2.32 & \cellcolor[RGB]{138,190,137}12.62 & \cellcolor[RGB]{130,188,139}0.754 \\

\textbf{FPCA-ARX} & \cellcolor[RGB]{124,187,140}1.92 & \cellcolor[RGB]{123,187,140}2.47 & \cellcolor[RGB]{119,186,140}9.82 & \cellcolor[RGB]{124,187,140}0.619 & \cellcolor[RGB]{163,196,133}1.88 & \cellcolor[RGB]{156,194,134}2.41 & \cellcolor[RGB]{158,195,134}12.96 & \cellcolor[RGB]{163,196,133}0.786 \\

\textbf{FPCA-VARX} & \cellcolor[RGB]{139,190,137}1.99 & \cellcolor[RGB]{136,190,138}2.56 & \cellcolor[RGB]{137,190,138}10.21 & \cellcolor[RGB]{139,190,137}0.641 & \cellcolor[RGB]{196,203,128}1.95 & \cellcolor[RGB]{185,201,130}2.51 & \cellcolor[RGB]{201,204,127}13.68 & \cellcolor[RGB]{196,203,128}0.817 \\

\textbf{FPCA-fARX} & \cellcolor[RGB]{113,185,142}1.87 & \cellcolor[RGB]{113,185,142}2.40 & \cellcolor[RGB]{113,185,142}9.66 & \cellcolor[RGB]{113,185,142}0.602 & \cellcolor[RGB]{113,185,142}1.76 & \cellcolor[RGB]{113,185,142}2.26 & \cellcolor[RGB]{113,185,142}12.19 & \cellcolor[RGB]{113,185,142}0.736 \\

\textbf{FPCA-fVARX} & \cellcolor[RGB]{123,187,140}1.91 & \cellcolor[RGB]{122,187,140}2.47 & \cellcolor[RGB]{123,187,140}9.90 & \cellcolor[RGB]{123,187,140}0.618 & \cellcolor[RGB]{127,188,139}1.79 & \cellcolor[RGB]{127,188,139}2.31 & \cellcolor[RGB]{134,189,138}12.55 & \cellcolor[RGB]{127,188,139}0.750 \\
\bottomrule
\end{tabular}
\caption{\rev{Functional error metrics for supply and demand curves prediction (\textbf{EPEX-FR, unrestricted domain}). The green-yellow-red colormap is applied market-wise.}}
\label{tab:epex_fr_supply_demand_unres}
\end{table}


\begin{table}[h!]
\centering
\footnotesize
\begin{tabular}{lrccc|ccc|ccc}
\toprule
& &
\multicolumn{3}{c}{GME} &
\multicolumn{3}{c}{EPEX-DE-LU} &
\multicolumn{3}{c}{EPEX-FR} \\
\cmidrule(lr){3-5}   
\cmidrule(lr){6-8}   
\cmidrule(lr){9-11}  
& &
\textbf{MAE} & \textbf{RMSE} & \multicolumn{1}{c}{\textbf{rMAE}} & 
\textbf{MAE} & \textbf{RMSE} & \multicolumn{1}{c}{\textbf{rMAE}} & 
\textbf{MAE} & \textbf{RMSE} & \textbf{rMAE} \\
\midrule

\multirow{4}{*}{\textit{price-based}}

& \textbf{Naive}
& \cellcolor[RGB]{218,134,118}11.32
& \cellcolor[RGB]{218,134,118}17.23
& \cellcolor[RGB]{218,134,118}1.000
& \cellcolor[RGB]{218,134,118}27.17 
& \cellcolor[RGB]{239,193,119}43.22 
& \cellcolor[RGB]{218,134,118}1.000
& \cellcolor[RGB]{218,134,118}21.48 
& \cellcolor[RGB]{164,196,133}29.67 
& \cellcolor[RGB]{218,134,118}1.000 \\

& \textbf{ARX}
& \cellcolor[RGB]{140,191,137}8.32
& \cellcolor[RGB]{134,189,138}11.87
& \cellcolor[RGB]{140,191,137}0.735
& \cellcolor[RGB]{137,190,138}14.64 
& \cellcolor[RGB]{113,185,141}26.65 
& \cellcolor[RGB]{137,190,138}0.539
& \cellcolor[RGB]{165,196,133}15.85 
& \cellcolor[RGB]{119,186,141}20.22 
& \cellcolor[RGB]{165,196,133}0.738 \\

& \textbf{fARX} 
& \cellcolor[RGB]{113,185,142}7.98
& \cellcolor[RGB]{113,185,142}11.39
& \cellcolor[RGB]{113,185,142}0.705
& \cellcolor[RGB]{126,187,139}14.07 
& \cellcolor[RGB]{116,185,141}26.95 
& \cellcolor[RGB]{126,187,139}0.518
& \cellcolor[RGB]{117,185,141}14.58 
& \cellcolor[RGB]{113,185,142}18.95 
& \cellcolor[RGB]{117,185,141}0.679 \\

& \textbf{LEAR}
& \cellcolor[RGB]{135,190,138}8.26
& \cellcolor[RGB]{133,189,138}11.84
& \cellcolor[RGB]{135,190,138}0.730
& \cellcolor[RGB]{113,185,142}13.40 
& \cellcolor[RGB]{113,185,142}26.61 
& \cellcolor[RGB]{113,185,142}0.493
& \cellcolor[RGB]{118,186,141}14.62 
& \cellcolor[RGB]{114,185,141}19.23 
& \cellcolor[RGB]{118,186,141}0.681 \\

\midrule
\multirow{8}{*}{\textit{curve-based}}

& \textbf{ZST-ARX}
& \cellcolor[RGB]{180,200,130}8.82
& \cellcolor[RGB]{179,199,131}12.83
& \cellcolor[RGB]{180,200,130}0.779
& \cellcolor[RGB]{183,200,130}16.99 
& \cellcolor[RGB]{189,202,129}34.03 
& \cellcolor[RGB]{183,200,130}0.625
& \cellcolor[RGB]{176,199,131}16.12 
& \cellcolor[RGB]{125,187,140}21.47 
& \cellcolor[RGB]{176,199,131}0.751 \\

& \textbf{ZST-VARX}
& \cellcolor[RGB]{189,202,129}8.93
& \cellcolor[RGB]{194,203,128}13.16
& \cellcolor[RGB]{189,202,129}0.789
& \cellcolor[RGB]{187,201,129}17.20 
& \cellcolor[RGB]{198,204,127}34.93 
& \cellcolor[RGB]{187,201,129}0.633
& \cellcolor[RGB]{171,197,132}15.98 
& \cellcolor[RGB]{123,187,140}21.15 
& \cellcolor[RGB]{171,197,132}0.744 \\

& \textbf{ZST-fARX}
& \cellcolor[RGB]{215,207,125}9.24
& \cellcolor[RGB]{215,207,125}13.61
& \cellcolor[RGB]{215,207,125}0.817
& \cellcolor[RGB]{238,189,119}22.45 
& \cellcolor[RGB]{218,134,118}52.76 
& \cellcolor[RGB]{238,189,119}0.826
& \cellcolor[RGB]{233,176,119}19.63 
& \cellcolor[RGB]{218,134,118}75.29 
& \cellcolor[RGB]{233,176,119}0.914 \\

& \textbf{ZST-fVARX}
& \cellcolor[RGB]{206,205,126}9.13
& \cellcolor[RGB]{217,208,124}13.65
& \cellcolor[RGB]{206,205,126}0.807
& \cellcolor[RGB]{247,214,120}20.24 
& \cellcolor[RGB]{231,170,118}46.80 
& \cellcolor[RGB]{247,214,120}0.745
& \cellcolor[RGB]{210,206,126}17.01 
& \cellcolor[RGB]{135,189,138}23.56 
& \cellcolor[RGB]{210,206,126}0.792 \\

& \textbf{FPCA-ARX}
& \cellcolor[RGB]{123,187,140}8.10
& \cellcolor[RGB]{119,186,140}11.53
& \cellcolor[RGB]{123,187,140}0.716
& \cellcolor[RGB]{158,195,134}15.70 
& \cellcolor[RGB]{144,192,136}29.67 
& \cellcolor[RGB]{158,195,134}0.578
& \cellcolor[RGB]{113,185,142}14.47 
& \cellcolor[RGB]{114,185,141}19.25 
& \cellcolor[RGB]{113,185,142}0.674 \\

& \textbf{FPCA-VARX}
& \cellcolor[RGB]{129,188,139}8.18
& \cellcolor[RGB]{131,189,138}11.80
& \cellcolor[RGB]{129,188,139}0.722
& \cellcolor[RGB]{162,196,133}15.95 
& \cellcolor[RGB]{136,190,138}28.89 
& \cellcolor[RGB]{162,196,133}0.587
& \cellcolor[RGB]{155,194,135}15.58 
& \cellcolor[RGB]{121,186,140}20.78 
& \cellcolor[RGB]{155,194,135}0.725 \\

& \textbf{FPCA-fARX}
& \cellcolor[RGB]{134,189,138}8.25
& \cellcolor[RGB]{129,188,139}11.74
& \cellcolor[RGB]{134,189,138}0.729
& \cellcolor[RGB]{206,205,126}18.16 
& \cellcolor[RGB]{247,214,119}39.78 
& \cellcolor[RGB]{206,205,126}0.668
& \cellcolor[RGB]{130,188,139}14.93 
& \cellcolor[RGB]{117,186,141}19.95 
& \cellcolor[RGB]{130,188,139}0.695 \\

& \textbf{FPCA-fVARX}
& \cellcolor[RGB]{149,193,136}8.43
& \cellcolor[RGB]{155,194,135}12.31
& \cellcolor[RGB]{149,193,136}0.745
& \cellcolor[RGB]{193,202,128}17.51 
& \cellcolor[RGB]{212,207,125}36.21 
& \cellcolor[RGB]{193,202,128}0.645
& \cellcolor[RGB]{129,188,139}14.89 
& \cellcolor[RGB]{118,186,141}20.15 
& \cellcolor[RGB]{129,188,139}0.693 \\

\bottomrule
\end{tabular}
\caption{\rev{Clearing price prediction performance (\textbf{unrestricted domain}). MAE and RMSE are expressed in \euro/MWh. The green-yellow-red colormap is applied column-wise.}}
\label{tab:price_all_markets_unres}
\end{table}

\clearpage
\section{FPCA results} \label{sec:fpca-res}

\begin{figure*}[h]
     \centering
     \begin{subfigure}[b]{\textwidth}
         \includegraphics[width=\textwidth]{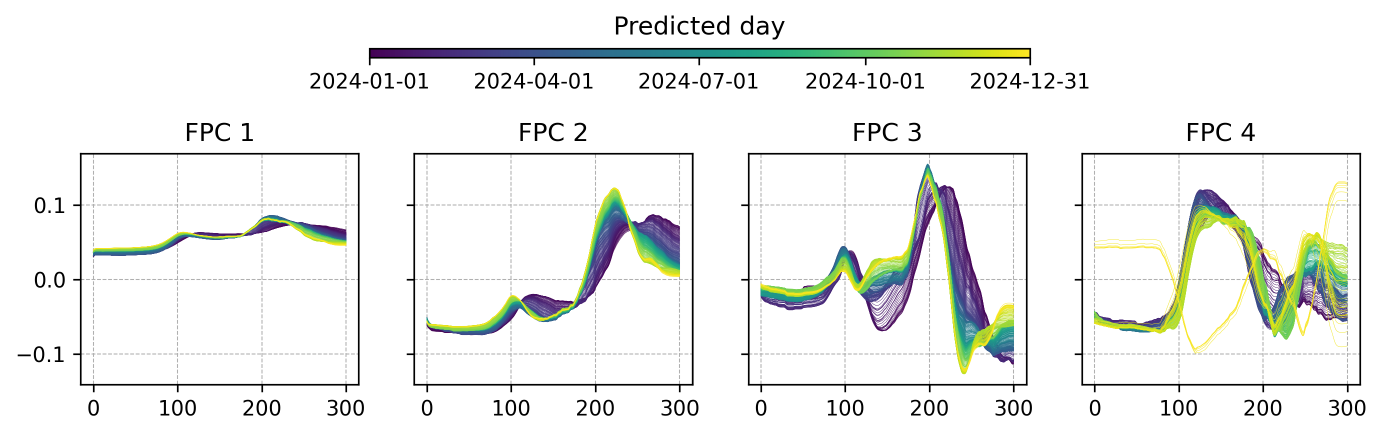}
         \caption{Supply}
     \end{subfigure}
     \begin{subfigure}[b]{\textwidth}
         \includegraphics[width=\textwidth]{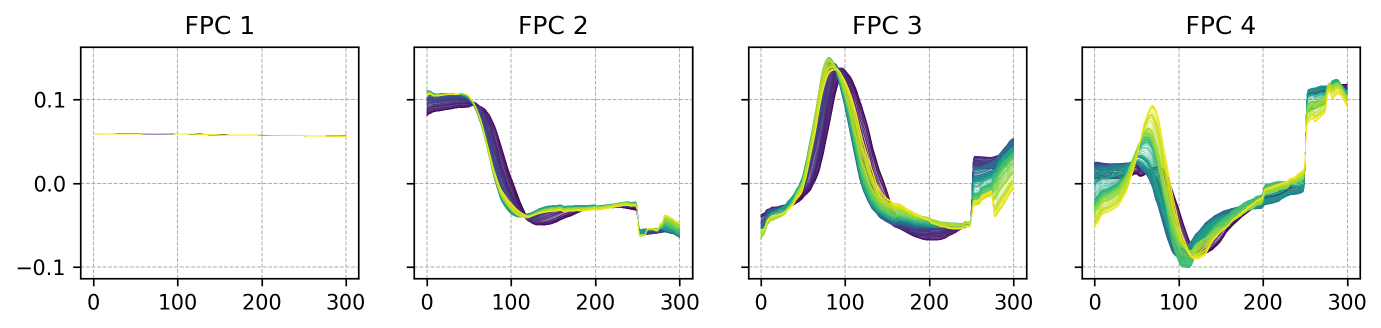}
         \caption{Demand}
     \end{subfigure}
        \caption{(\textit{Color optional}) Functional principal components across the test period (\textbf{GME})}
        \label{fig:fpc_gme}
\end{figure*}

\begin{figure*}[h]
     \centering
     \begin{subfigure}[b]{\textwidth}
         \includegraphics[width=\textwidth]{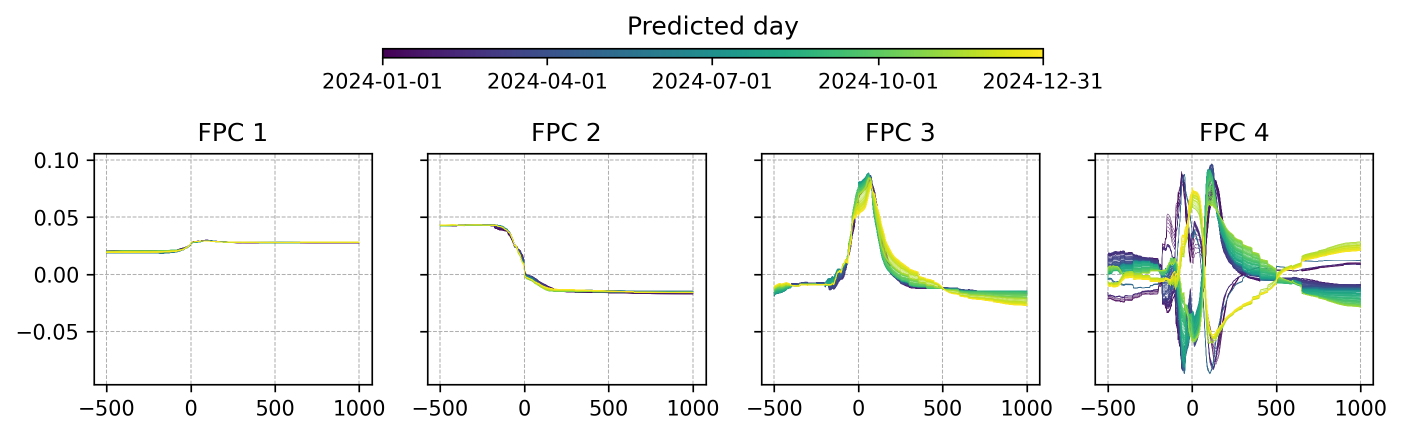}
         \caption{Supply}
     \end{subfigure}
     \begin{subfigure}[b]{\textwidth}
         \includegraphics[width=\textwidth]{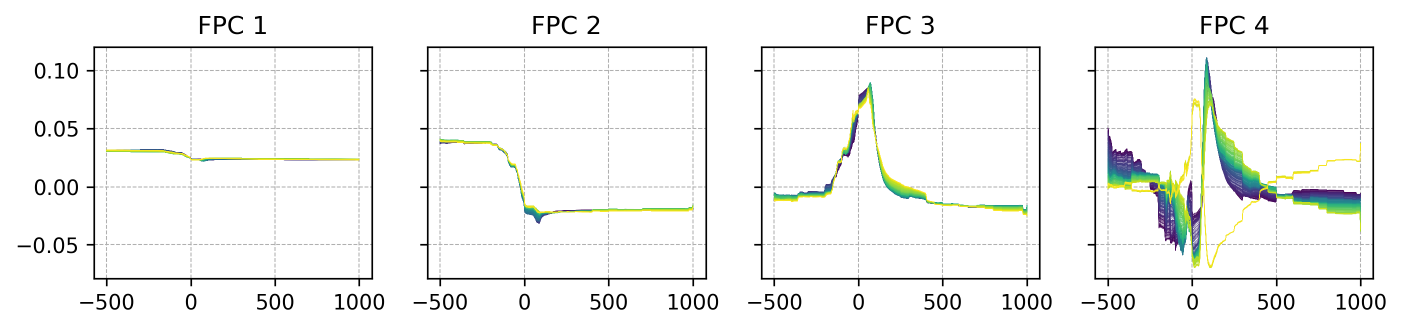}
         \caption{Demand}
     \end{subfigure}
        \caption{\rev{(\textit{Color optional}) Functional principal components across the test period (\textbf{EPEX-DE-LU})}}
        \label{fig:fpc_epex-de-lu}
\end{figure*}

\begin{figure*}[h]
     \centering
     \begin{subfigure}[b]{\textwidth}
         \includegraphics[width=\textwidth]{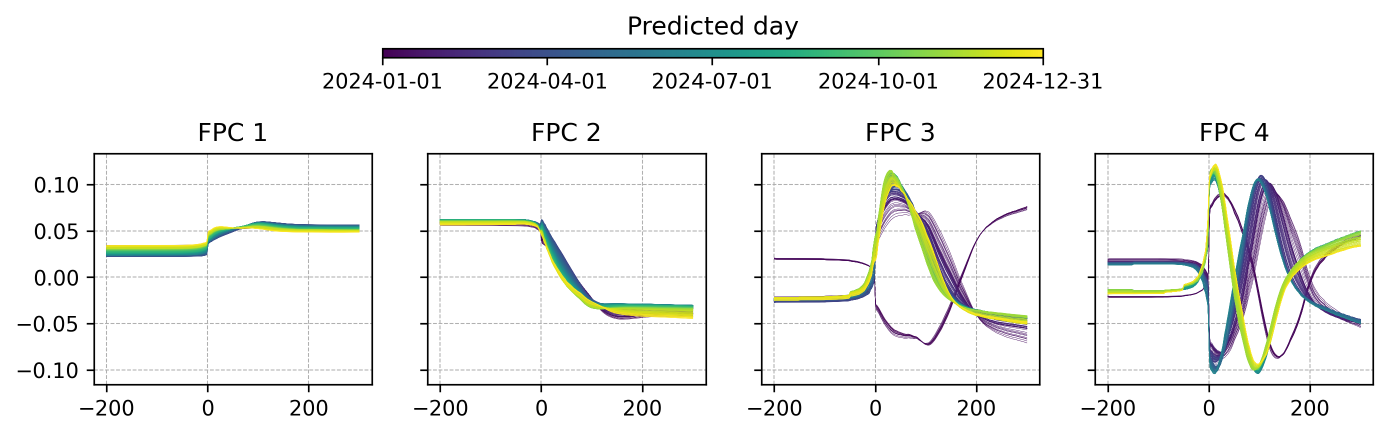}
         \caption{Supply}
     \end{subfigure}
     \begin{subfigure}[b]{\textwidth}
         \includegraphics[width=\textwidth]{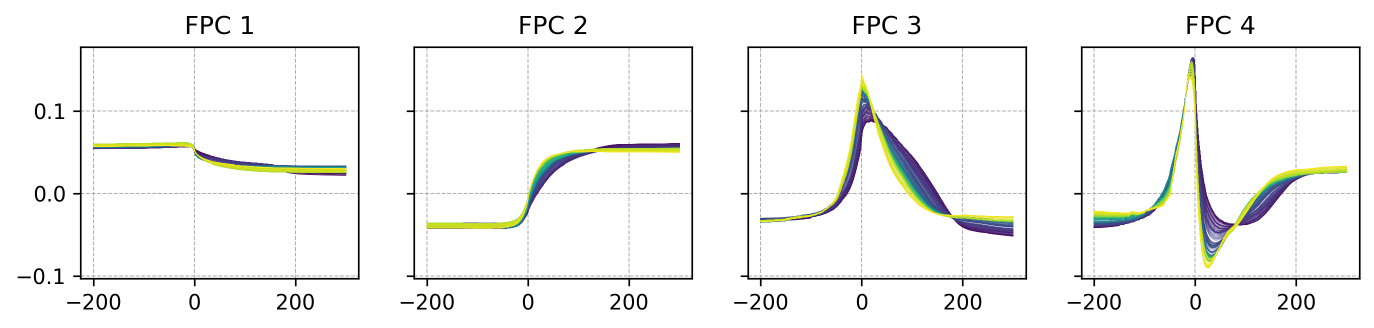}
         \caption{Demand}
     \end{subfigure}
        \caption{\rev{(\textit{Color optional}) Functional principal components across the test period (\textbf{EPEX-FR})}}
        \label{fig:fpc_epex-fr}
\end{figure*}

\begin{figure*}[h]
     \centering
     \begin{subfigure}[b]{0.8\textwidth}
         \includegraphics[width=\textwidth]{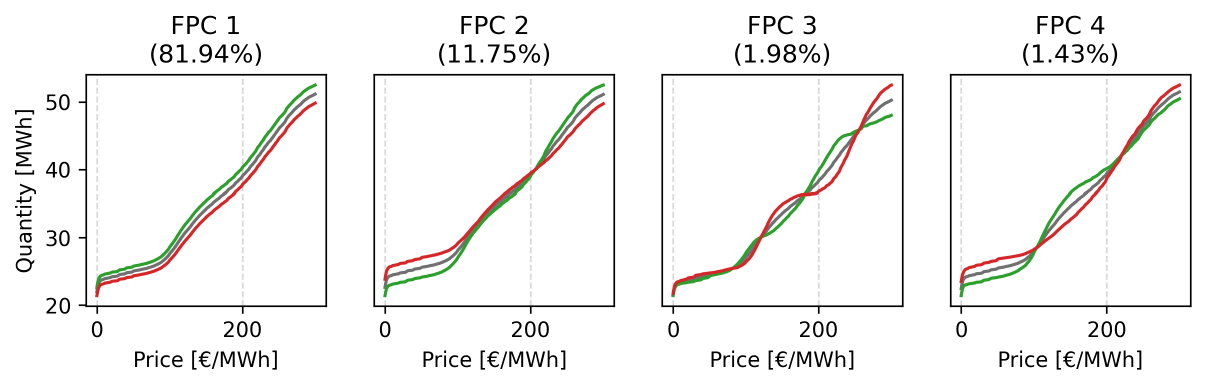}
         \caption{\textbf{GME}}
     \end{subfigure}
     \begin{subfigure}[b]{0.8\textwidth}
         \includegraphics[width=\textwidth]{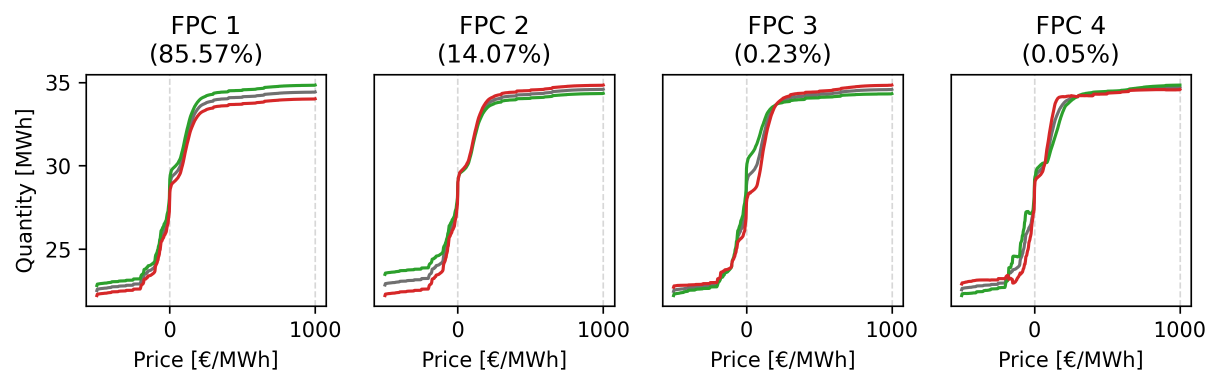}
         \caption{\textbf{EPEX-DE-LU}}
     \end{subfigure}
     \begin{subfigure}[b]{0.8\textwidth}
         \includegraphics[width=\textwidth]{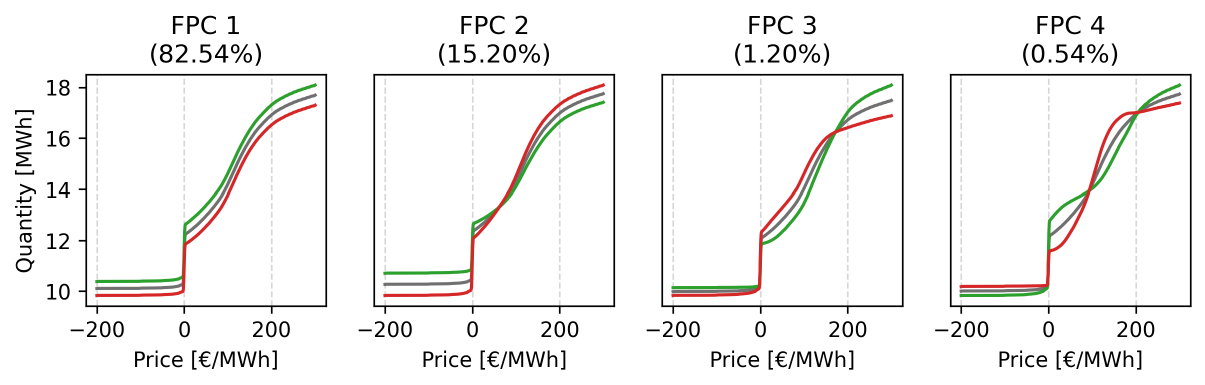}
         \caption{\textbf{EPEX-FR}}
     \end{subfigure}
        \caption{\rev{(\textit{Color optional}) Visualization of the effects of the first 4 \textit{supply} functional principal components (FPCs) on the mean curve (in grey) estimated for prediction of 2024-07-01. The green curve corresponds to mean curve to which a multiple of the FPC is added, while the red curve corresponds to the mean curve to which a multiple of the FPC is subtracted. The number in parenthesis above each plot is the proportion of variance explained by the FPC.}}
        \label{fig:fpc_effect_supply}
\end{figure*}

\begin{figure*}[h]
     \centering
     \begin{subfigure}[b]{0.8\textwidth}
         \includegraphics[width=\textwidth]{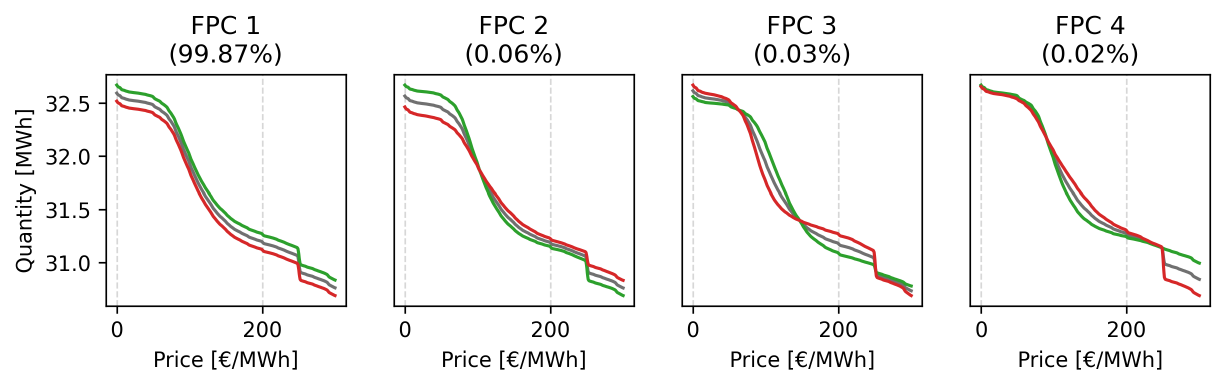}
         \caption{\textbf{GME}}
     \end{subfigure}
     \begin{subfigure}[b]{0.8\textwidth}
         \includegraphics[width=\textwidth]{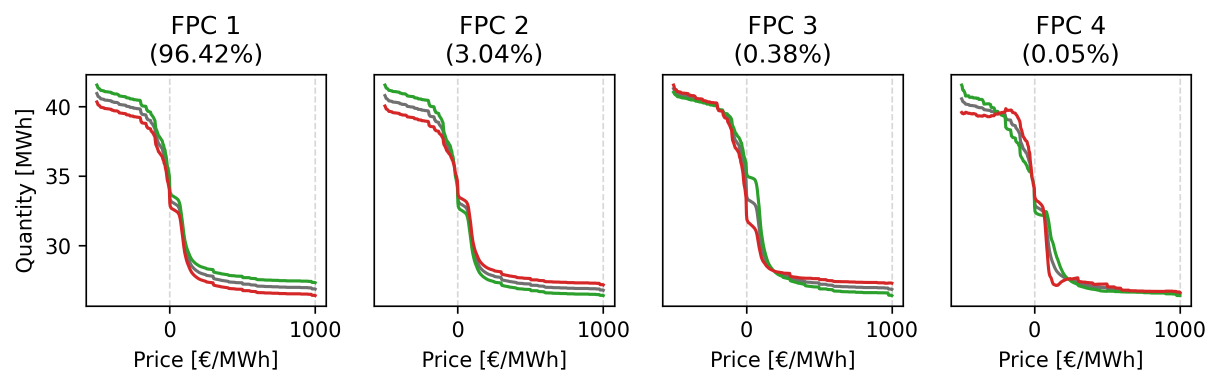}
         \caption{\textbf{EPEX-DE-LU}}
     \end{subfigure}
     \begin{subfigure}[b]{0.8\textwidth}
         \includegraphics[width=\textwidth]{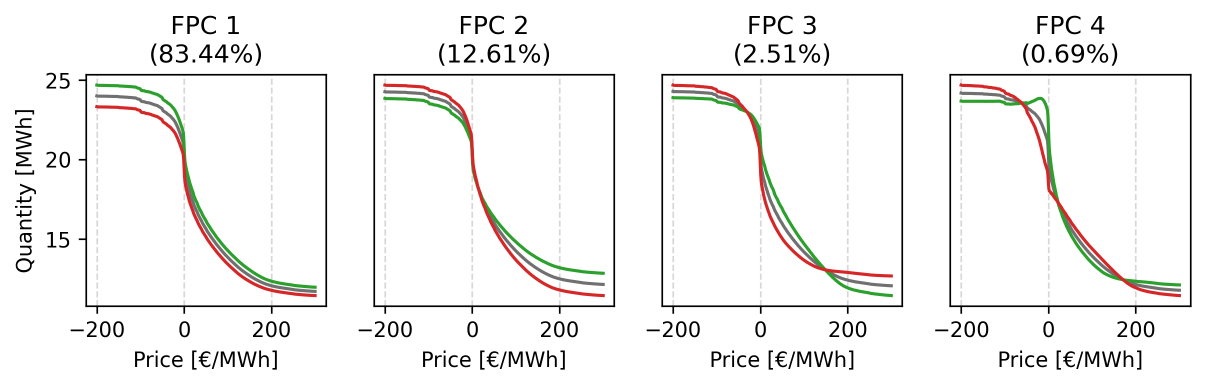}
         \caption{\textbf{EPEX-FR}}
     \end{subfigure}
        \caption{\rev{(\textit{Color optional}) Visualization of the effects of the first 4 \textit{demand} functional principal components (FPCs) on the mean curve (in grey) estimated for prediction of 2024-07-01. The green curve corresponds to mean curve to which a multiple of the FPC is added, while the red curve corresponds to the mean curve to which a multiple of the FPC is subtracted. The number in parenthesis above each plot is the proportion of variance explained by the FPC.}}
        \label{fig:fpc_effect_demand}
\end{figure*}

\clearpage
\section{DM tests for curves forecasts performance}
\label{sec:dm_curves}

\begin{figure*}[h!]
     \centering
     \begin{subfigure}[b]{0.4\textwidth}
         \includegraphics[width=\textwidth]{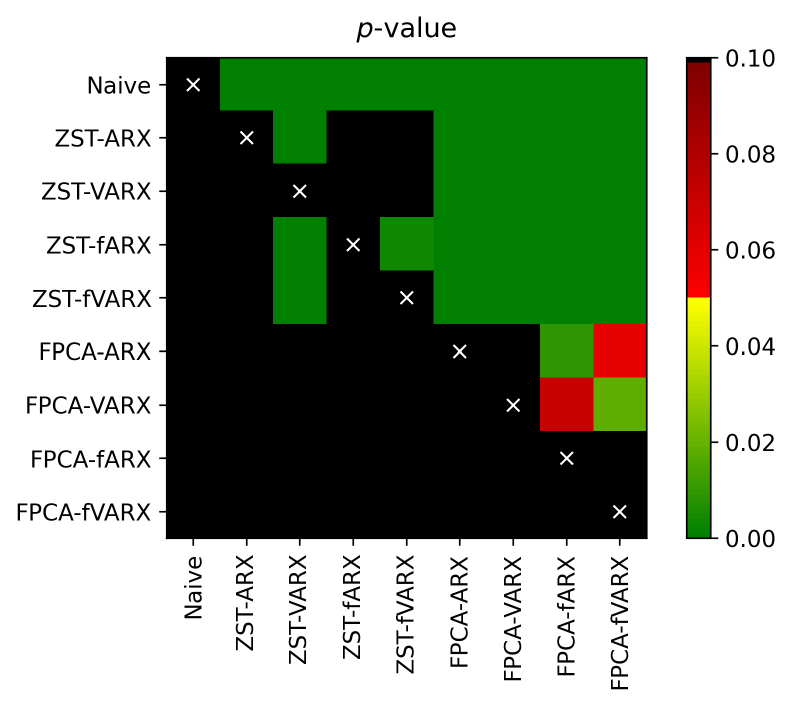}
         \caption{Supply (\textbf{GME})}
         \label{fig:dm_curve:gme_supply}
     \end{subfigure}
     \begin{subfigure}[b]{0.4\textwidth}
         \includegraphics[width=\textwidth]{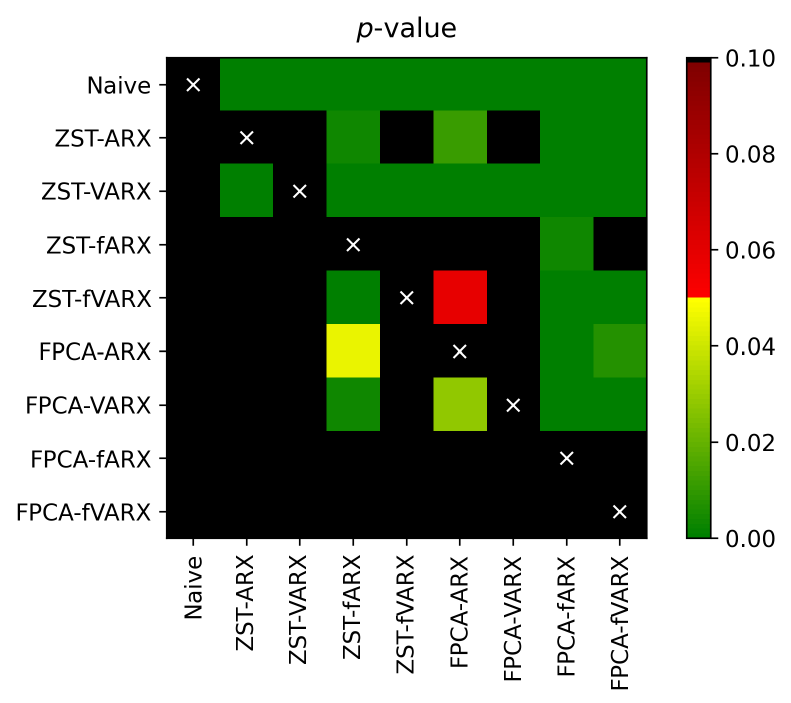}
         \caption{Demand (\textbf{GME})}
         \label{fig:dm_curve:gme_demand}
     \end{subfigure}
     \begin{subfigure}[b]{0.4\textwidth}
         \includegraphics[width=\textwidth]{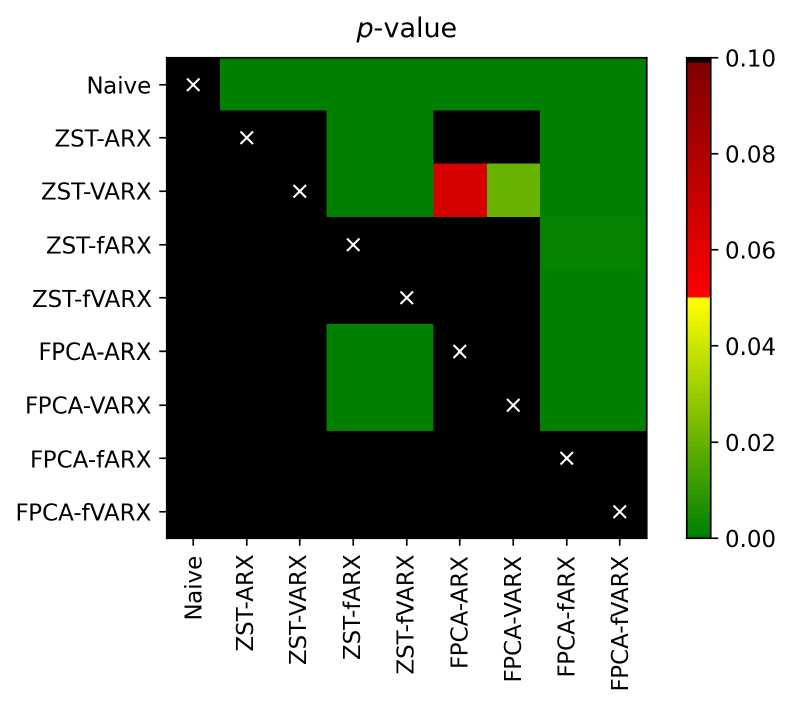}
         \caption{Supply (\textbf{EPEX-DE-LU})}
         \label{fig:dm_curve:epex-de-lu_supply}
     \end{subfigure}
     \begin{subfigure}[b]{0.4\textwidth}
         \includegraphics[width=\textwidth]{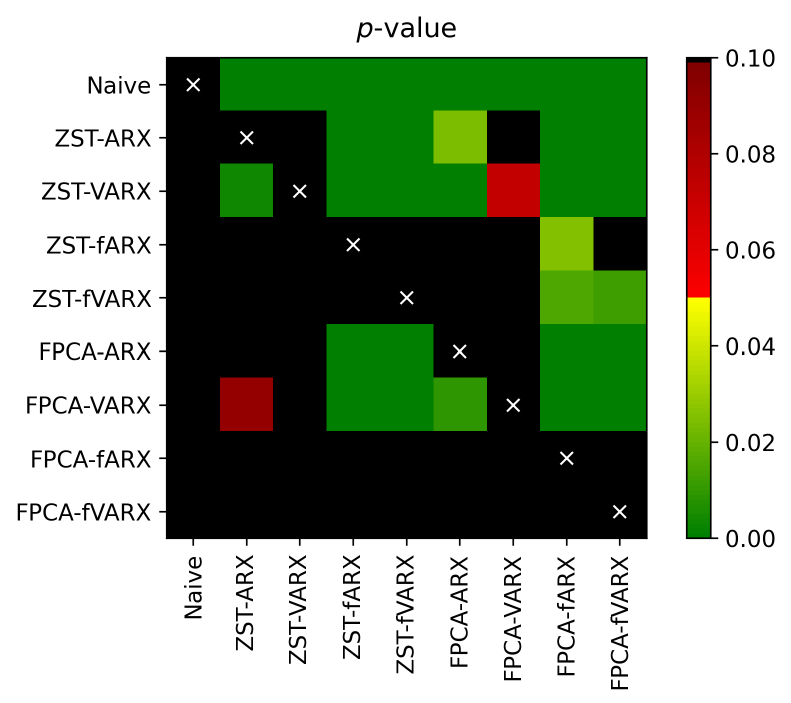}
         \caption{Demand (\textbf{EPEX-DE-LU})}
         \label{fig:dm_curve:epex-de-lu_demand}
     \end{subfigure}
     \begin{subfigure}[b]{0.4\textwidth}
         \includegraphics[width=\textwidth]{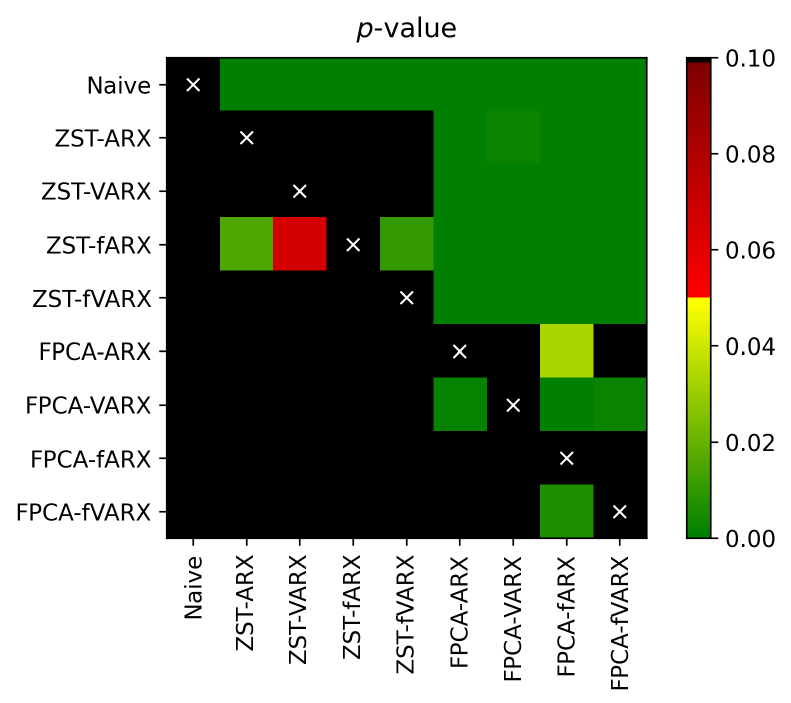}
         \caption{Supply (\textbf{EPEX-FR})}
         \label{fig:dm_curve:epex-fr_supply}
     \end{subfigure}
     \begin{subfigure}[b]{0.4\textwidth}
         \includegraphics[width=\textwidth]{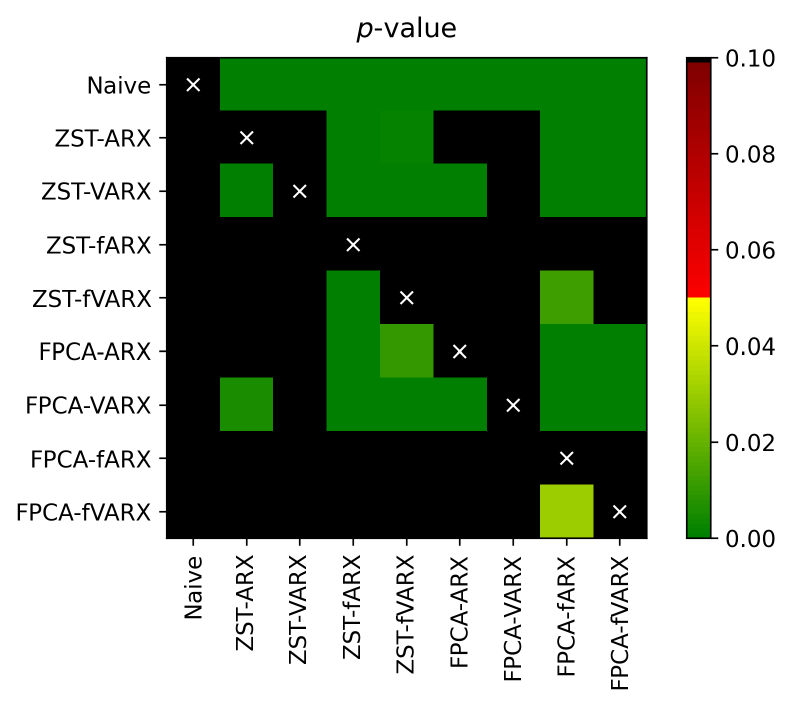}
         \caption{Demand (\textbf{EPEX-FR})}
         \label{fig:dm_curve:epex-fr_demand}
     \end{subfigure}
        \caption{\rev{(\textit{Color optional}) Results of the Diebold-Mariano test for the difference in curve forecasting performance on average daily absolute errors. The alternative hypothesis is that models on the $x$-axis outperform those on the $y$-axis (one-sided test).}}
        \label{fig:dm_curve}
\end{figure*}

\clearpage
\section{Example of curves and clearing price prediction}
\label{sec:preds}

\begin{figure}[h]
    \centering
    \begin{subfigure}[b]{0.32\textwidth}
        \includegraphics[width=\textwidth]{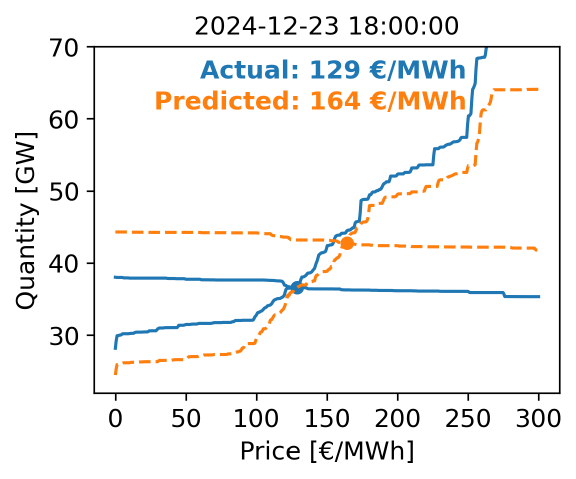}
        \caption{\textbf{Naive}}
    \end{subfigure}
    \begin{subfigure}[b]{0.32\textwidth}
        \includegraphics[width=\textwidth]{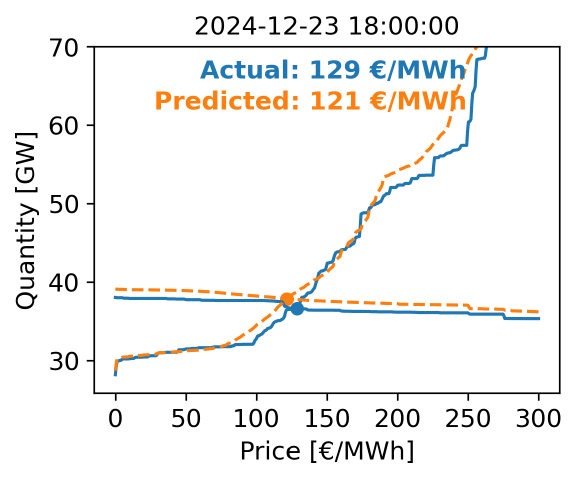}
        \caption{\textbf{ZST-ARX}}
    \end{subfigure}
    \begin{subfigure}[b]{0.32\textwidth}
        \includegraphics[width=\textwidth]{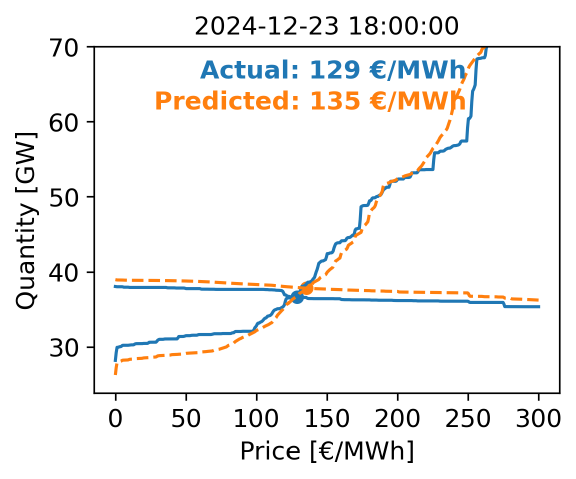}
        \caption{\textbf{ZST-VARX}}
    \end{subfigure}

    \vspace{0.3cm}
        
    \hfill
    \begin{subfigure}[b]{0.32\textwidth}
        \includegraphics[width=\textwidth]{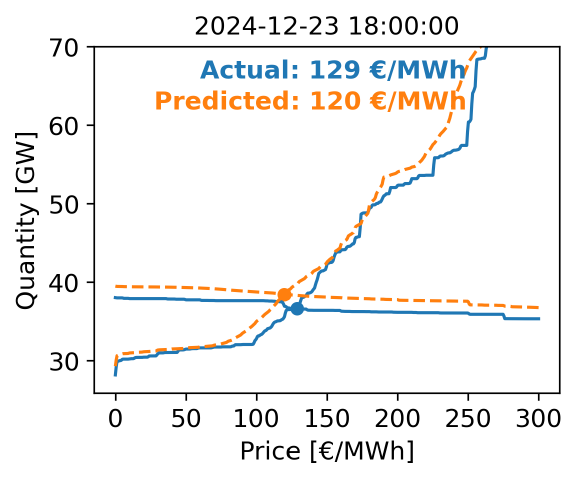}
        \caption{\textbf{ZST-fARX}}
    \end{subfigure}
    \hfill
    \begin{subfigure}[b]{0.32\textwidth}
        \includegraphics[width=\textwidth]{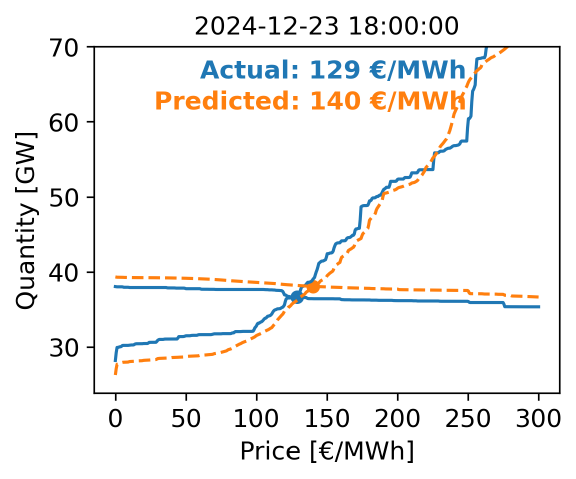}
        \caption{\textbf{ZST-fVARX}}
    \end{subfigure}
    \hfill
    \begin{subfigure}[b]{0.32\textwidth}
        \includegraphics[width=\textwidth]{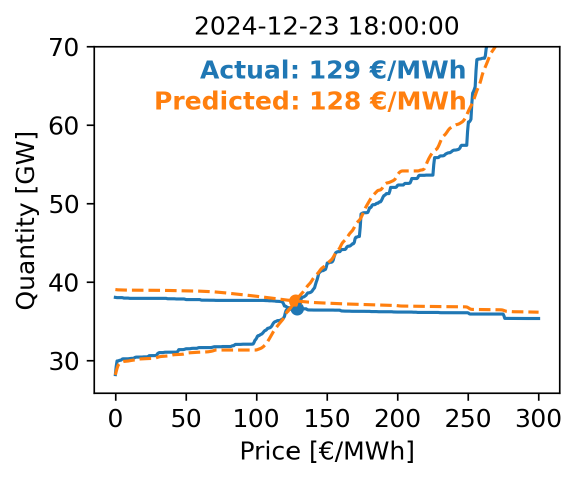}
        \caption{\textbf{FPCA-ARX}}
    \end{subfigure}
    
    \vspace{0.3cm}
    
    \hfill
    \begin{subfigure}[b]{0.32\textwidth}
        \includegraphics[width=\textwidth]{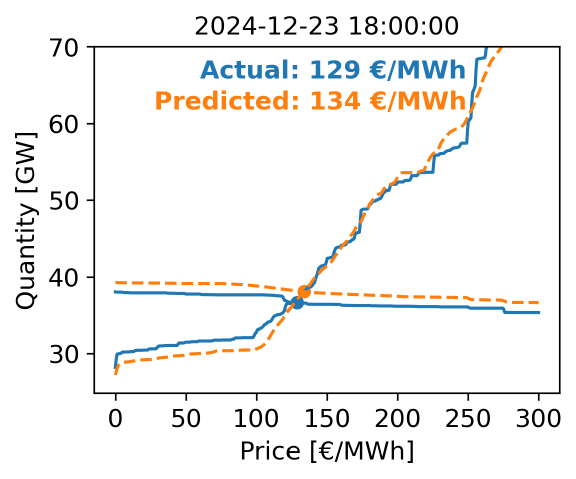}
        \caption{\textbf{FPCA-VARX}}
    \end{subfigure}
    \begin{subfigure}[b]{0.32\textwidth}
        \includegraphics[width=\textwidth]{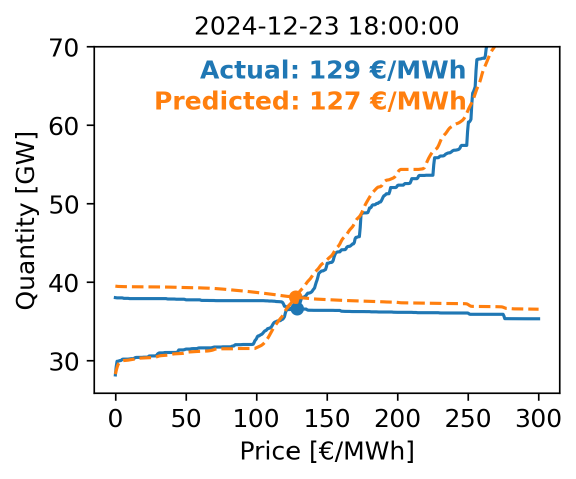}
        \caption{\textbf{FPCA-fARX}}
    \end{subfigure}
    \hfill
    \begin{subfigure}[b]{0.32\textwidth}
        \includegraphics[width=\textwidth]{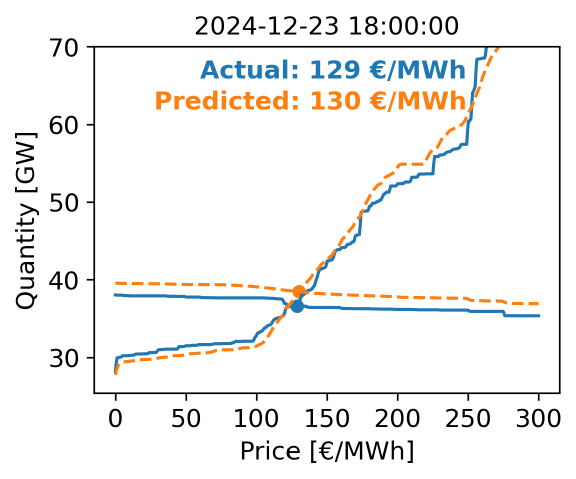}
        \caption{\textbf{FPCA-fVARX}}
    \end{subfigure}
    \caption{(\textit{Color optional}) Curves and clearing price predictions for the 18:00-19:00 interval of December 23\textsuperscript{rd}, 2024. (\textbf{GME})}
    \label{fig:preds_gme}
\end{figure}

\begin{figure}[h]
    \centering
    \begin{subfigure}[b]{0.32\textwidth}
        \includegraphics[width=\textwidth]{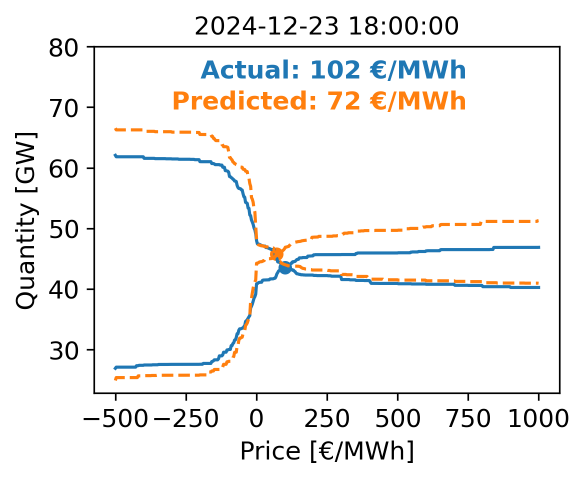}
        \caption{\textbf{Naive}}
    \end{subfigure}
    \begin{subfigure}[b]{0.32\textwidth}
        \includegraphics[width=\textwidth]{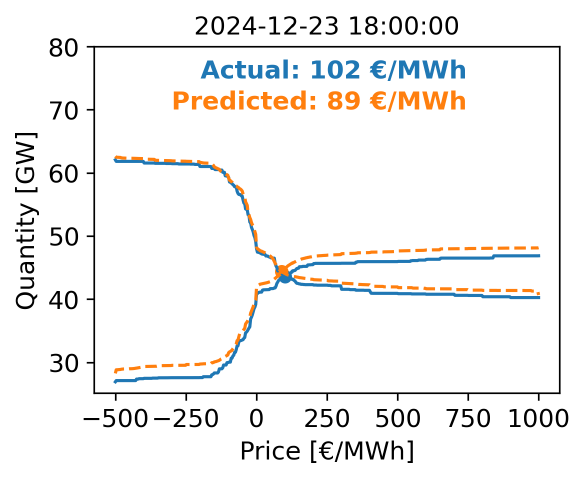}
        \caption{\textbf{ZST-ARX}}
    \end{subfigure}
    \begin{subfigure}[b]{0.32\textwidth}
        \includegraphics[width=\textwidth]{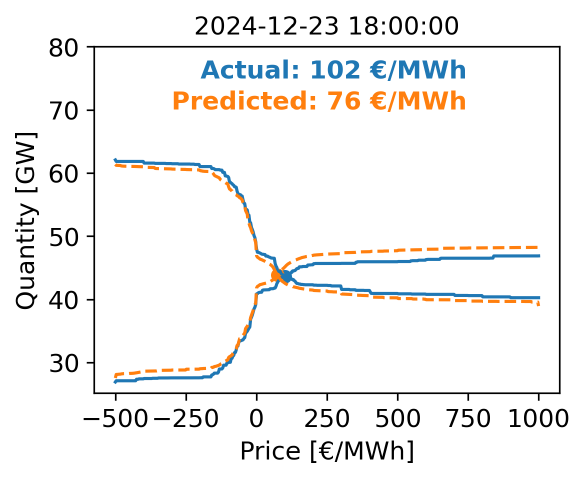}
        \caption{\textbf{ZST-VARX}}
    \end{subfigure}

    \vspace{0.3cm}
        
    \hfill
    \begin{subfigure}[b]{0.32\textwidth}
        \includegraphics[width=\textwidth]{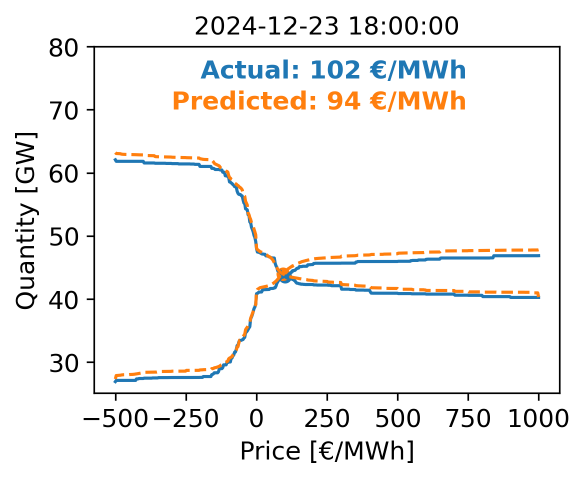}
        \caption{\textbf{ZST-fARX}}
    \end{subfigure}
    \hfill
    \begin{subfigure}[b]{0.32\textwidth}
        \includegraphics[width=\textwidth]{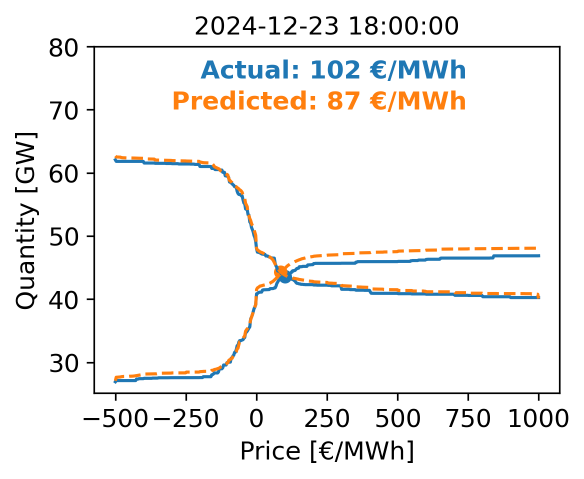}
        \caption{\textbf{ZST-fVARX}}
    \end{subfigure}
    \hfill
    \begin{subfigure}[b]{0.32\textwidth}
        \includegraphics[width=\textwidth]{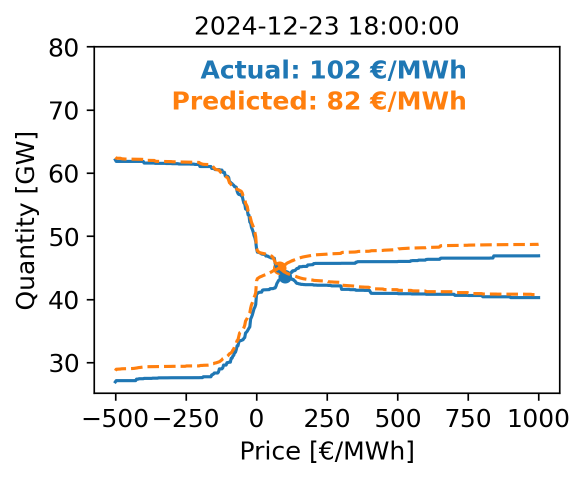}
        \caption{\textbf{FPCA-ARX}}
    \end{subfigure}
    
    \vspace{0.3cm}
    
    \hfill
    \begin{subfigure}[b]{0.32\textwidth}
        \includegraphics[width=\textwidth]{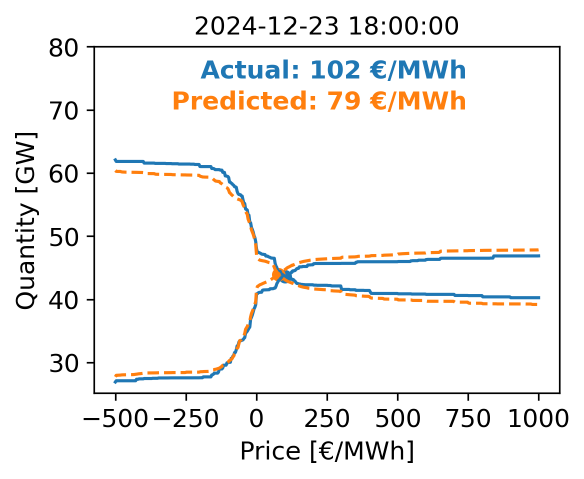}
        \caption{\textbf{FPCA-VARX}}
    \end{subfigure}
    \begin{subfigure}[b]{0.32\textwidth}
        \includegraphics[width=\textwidth]{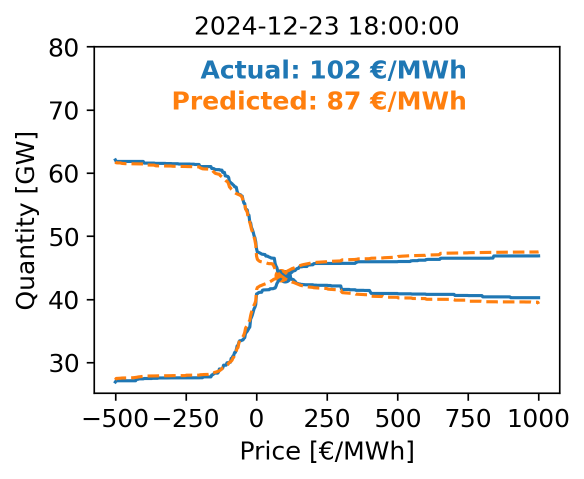}
        \caption{\textbf{FPCA-fARX}}
    \end{subfigure}
    \hfill
    \begin{subfigure}[b]{0.32\textwidth}
        \includegraphics[width=\textwidth]{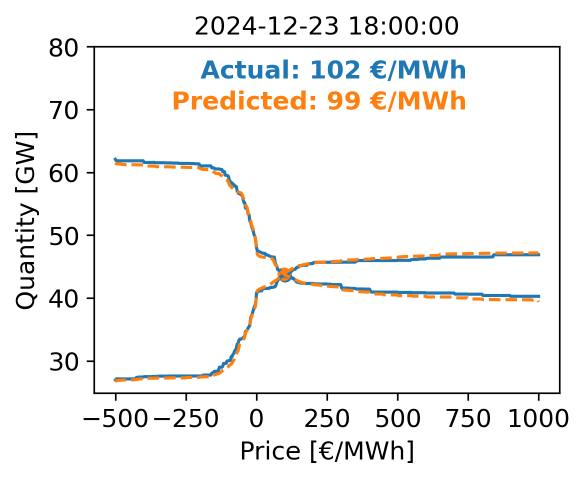}
        \caption{\textbf{FPCA-fVARX}}
    \end{subfigure}
    \caption{\rev{(\textit{Color optional}) Curves and clearing price predictions for the 18:00-19:00 interval of December 23\textsuperscript{rd}, 2024. (\textbf{EPEX-DE-LU})}}
    \label{fig:preds_epex-de-lu}
\end{figure}

\begin{figure}[h]
    \centering
    \begin{subfigure}[b]{0.32\textwidth}
        \includegraphics[width=\textwidth]{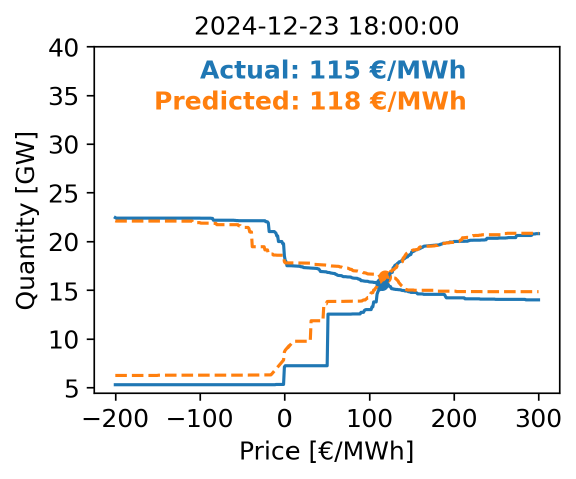}
        \caption{\textbf{Naive}}
    \end{subfigure}
    \begin{subfigure}[b]{0.32\textwidth}
        \includegraphics[width=\textwidth]{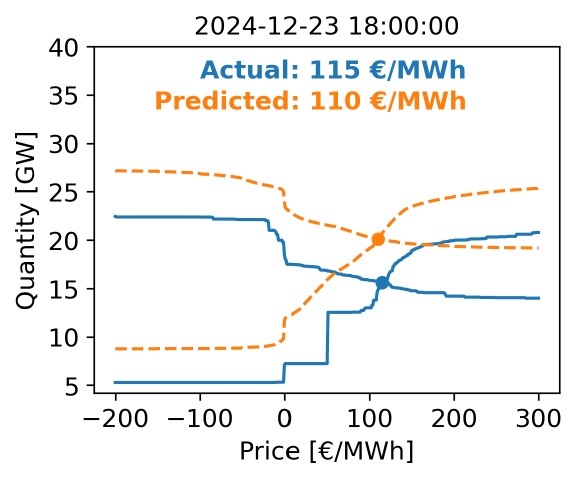}
        \caption{\textbf{ZST-ARX}}
    \end{subfigure}
    \begin{subfigure}[b]{0.32\textwidth}
        \includegraphics[width=\textwidth]{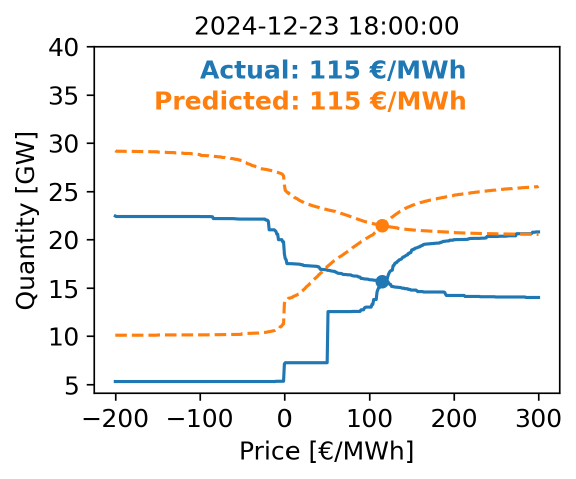}
        \caption{\textbf{ZST-VARX}}
    \end{subfigure}

    \vspace{0.3cm}
        
    \hfill
    \begin{subfigure}[b]{0.32\textwidth}
        \includegraphics[width=\textwidth]{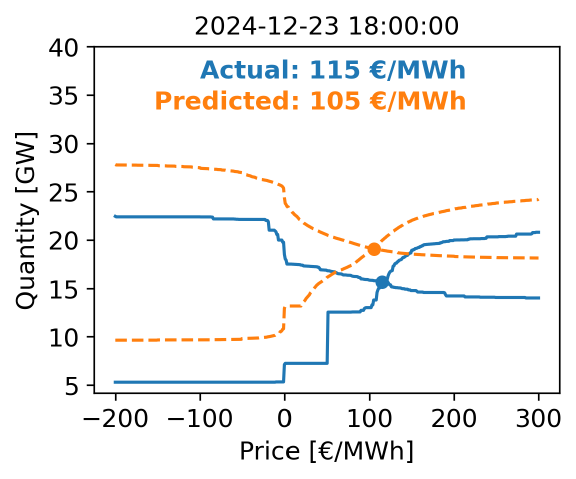}
        \caption{\textbf{ZST-fARX}}
    \end{subfigure}
    \hfill
    \begin{subfigure}[b]{0.32\textwidth}
        \includegraphics[width=\textwidth]{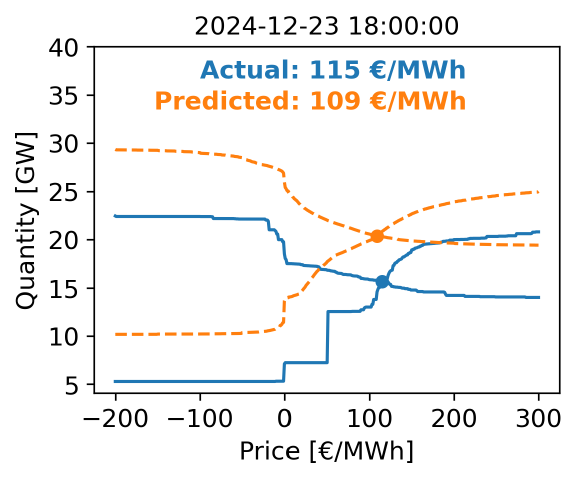}
        \caption{\textbf{ZST-fVARX}}
    \end{subfigure}
    \hfill
    \begin{subfigure}[b]{0.32\textwidth}
        \includegraphics[width=\textwidth]{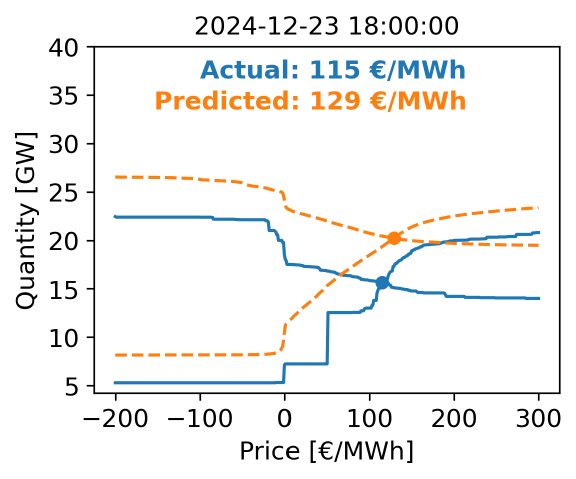}
        \caption{\textbf{FPCA-ARX}}
    \end{subfigure}
    
    \vspace{0.3cm}
    
    \hfill
    \begin{subfigure}[b]{0.32\textwidth}
        \includegraphics[width=\textwidth]{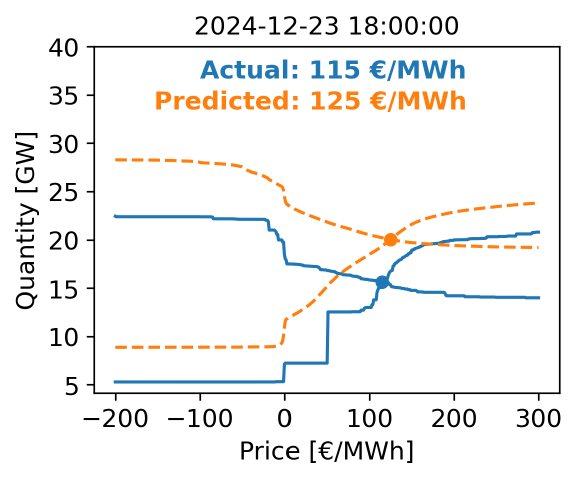}
        \caption{\textbf{FPCA-VARX}}
    \end{subfigure}
    \begin{subfigure}[b]{0.32\textwidth}
        \includegraphics[width=\textwidth]{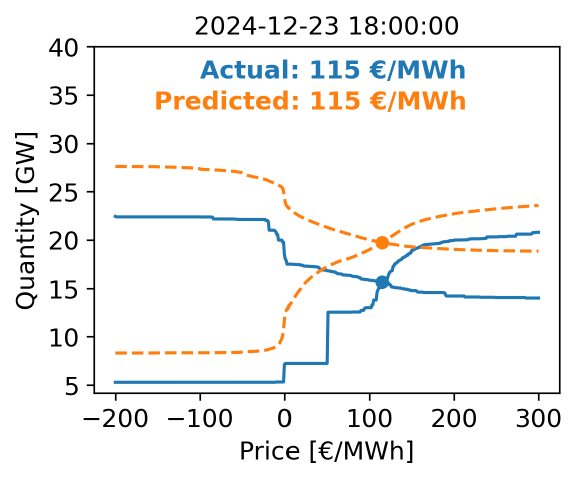}
        \caption{\textbf{FPCA-fARX}}
    \end{subfigure}
    \hfill
    \begin{subfigure}[b]{0.32\textwidth}
        \includegraphics[width=\textwidth]{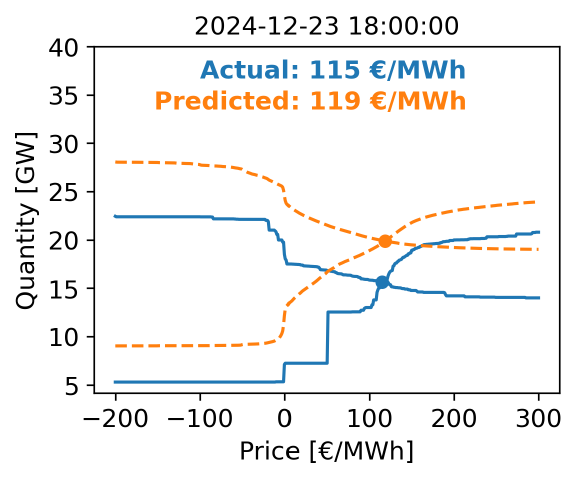}
        \caption{\textbf{FPCA-fVARX}}
    \end{subfigure}
    \caption{\rev{(\textit{Color optional}) Curves and clearing price predictions for the 18:00-19:00 interval of December 23\textsuperscript{rd}, 2024. (\textbf{EPEX-FR})}}
    \label{fig:preds_epex-fr}
\end{figure}

\end{document}